\documentclass[10pt]{article}
\usepackage{amssymb}
\usepackage{color}

{\hspace*{\fill}$\rule{.3\baselineskip}{.35\baselineskip}$\end{trivlist}}

\makeatletter
\@addtoreset{equation}{section}
\makeatother

\newfam\bifam
\font\tenbi=cmmib10 scaled \magstep1
\font\sevenbi=cmmib10 at 11pt
\font\fivebi=cmmib10 at 6pt
\textfont\bifam = \tenbi
\scriptfont\bifam= \sevenbi
\scriptscriptfont\bifam= \fivebi

\usepackage[dvips]{epsfig}
\usepackage{graphicx}

\begin{document}

\title{\bf Spinor Bose-Einstein condensates in double well potentials}

\author{C. Wang$^1$, P.G.\ Kevrekidis$^1$, N. Whitaker$^1$, \\
  T.J. Alexander$^2$,  D.J. Frantzeskakis$^3$ and P. Schmelcher$^{4,5}$ \\
$^1$ Department of Mathematics and Statistics, University of Massachusetts, \\
Amherst MA 01003-4515, USA \\
$^2$ Nonlinear Physics Center,  Research School of Physical Sciences \\
and
Engineering, Australian National University, Canberra ACT 0200,
Australia \\
$^3$ Department of Physics, University of Athens, Panepistimiopolis, \\
Zografos, Athens 157 84, Greece \\
$^4$ Theoretische Chemie, Institut f\"ur Physikalische Chemie, Im Neuenheimer Feld 229, \\
Universit\"at Heidelberg, 69120 Heidelberg, Germany\\
$^5$ Physikalisches Institut, Philosophenweg 12, Universit\"at Heidelberg\\
69120 Heidelberg, Germany\\
}






\date{\today}
\maketitle

\begin{abstract}
We consider the statics and dynamics of $F=1$ spinor Bose-Einstein 
condensates (BECs) confined in 
double well potentials. We use  
a two-mode Galerkin-type quasi-analytical approximation 
to describe the stationary states of the system. This way, 
we are able to obtain not only earlier results based 
on the {\it single mode approximation} (SMA) frequently used in studies of spinor BECs, 
but also additional 
modes that involve either two or all three spinor components of the $F=1$ spinor BEC.
The results based on this Galerkin-type decomposition  
are in good agreement with the analysis of the full system. 
We subsequently analyze the stability of these 
multi-component states, as well as their dynamics when we find them
to be unstable. The instabilities of the symmetric or anti-symmetric
states exhibit symmetry-breaking and recurrent asymmetric 
patterns. Our results yield qualitatively similar bifurcation diagrams both 
for polar (such as $^{23}$Na) and ferromagnetic (such as $^{87}$Rb) spinor BECs.
\end{abstract}


\section{Introduction}

In the past few years, there has been a remarkable amount
of progress in the study of Bose-Einstein condensates (BECs)
\cite{book1,book2} and an intense investigation of the localized 
nonlinear 
states that can be formed therein \cite{ourbook}. 
%
%
From the nonlinear dynamics point of view,  
one of the features that makes this 
system a particularly relevant and interesting one 
is associated with the presence
of a diverse range of external potentials which are used to
trap the atoms magnetically, optically or electrically (or
through combinations thereof) \cite{book1,book2,ourbook}.
Hence, monitoring the existence, stability and dynamical properties
of nonlinear localized modes within these diverse potentials has become
a principal theme of research effort, especially within the
realm of the prototypical mean-field model, namely the
Gross-Pitaevskii (GP) equation (which is a variant of the 
nonlinear Schr{\"o}dinger (NLS) equation extensively used 
in nonlinear optics \cite{kivshar}).

While most of the work in BECs has centered around one-component
systems, more recently
far-off-resonant optical techniques for trapping ultracold
atomic gases \cite{ket0}, 
have produced an intense focus on 
the study of {\it spinor} BECs \cite{ket1,cahn}, in which the spin 
degree of freedom (frozen in magnetic traps) emerges.
In this context, various phenomena absent in single-component
BECs may arise  
in the dynamical evolution of multi-component spinor condensates \cite{kronjaeger},  
including the formation of spin domains \cite{spindomain} and spin 
textures \cite{spintext},  
spin-mixing dynamics \cite{pulaw}, dynamic fragmentation \cite{yousun}, 
or dynamics of quantum phases \cite{yiyou}.
Moreover, macroscopic nonlinear states in the form of multi-component vector solitons 
of the bright \cite{wad1,boris,zh}, 
dark \cite{wad2} 
and gap \cite{dabr} 
type, along with 
more complex structures, such as bright-dark soliton 
complexes \cite{ourbd} and domain walls \cite{ourdw}, have also been studied.

On the other hand, as indicated above, one of the most attractive
traits of such ultracold systems is the possibility of experimental
realization of different potentials. Among them, 
one that has drawn
considerable attention due to its fundamental nature
is the double-well potential. One of the prototypical
realizations thereof comes from combining a strong harmonic trap
with a periodic lattice 
\cite{markus1}. 
In this context, 
the study of BECs loaded in double well potentials 
allows for the investigation of a variety of fundamental phenomena, 
including Josephson oscillations and
tunneling for a small number of atoms, or macroscopic quantum
self-trapping and an asymmetric partition of the atoms between
the wells for sufficiently large numbers of atoms \cite{markus1}.
Double well potentials have also spurred numerous theoretical insights
including, among others, finite-mode reductions, analytical results for
specially designed shapes of the potential and quantum depletion effects
\cite{smerzi,kiv2,mahmud,bam,Bergeman_2mode,infeld,todd,george,carr}.
It should be noted that such potentials have also been studied in 
the context of nonlinear optics, with relevant experimental results appearing, 
e.g., in twin-core self-guided laser beams in Kerr 
media \cite{HaeltermannPRL02} and optically
induced dual-core waveguiding structures in photorefractive crystals 
\cite{zhigang}.

The principal scope of the present work is to combine these two interesting 
developments, namely to examine spinor BECs 
in the presence of double well potentials. 
In particular, our aim is to present a systematic classification
of the states that are possible for $F=1$ spinor condensates 
confined in a double well potential. 
This classification is formulated on the basis of a two-mode Galerkin-type 
approximation of the stationary states of the system, which involves 
the decomposition of the spatial part of the solutions into a linear combination of the eigenfunctions
of the underlying linear operator; the relevant two-mode reduction used
here is in the spirit of Galerkin truncations in finite element and
similar methods (see, e.g., Ref. \cite{becker}). Such an approach 
can provide a detailed analytical handle on 
the dynamics of this multi-component BEC system (see, e.g., the earlier
work on one-component 
\cite{george} and two-component BECs \cite{chenyu} in 
double well potentials). 
%
%
It should be noted here that such 
attempts have been made before, most notably in 
Refs. \cite{liyou1} 
and \cite{liyou2}:  
the first of these works examined magnetization oscillations and beats
among other dynamical states, while the second one 
(which provided a description beyond mean-field theory) examined
quantum entanglement and pseudo-spin-squeezing properties.
However, these works were constrained within the commonly used {\it single
mode approximation} (SMA) 
for spinor condensates (see, e.g., \cite{pulaw,ofy2,osgur}) in  
which 
the wavefunction of each of the three components 
is taken to be a time-dependent multiple of a single 
stationary state.  
One of the particularly interesting features in the present setting
is that the two-mode Galerkin-type approximation reveals a considerable
wealth of states, many of which can not be described in the framework of SMA. 
In particular, as we will see below, there is a multitude of states 
which are symmetric in some of the components, while they are anti-symmetric 
in others, and out of these states also bifurcate further states which are asymmetric 
(but often in a non-single-mode way) in the different components.
Many of these novel states have a good chance to be observed 
in experiments with 
quasi one-dimensional (1D) spinor BECs
since, by performing their linear stability
analysis, we find them to be stable. Finally, we examine the
dynamical evolution of both two-component and three-component states 
that we find to be unstable; 
we observe that the manifestation of the associated instabilities is
exhibited typically through symmetry-breaking between the components
and is also associated with recurring  asymmetric patterns.

The paper is structured as follows. In 
section 2, we 
present the model, and provide the analytical approach. In section 3,
we present our numerical results. In particular,
we obtain the complete bifurcation diagram of the possible stationary states,
both for the full three-component system, as well as for the two-mode 
Galerkin-type approximation 
(in very good agreement between the two). In that section, we also 
illustrate the spatial profiles of the nonlinear modes arising in the
spinor 
condensate and quantify their stability. Lastly,
we corroborate these stability predictions by performing dynamical
simulations of the unstable modes. Finally, in section 4, we summarize our
findings and present our conclusions, as well as some topics for future 
study.

\section{The model and the analytical approach}


In the framework of mean-field theory, the wavefunctions $\psi_{\pm 1,0}(x,t)$ of 
the three hyperfine components ($m_{F}= \pm 1,0$) 
of a quasi-1D spinor $F=1$ condensate is governed by the following system 
of coupled normalized GP equations \cite{dabr}:
\begin{eqnarray}
i \partial_t \psi_{\pm 1} &=& {\cal L} \psi_{\pm 1}
+ \nu_s S \psi_{\pm 1}  + \nu_a (S-2 |\psi_{\mp 1}|^2) \psi_{\pm 1}
+ \nu_a \psi_0^2 \psi_{\mp 1}^{*}
\label{eq012}
\\
i \partial_t \psi_0 &=& {\cal L} \psi_{0}
+ \nu_s S \psi_{0}  + \nu_a (S-|\psi_{0}|^2) \psi_{0}
+ 2 \nu_a \psi_0^{*} \psi_{1} \psi_{-1}.
\label{eq0}
\end{eqnarray}
In these expressions $S= |\psi_{-1}|^2 + |\psi_0|^2 + |\psi_1|^2$ represents the total normalized 
density, while 
the coupling coefficients $\nu_s$
and $\nu_a$ represent, respectively, the symmetric spin-independent
and the antisymmetric spin-dependent interaction 
strengths (see, e.g., \cite{dabr} for details).  Additionally, 
\begin{eqnarray}
{\cal L} \psi_j=-\frac{1}{2} \partial_x^2 \psi_j + V(x) \psi_j ,
\label{eq3} 
\end{eqnarray}
($j=-1,0,1$) represents the 
single-particle operator with a confining potential $V(x)$ assumed to 
be of the form
\begin{eqnarray}
V(x)=\frac{1}{2} \Omega^2 x^2 + V_0 {\rm sech}^2(x/w). 
\label{eq4}
\end{eqnarray}
This is 
a double well potential consisting of a parabolic 
potential of strength $\Omega$ 
emulating, e.g., a 
usual harmonic trap, and a localized barrier potential of strength $V_0$ and width $w$, 
representing, e.g., a blue-detuned laser beam that repels atoms from 
the harmonic trap center. Notice that 
the main qualitative features of our results (i.e., the nature of
the bifurcation diagram near the linear limit, 
the emergence of the symmetry-breaking
asymmetric states, and their dynamical manifestations) 
do not rely on the specific form of the double well potential and
should be expected to arise more broadly within this class of models (spinor
condensates in double well potentials). 

We now seek stationary solutions of Eqs. (\ref{eq012})-(\ref{eq0}) in the form 
$\psi_j=u_j \exp(-i \mu_j t) \exp(i \theta_j)$, where $\mu_j$ and $\theta_j$ 
correspond, respectively, to the chemical potentials and 
phases of the wavefunctions 
obeying the usual constraints (see, e.g., \cite{ofy2,ourdw}), namely, 
$2 \mu_0=\mu_1 + \mu_{-1}$ 
and $\Delta \theta=2 \theta_0 - (\theta_1+\theta_{-1})=0$ or $\pi$.
These are the so-called phase matching conditions relevant for 
stationary states typically in systems with parametric interactions;
see, e.g., Ref. \cite{kiv_optics} for the case of four-wave mixing 
in nonlinear optics.
This ansatz results in the following 
equations for the stationary states $u_j$:
\begin{eqnarray}
\mu_{\pm 1} u_{\pm 1}
&=& {\cal L} u_{\pm 1} + \nu_s S u_{\pm 1} + \nu_a
(S- 2 u_{\mp 1}^2) u_{\pm 1} \pm \nu_a u_0^2 u_{\mp 1}, 
\label{eq5}
\\
\mu_{0} u_{0}
&=& {\cal L} u_{0} + \nu_s S u_{0} + \nu_a
(S- u_{0}^2) u_{0} \pm 2 \nu_a u_0 u_{1} u_{-1},
\label{eq6}
\end{eqnarray}
where the $\pm$ sign in the last terms of the above equations corresponds, 
respectively, to the cases of $\Delta \theta=0$ and $\Delta \theta=\pi$.

The basic idea of the two mode Galerkin-type approximation 
\cite{kiv2,Bergeman_2mode,todd,george} that we will employ
to approximate the solutions of Eqs. (\ref{eq5})-(\ref{eq6})
is the following: 
we assume that in 
the vicinity of the linear limit of the system (and for appropriate
selection of the chemical potential) each of the states $u_{\pm 1,0}$
can be decomposed into a two-mode expansion, involving the 
symmetric ground state and the antisymmetric first excited state
of the underlying linear potential. We will denote those states
by $\phi_a$ and $\phi_b$, respectively, and accordingly decompose $u_j$ 
(where $j \in \{-1,0,1\}$) as follows,
\begin{eqnarray}
u_{j}(x,t) = c_a^{(j)} \phi_a(x)+ c_b^{(j)} \phi_b(x),
\label{eq7}
\end{eqnarray}
where $c_a^{(j)}$ and $c_b^{(j)}$ are unknown time-dependent complex prefactors. 
By substituting Eq. (\ref{eq7}) into 
Eqs. (\ref{eq5})-(\ref{eq6}), and upon projecting each of the three equations
in (\ref{eq5})-(\ref{eq6}) to $\phi_a$ and $\phi_b$, we obtain through tedious
but straightforward algebra, the following six nonlinear, {\it algebraic}
equations 
describing the stationary states of the original system \footnote{Note that 
since we are interested in stationary states herein, we will
consider $c_a^{(j)}$ and $c_b^{(j)}$ to be independent of time, 
although the expansion can also be used to probe the dynamics, by 
substituting the ansatz of
Eq. (\ref{eq7}) with time-dependent coefficients into Eqs. 
(\ref{eq012})-(\ref{eq0}) and projecting into 
$\phi_a$ and $\phi_b$.}
\begin{eqnarray}
\mu_{\pm 1} c_a^{(\pm 1)}
&=& \omega_a c_a^{(\pm 1)} + \nu_s \left[
\Gamma_a S_a c_a^{(\pm 1)} + \Gamma_{ab} S_b c_a^{(\pm 1)}
+\Gamma_{ab} S_{ab} c_b^{(\pm 1)} \right]
\nonumber
\\
&+& \nu_a \left[\nu_a (S_a-2 (c_a^{(\mp 1)})^2) c_a^{\pm 1}
+ \Gamma_{ab} (S_b-2 (c_b^{(\mp 1)})^2) c_a^{(\pm 1)}  + \Gamma_{ab} (S_{ab}
-4 c_a^{(\mp 1)} c_b^{(\mp 1)}) c_b^{(\pm 1)}  \right]
\nonumber
\\
&\pm& \nu_a \left[\Gamma_a (c_a^{(0)})^2 c_a^{(\mp 1)}
+ \Gamma_{ab} (c_b^{(0)})^2 c_a^{(\mp 1)} + 2 \Gamma_{ab} c_a^{(0)} 
c_b^{(0)} c_b^{(\mp 1)} \right],
\label{eq9} 
\\
\mu_{\pm 1} c_b^{(\pm 1)}
&=& \omega_b c_b^{(\pm 1)} + \nu_s \left[
\Gamma_b S_b c_b^{(\pm 1)} + \Gamma_{ab} S_a c_b^{(\pm 1)}
+\Gamma_{ab} S_{ab} c_a^{(\pm 1)} \right]
\nonumber
\\
&+& \nu_a \left[\Gamma_b (S_b-2 (c_b^{(\mp 1)})^2) c_b^{\pm 1}
+ \Gamma_{ab} (S_a-2 (c_a^{(\mp 1)})^2) c_b^{(\pm 1)} + \Gamma_{ab} (S_{ab}
-4 c_a^{(\mp 1)} c_b^{(\mp 1)}) c_a^{(\pm 1)}  \right]
\nonumber
\\
&\pm& \nu_a \left[\Gamma_b (c_b^{(0)})^2 c_b^{(\mp 1)}
+ \Gamma_{ab} (c_a^{(0)})^2 c_b^{(\mp 1)} + 2 \Gamma_{ab} c_a^{(0)} 
c_b^{(0)} c_a^{(\mp 1)} \right], 
\label{eq10} 
\\
\mu_{0} c_a^{(0)}
&=& \omega_a c_a^{(0)} + \nu_s \left[
\Gamma_a S_a c_a^{(0)} + \Gamma_{ab} S_b c_a^{(0)}
+\Gamma_{ab} S_{ab} c_b^{(0)} \right]
\nonumber
\\
&+& \nu_a \left[\Gamma_a (S_a-(c_a^{(0)})^2) c_a^{(0)}
+ \Gamma_{ab} (S_b-(c_b^{(0)})^2) c_a^{(0)}  + \Gamma_{ab} (S_{ab}
-2 c_a^{(0)} c_b^{(0)}) c_b^{(0)}  \right]
\nonumber
\\
&\pm& 2 \nu_a \left[\Gamma_a (c_a^{(0)}) c_a^{(1)} c_a^{(-1)}
+ \Gamma_{ab} (c_a^{(0)}) c_b^{(-1)} c_b^{(1)} + \Gamma_{ab} (c_a^{(-1)} 
c_b^{(1)} + c_b^{(-1)} c_a^{(1)}) c_b^{(0)} \right], 
\label{eq11} 
\\
\mu_{0} c_b^{(0)}
&=& \omega_b c_b^{(0)} + \nu_s \left[
\Gamma_b S_b c_b^{(0)} + \Gamma_{ab} S_a c_b^{(0)}
+\Gamma_{ab} S_{ab} c_a^{(0)} \right]
\nonumber
\\
&+& \nu_a \left[\Gamma_b (S_b-(c_b^{(0)})^2) c_b^{(0)}
+ \Gamma_{ab} (S_a-(c_a^{(0)})^2) c_b^{(0)}  + \Gamma_{ab} (S_{ab}
-2 c_a^{(0)} c_b^{(0)}) c_a^{(0)}  \right]
\nonumber
\\
&\pm& 2 \nu_a \left[\Gamma_b (c_b^{(0)}) c_b^{(1)} c_b^{(-1)}
+ \Gamma_{ab} (c_b^{(0)}) c_a^{(-1)} c_a^{(1)} + \Gamma_{ab} (c_a^{(-1)} 
c_b^{(1)} + c_b^{(-1)} c_a^{(1)}) c_a^{(0)} \right]. 
\label{eq12} 
\end{eqnarray}
In these equations, $\omega_{a,b}$ are the eigenvalues
that correspond to the linear eigenfunctions $\phi_{a,b}$, respectively, 
$S_a=\sum_j (c_a^{(j)})^2$, 
$S_b=\sum_j (c_b^{(j)})^2$, 
$S_{ab}=\sum_j 2 c_a^{(j)} c_b^{(j)}$,
while the coefficients 
$\Gamma_a=\int \phi_a^4 dx$, $\Gamma_b=\int \phi_b^4 dx$
and $\Gamma_{ab}=\int \phi_a^2 \phi_b^2 dx$ are constants that 
depend on the potential. 

By solving this algebraic set of equations, for appropriate 
parameters of the potential [e.g., in the normalization detailed below
(see Section 3) using the mean-field interaction energy, 
for $\Omega=0.2$, $V_0=1$
and $w=0.5$, $\omega_a=0.249$,
$\omega_b=0.315$, and $\Gamma_a=0.127$, $\Gamma_b=0.134$
and $\Gamma_{ab}=0.120$], we can extract all the potential
stationary states for a given combination of chemical
potentials in the two-parameter space $(\mu_0,\mu_1)$
[recall that $\mu_{-1}$ is fully determined by the above two
parameters]. Then, these can be used as 
initial guesses 
for identifying these solutions in the full nonlinear eigenvalue
problem of Eqs. (\ref{eq5})-(\ref{eq6}).

Some special cases of solutions 
can be found in the above algebraic equations. Since these
have been discussed in earlier works in one-component
systems (see, e.g., \cite{george}), two-component systems 
(see, e.g., \cite{chenyu}), and even in the full three-component
system but in the context of the SMA (see e.g., 
\cite{pulaw,dabr,ofy2,osgur}), we relegate the relevant detailed
analysis of these to Appendix A. Importantly, we note that
the principal building blocks of the single component setting
are a symmetric state, where only $c_a$ (of a particular component)
is non-zero, an anti-symmetric state with only $c_b \neq 0$,
as well as asymmetric states with both $c_a$ and $c_b \neq 0$.
The latter bifurcate beyond a critical point 
from the symmetric state if $\nu_s<0$, or from the antisymmetric
state if $\nu_s>0$ \cite{george}.

In the next section, 
we will offer the full set of solutions that can be constructed within our
two-mode Galerkin-type approximation and compare them to the detailed numerical
results of the original problem.

\section{Numerical Results}

Before proceeding further, it is necessary to provide here values of 
the physical parameters that we will use. First we note that  
for simplicity in our computations we use the rescaling of $\nu_s$ 
to $1$ and of $\nu_a$ to $\nu_a/\nu_s \equiv \delta$. 
Notice that in the relevant cases of the ferromagnetic $^{87}$Rb and 
polar $^{23}$Na spinor ($F=1$) condensates, the parameter $\delta$ takes 
the values $\delta =-4.66\times 10^{-3}$ \cite{kemp} and 
$\delta =+3.14\times 10^{-2}$ \cite{greene}, respectively. 
Under this rescaling, the connection between physical values and 
dimensionless parameters is as described in detail in Refs. 
\cite{ourbd,ourdw}. We will also fix the double well potential 
parameters to the values $\Omega=0.2$, $V_0=1$ and $w=0.5$ (other values 
lead to qualitatively similar results). Taking into regard that the 
normalized chemical potentials will take values $\mu_j \approx 0.5$, this choice 
may correspond, e.g., to a spinor condensate of sodium atoms  
confined in a highly asymmetric trap with frequencies $\omega _{\perp }=3\omega _{x}=2\pi 
\times 230~$Hz, with peak 1D density $n_{0}\simeq 10^{8}$ m$^{-1}$ and number 
of atoms of order $O(10^3)$. In this case, the time and space units in the numerical 
results that will be presented below are $1.2$ ms and $1.8~\mu $m, respectively.


\begin{figure}[tbhp!]
\centering
\includegraphics[width=.4\textwidth]{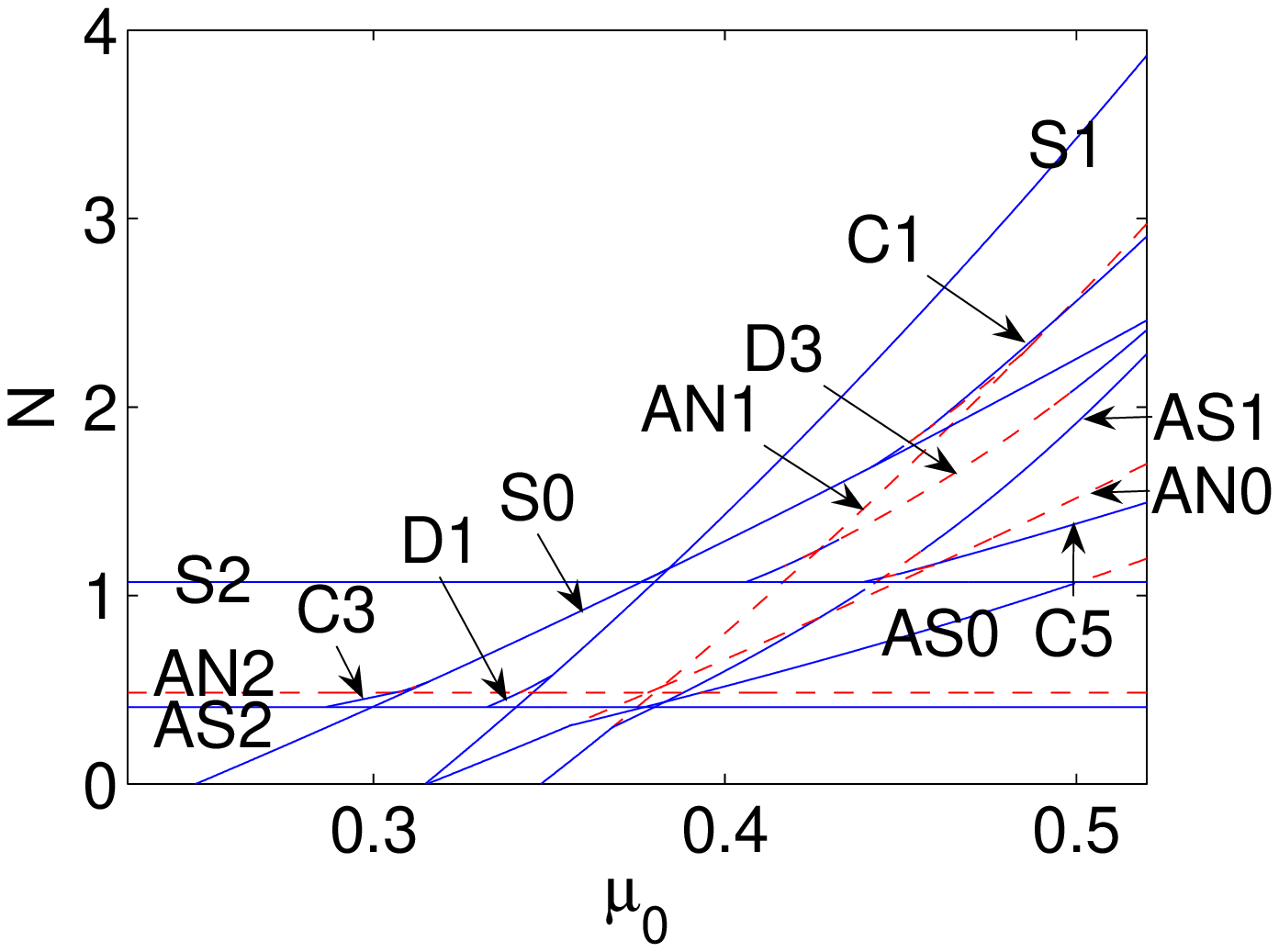}
\includegraphics[width=.4\textwidth]{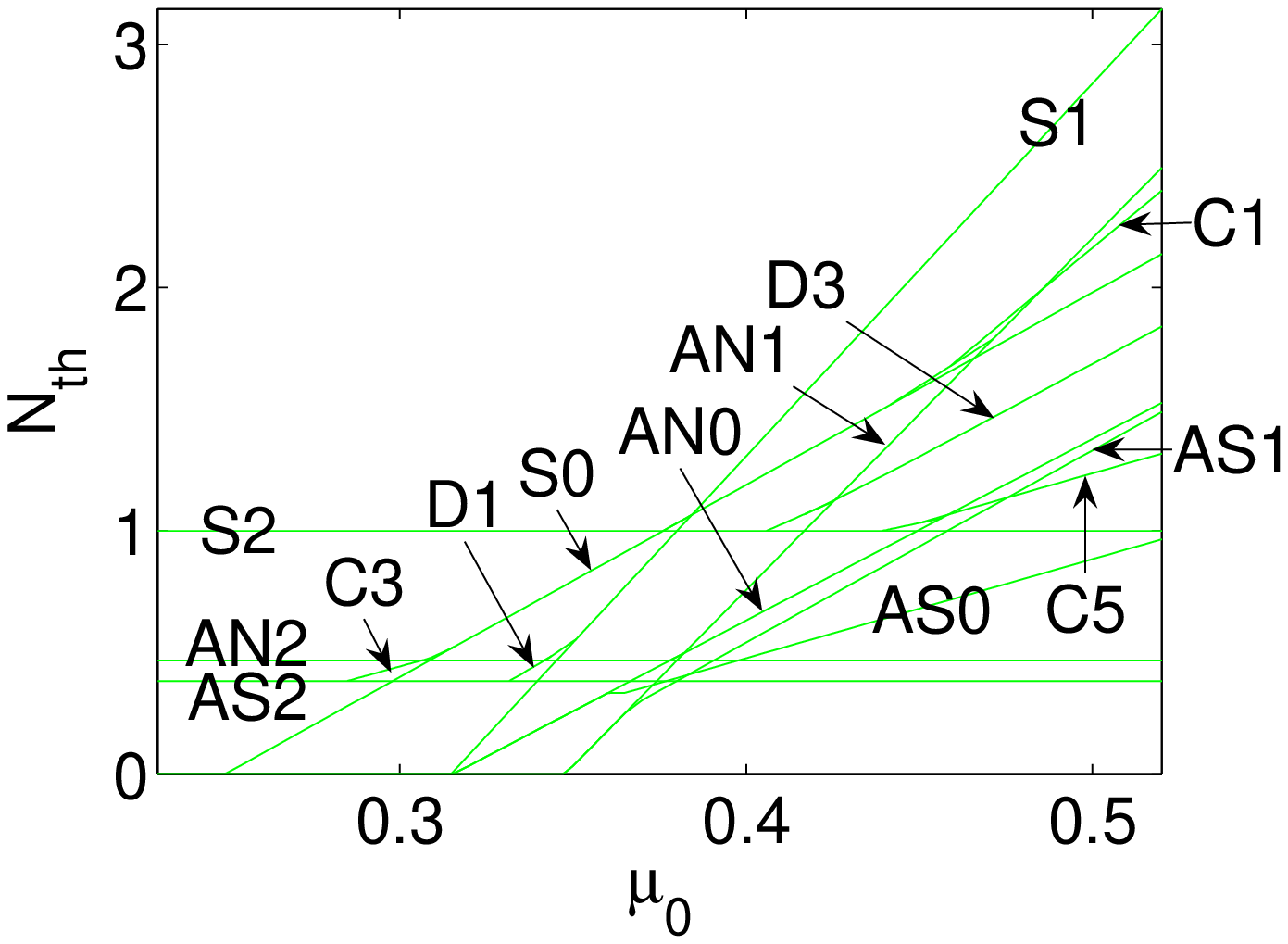}\\
\includegraphics[width=.24\textwidth]{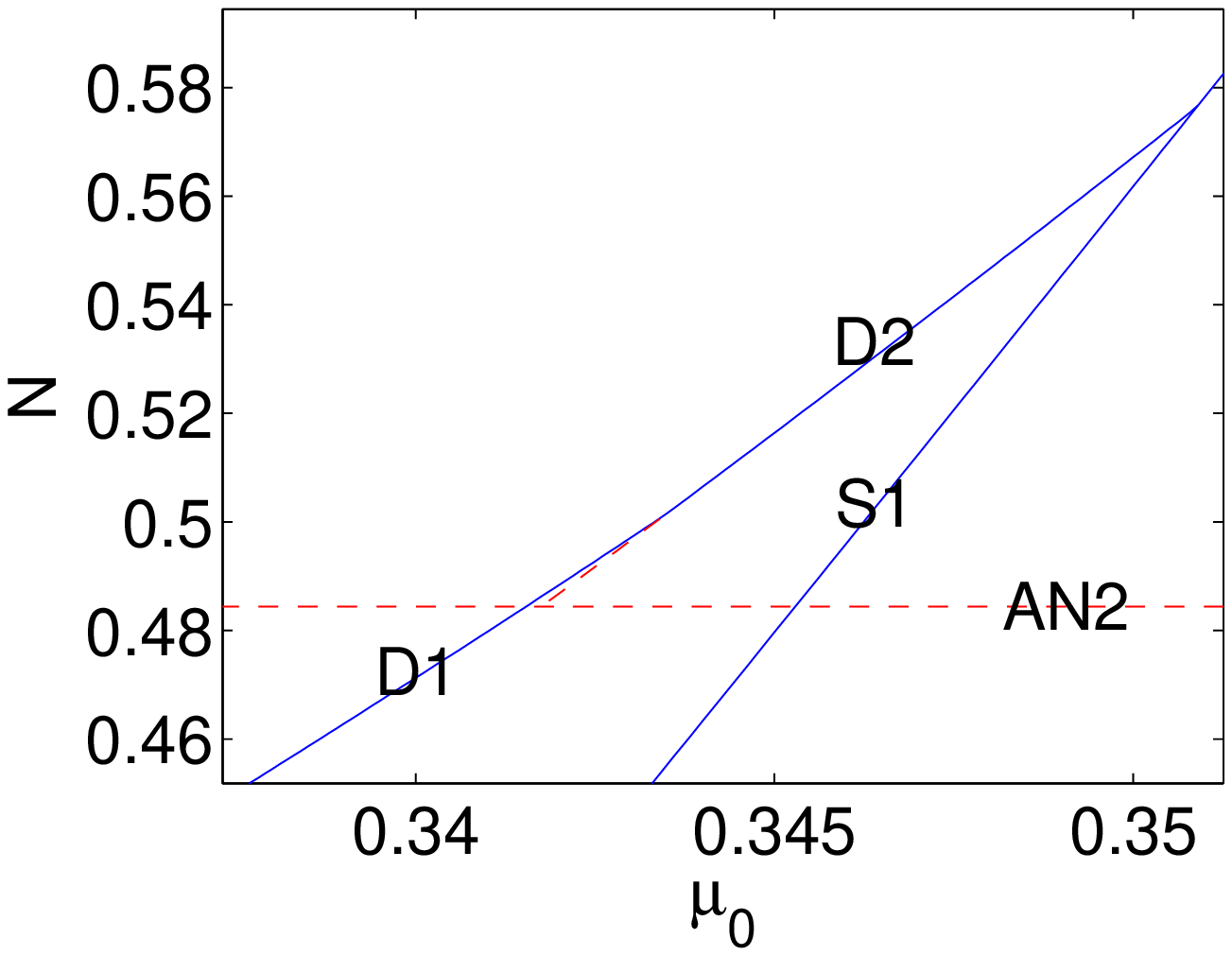}
\includegraphics[width=.24\textwidth]{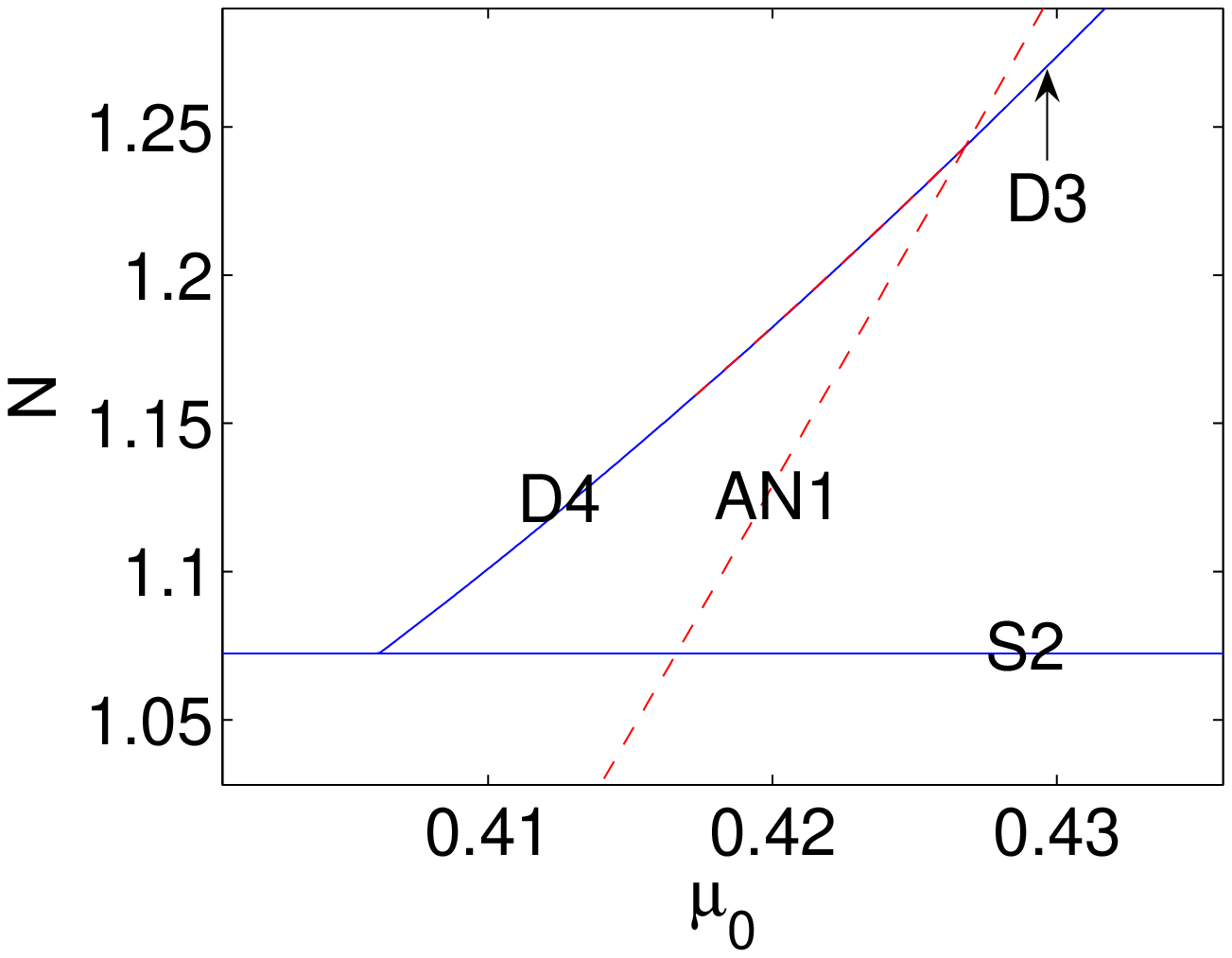}
\includegraphics[width=.24\textwidth]{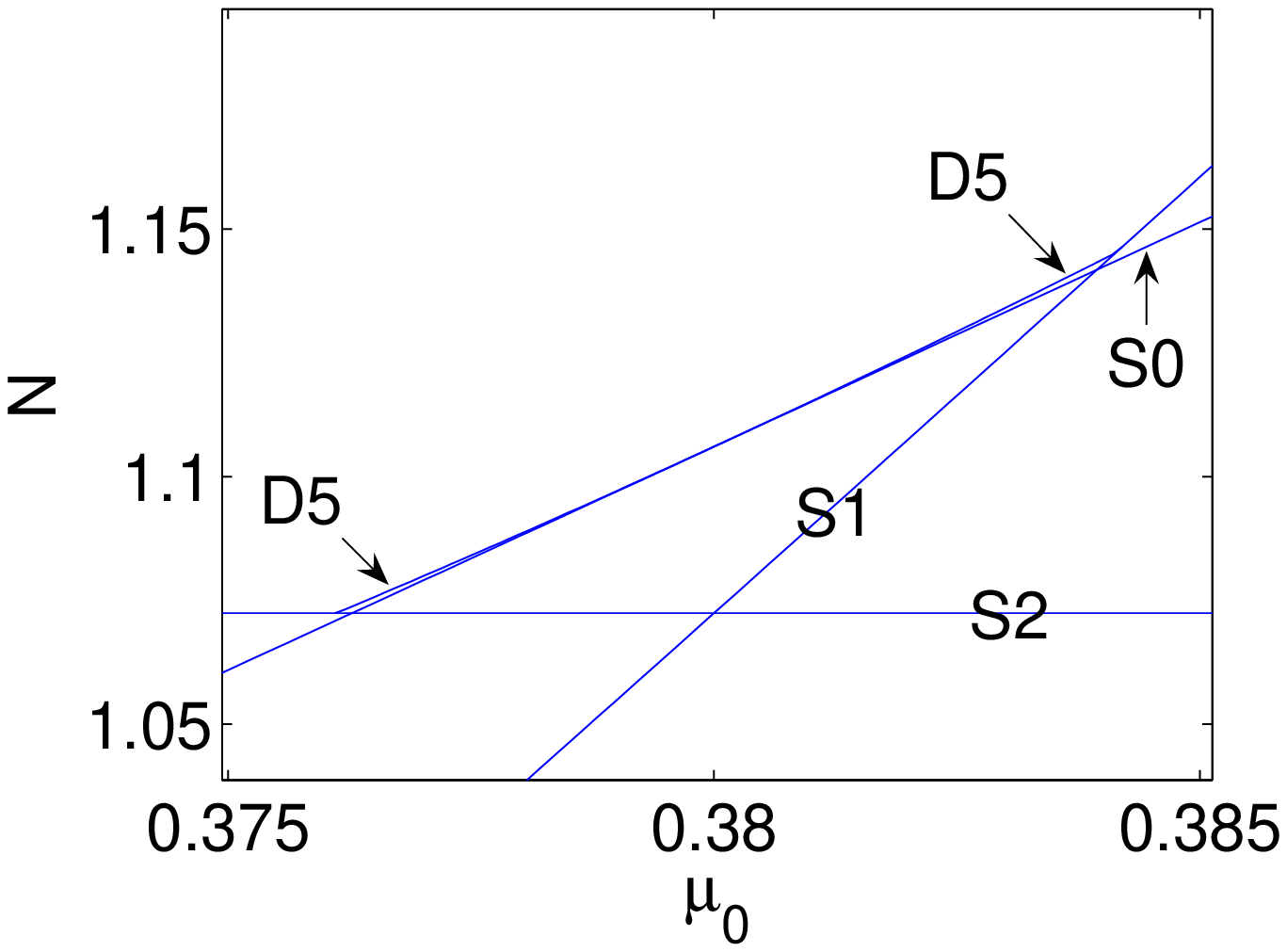}
\includegraphics[width=.24\textwidth]{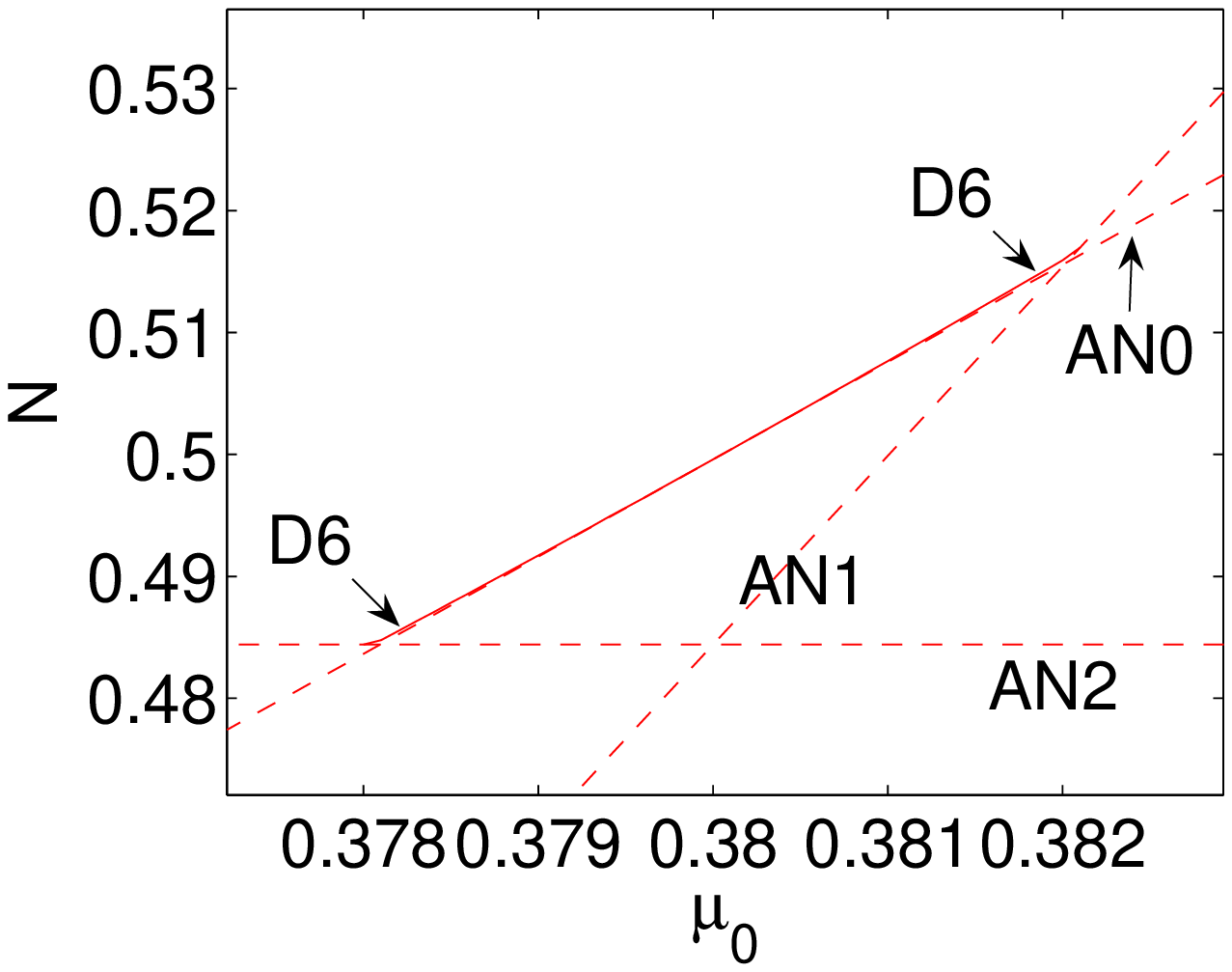}\\
\includegraphics[width=.24\textwidth]{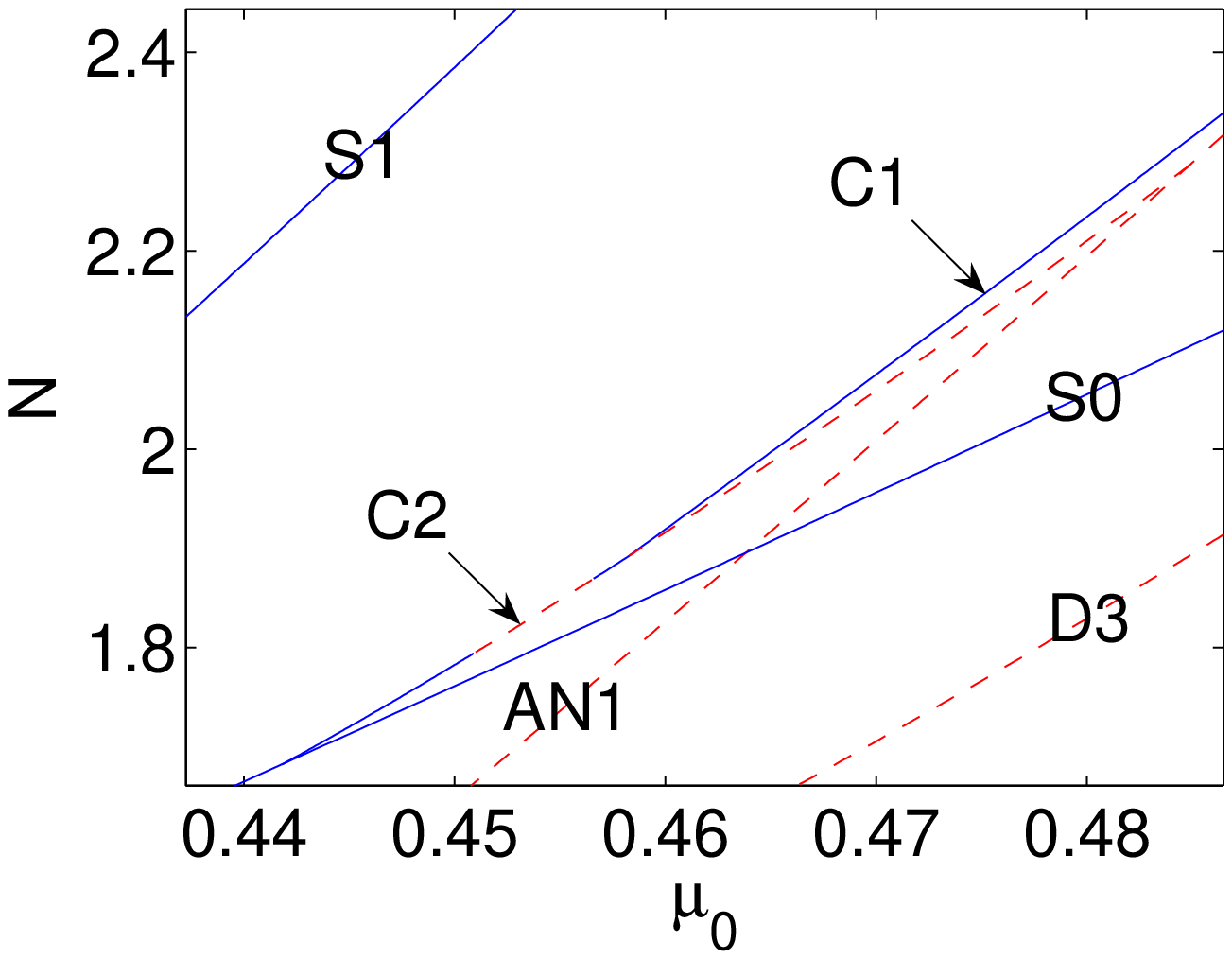}
\includegraphics[width=.24\textwidth]{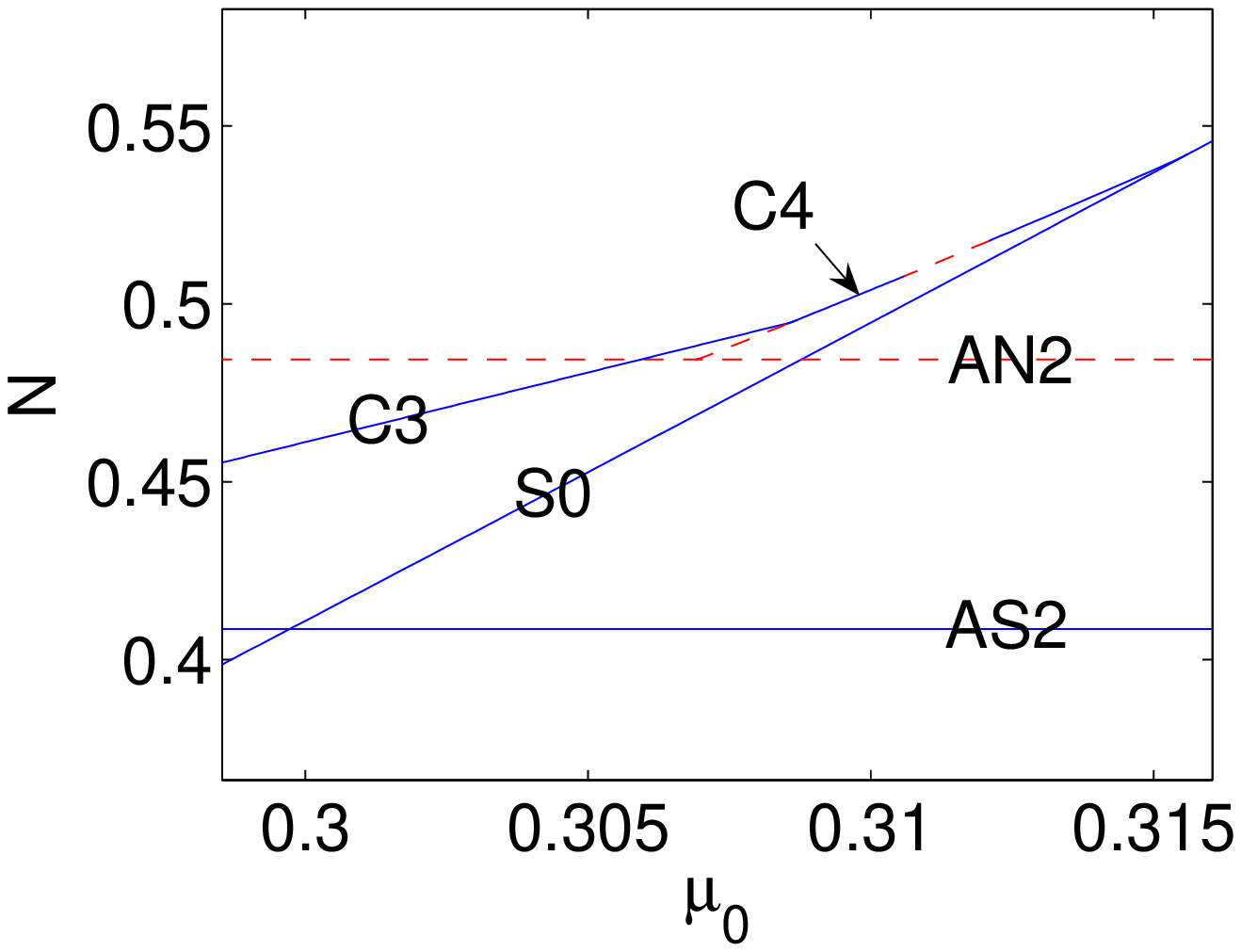}
\includegraphics[width=.24\textwidth]{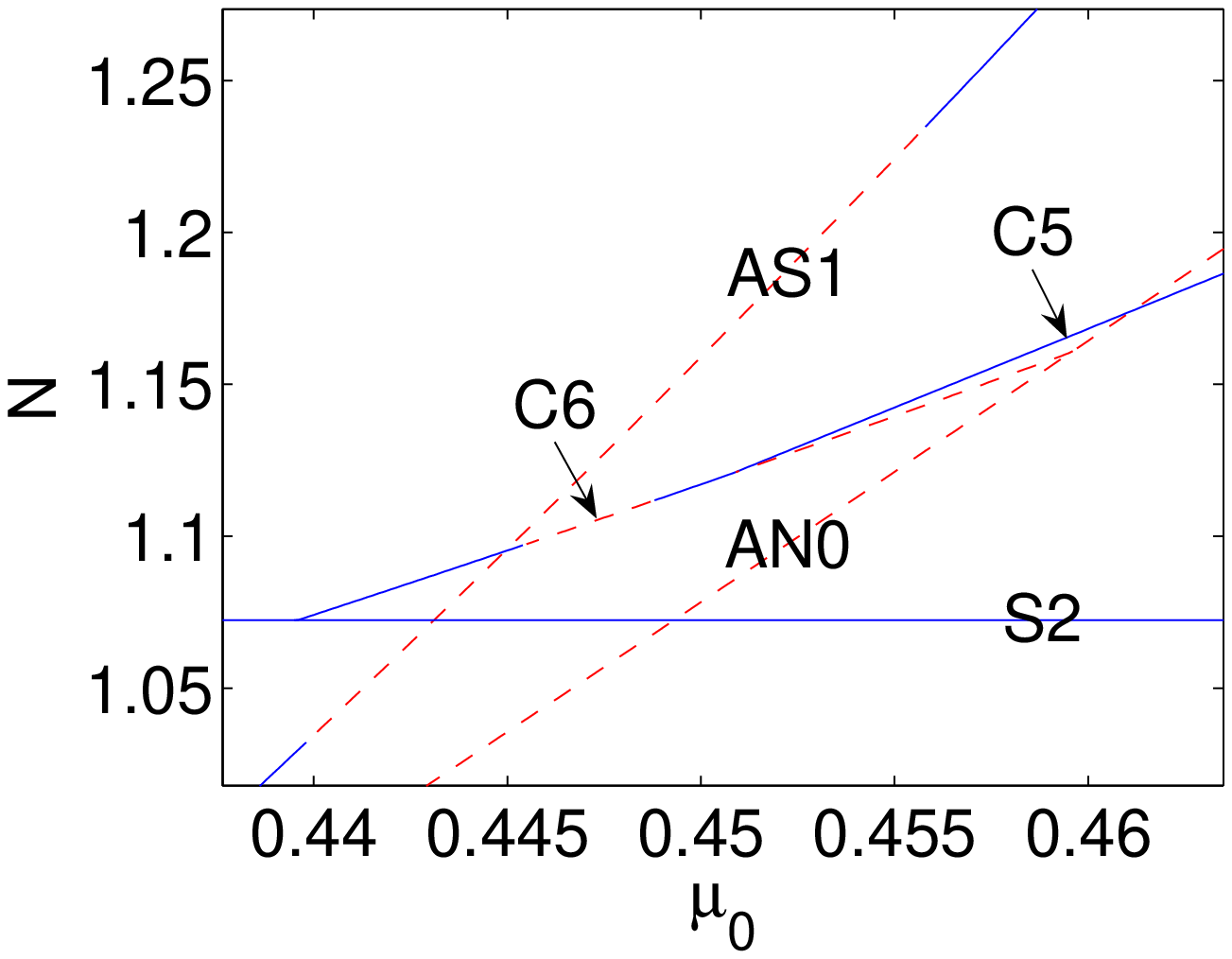}
\includegraphics[width=.24\textwidth]{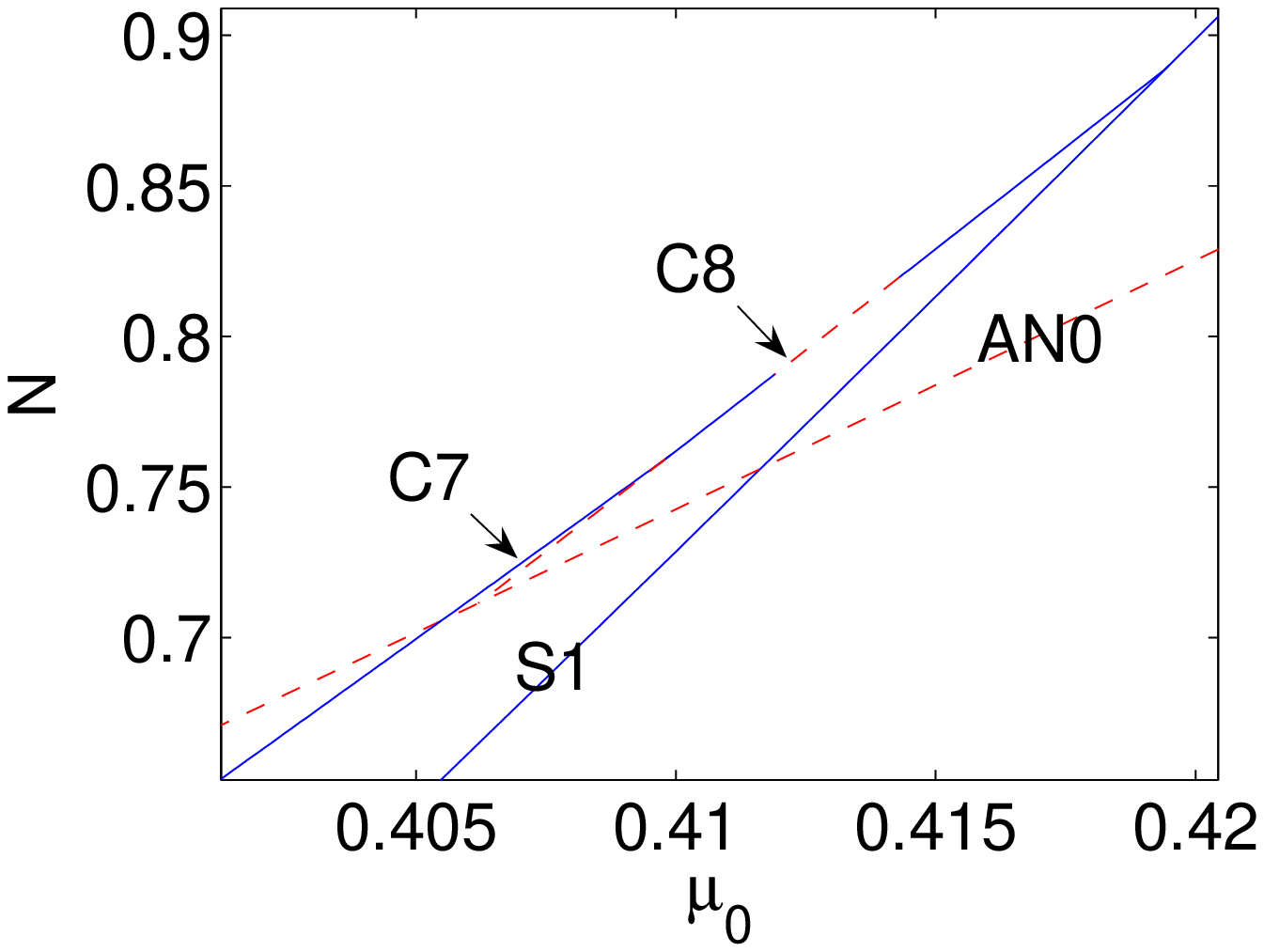}\\
\caption{(Color online) Top panels: the norm $N$ (normalized number of atoms) of the numerically found solutions 
of Eqs. (\protect\ref{eq012})-(\protect\ref{eq0}) (left) and their counterparts predicted by the two-mode 
Galerkin-type approximation (right) for the case of the $^{23}$Na spinor BEC, as a function of $\mu_0$ 
with fixed $\mu_{-1}=0.38$. Middle and bottom panels: blowups of segments in the top left panel 
where new combined solutions emerge at the relevant bifurcation points. 
All blowups, except the last one, are dedicated to the case of $\mu_{-1}=0.38$ 
as in the top left panel. The last one, showing the branches C7 and C8 (see text) 
corresponds to the case of $\mu_{-1}=0.48$ (as in the top left panel of Fig. \protect\ref{figEx} below). 
The branches D1 and D3 are generated from D2 and D4, respectively. 
Branches labeled by D (D1-D6 in the different panels of the figure)
correspond to waveforms with two nonzero components, $u_1$ and $u_{-1}$;
see the discussion of 
section \ref{two_comp} and Figs. \ref{figNaD34} -\ref{figNaD56} for more
details.
The branches C1, C3, C5 and C7 are generated from C2, C4, C6 and C8, 
respectively.  Branches labeled by C (C1-C8 in the different panels of the 
figure) correspond to waveforms with all three components nonzero;
see the discussion of 
section \ref{three_comp} and Figs. \ref{figNaC12}-\ref{figNaC56} for more
details.
Solid (blue) lines and dashed (red) lines denote stable and unstable solutions, respectively.}
\label{figNa}
\end{figure}

\begin{figure}[tbhp!]
\centering
\includegraphics[width=.3\textwidth]{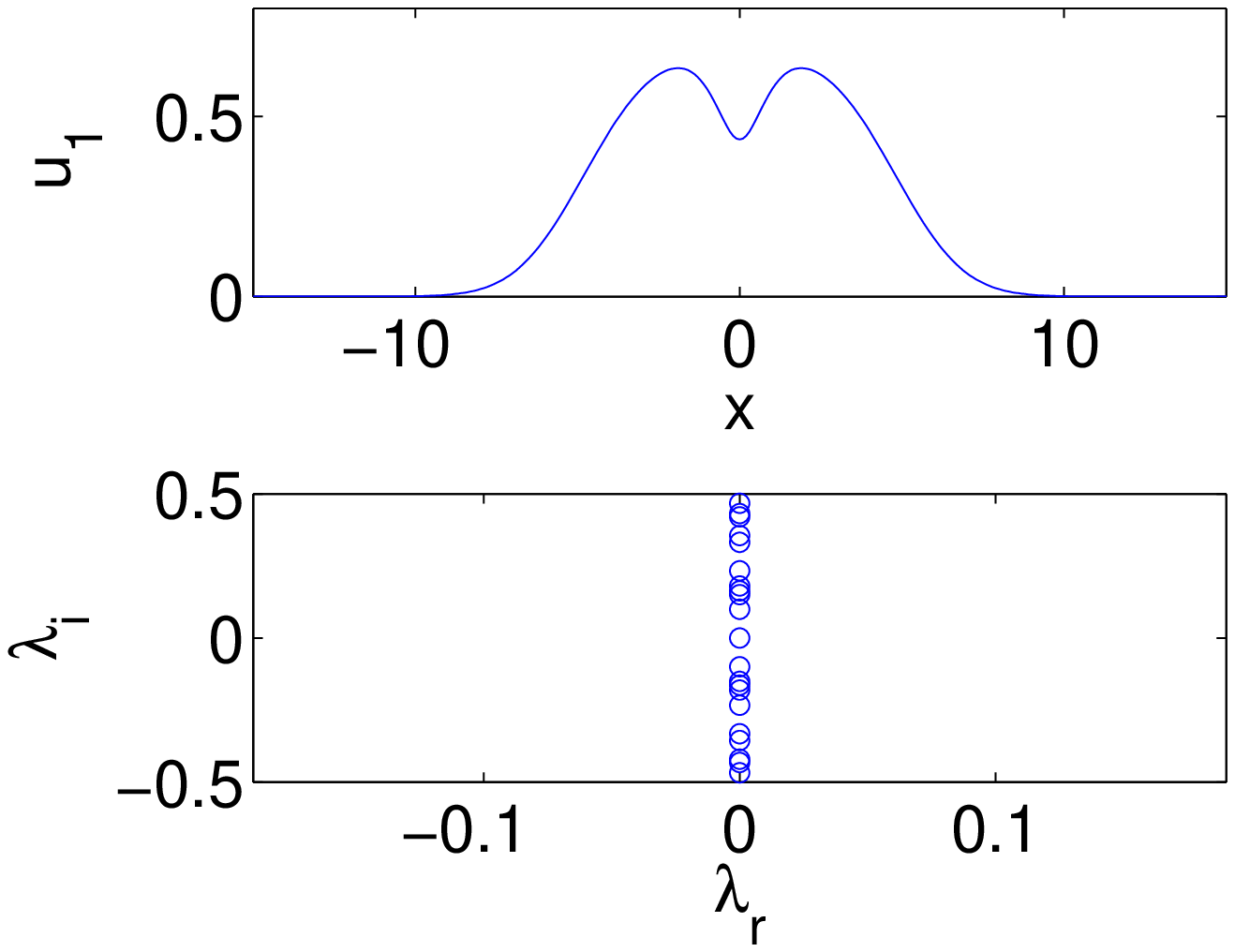}  
\includegraphics[width=.3\textwidth]{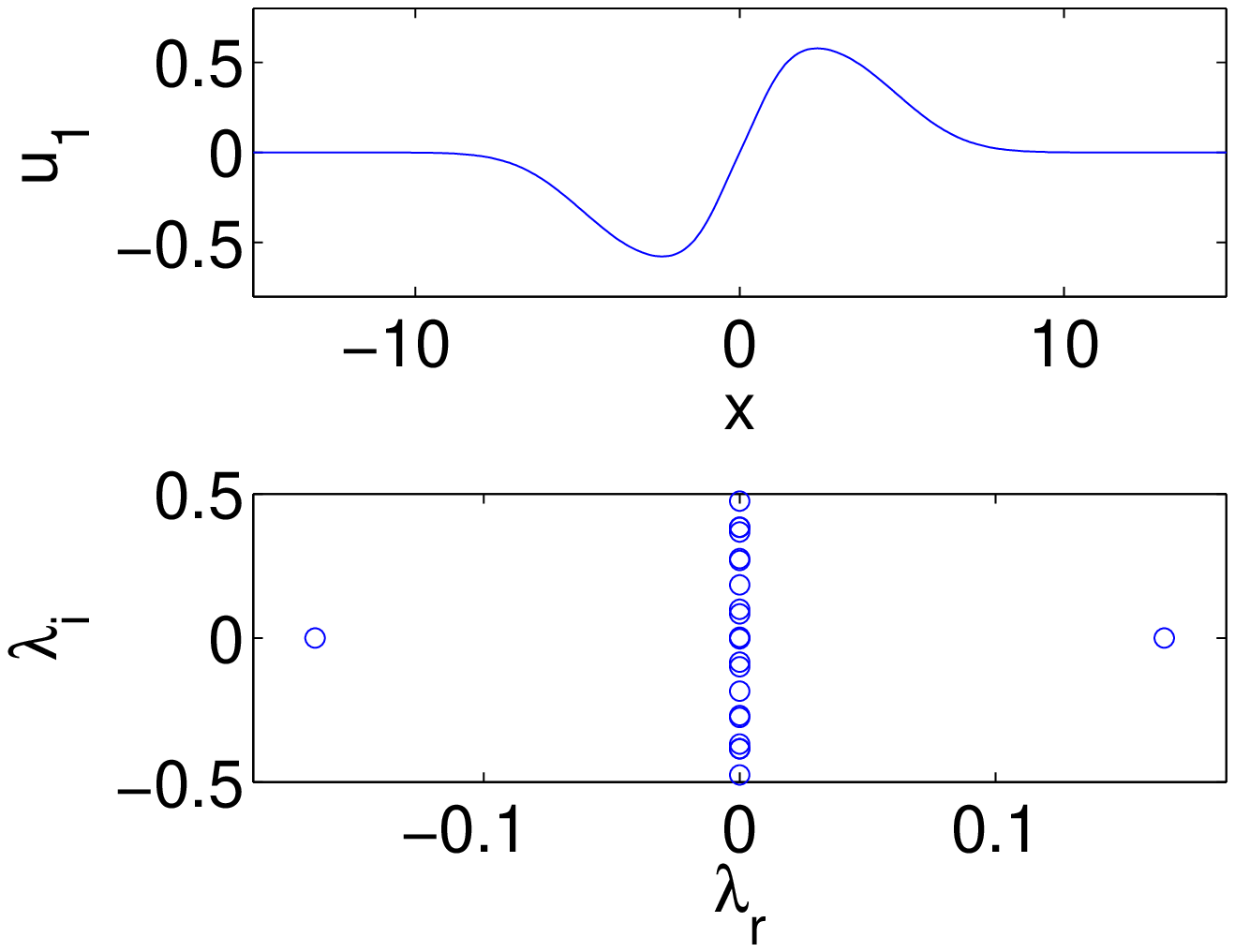}\\ 
\includegraphics[width=.3\textwidth]{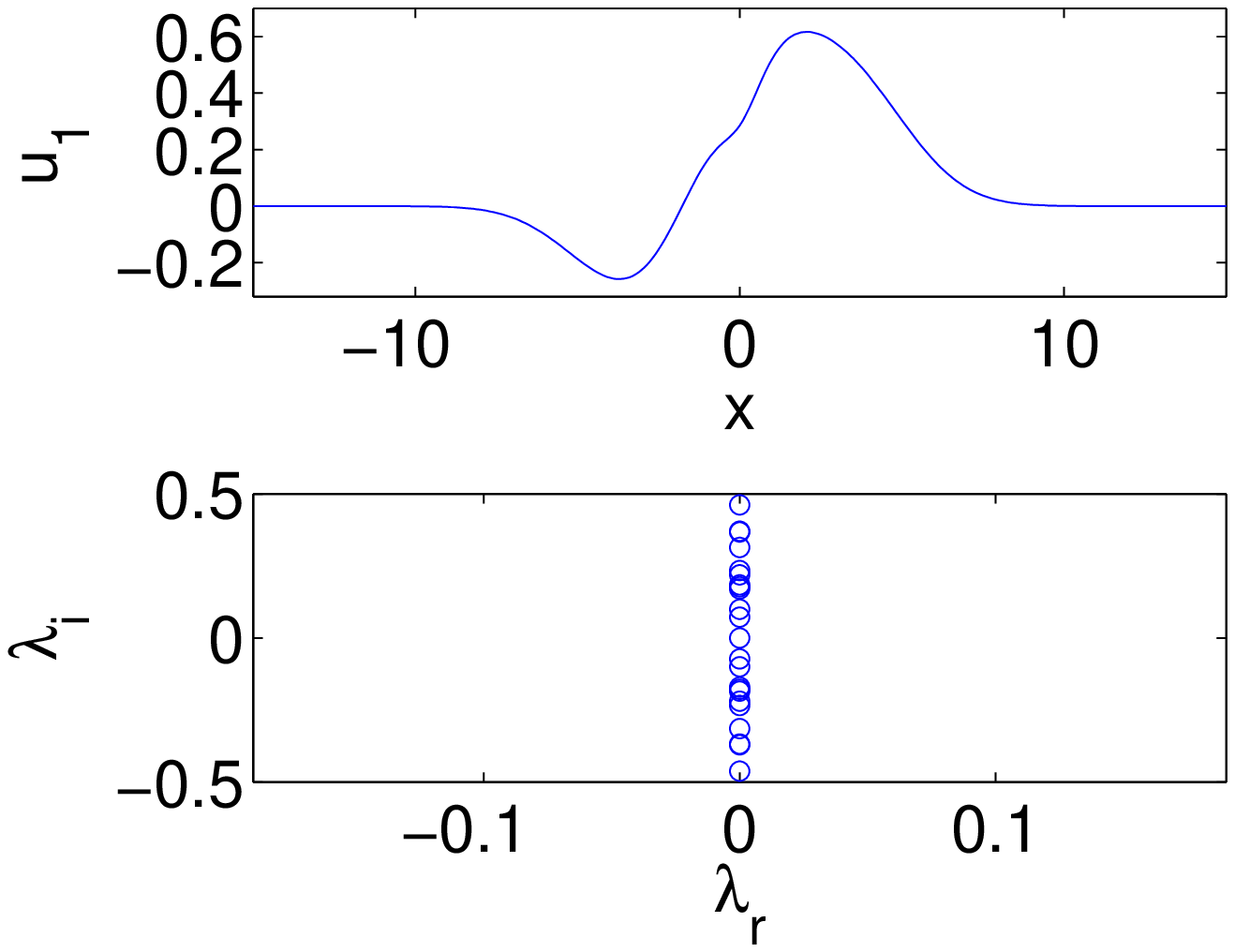}
\includegraphics[width=.3\textwidth]{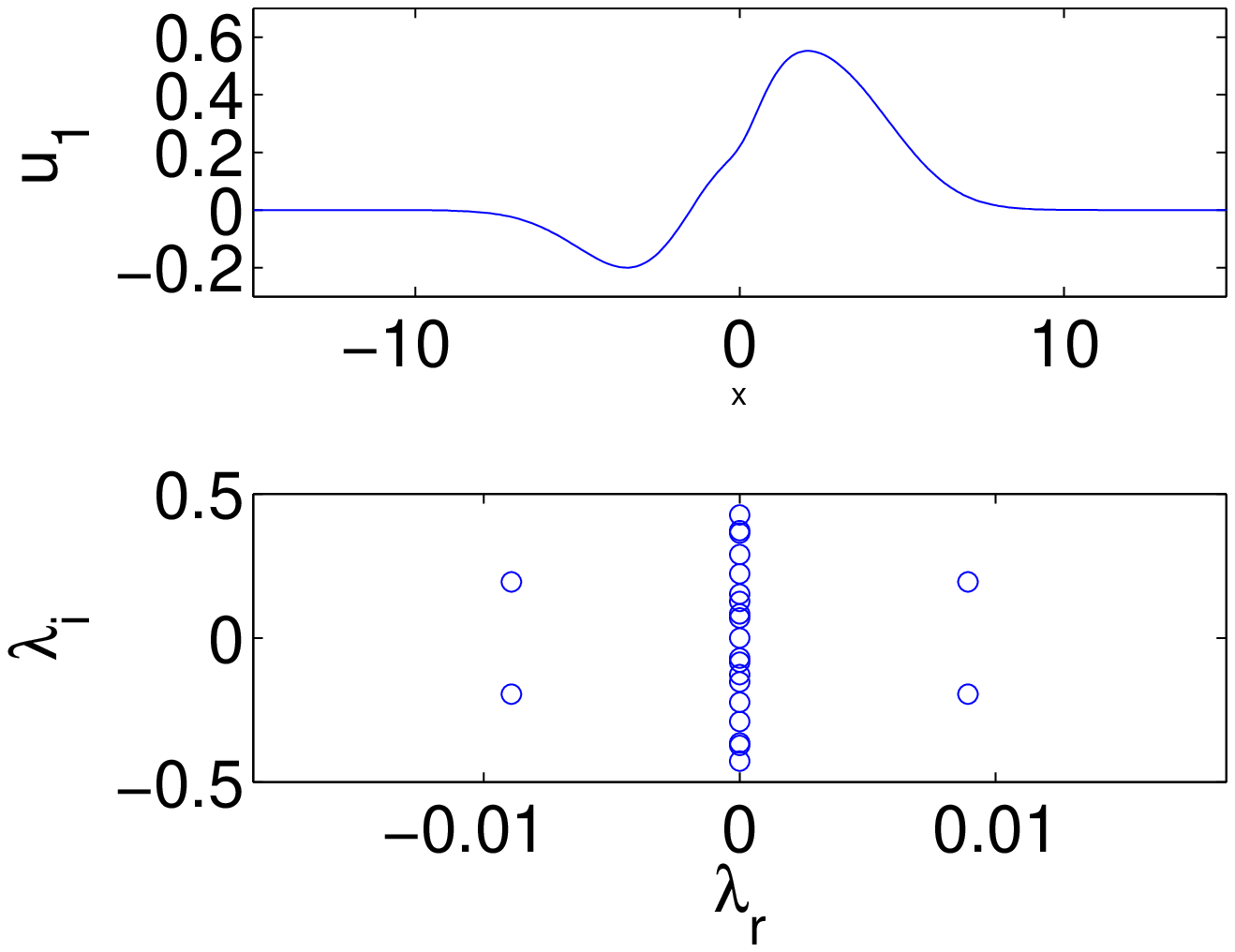}\\
\caption{(Color online) Profiles of wave functions $u_1$ (first and third rows) and stability eigenvalues (second and fourth rows) 
corresponding to S1, AN1 and AS1 (branches of these solutions are shown in Fig. \ref{figNa}): 
S1 for $\mu_{0}=0.48$ (top left), AN1 for $\mu_{0}=0.48$ (top right), AS1 for $\mu_{0}=0.48$ 
(third row left) and AS1 for $\mu_{0}=0.45$ 
(third row right).}
\label{figNaSim1}
\end{figure}

Perhaps the cornerstone of the present work is encapsulated within 
Fig. \ref{figNa}, generated for the case of a $^{23}$Na spinor condensate.
This, admittedly, rather busy 
diagram encompasses {\it all} possible
states near the linear limit of a spinor BEC loaded in a double well potential.
Among them, one can discern one-component states that are well-known
from the considerations of single-component double well settings
\cite{markus1,smerzi,Bergeman_2mode,george} for each one of the three
components (labeled by S for symmetric, AN for antisymmetric and
AS for asymmetric realizations of each component); see also 
Fig. \ref{figNaSim1}.
They are labeled by $1$ when they
belong to component $u_1$, by $0$ when they belong to $u_0$ and
by $2$ when they belong to $u_{-1}$. Notice that the AS state
bifurcates from the S state in the attractive case and from
the AN state in the repulsive case, inheriting their stability
(and rendering them unstable, through a pitchfork bifurcation) \cite{george}.

Importantly, the graph also contains two-component states involving components 
with subscripts $1$ and $-1$ (and with a vanishing $0$ component),
denoted collectively by D and analyzed in Section \ref{two_comp}
below (cf. also \cite{chenyu}). These are branches involving the symmetric
and/or antisymmetric components of $u_1$ and $u_{-1}$; interestingly,
out of these branches bifurcate new asymmetric solutions involving
the same two components.

Finally, branches with contributions from all three components
are also identified and are collectively labeled by C; these are
examined more systematically in Section \ref{three_comp} below.
These predominantly consist of either components $u_1$ and $u_0$
(with a small component in $u_{-1}$) or $u_{-1}$ and $u_0$ (with
a small component $u_1$). Again different combinations of 
symmetric and/or antisymmetric configurations are
possible among these components, and additional asymmetric branches
are observed to bifurcate from them. We now turn to a more detailed
explanation of our findings.



To analyze the stability of a particular state (in this case and in the following ones), 
we perform linearization around the unperturbed 
state $u_j$, assuming a perturbed solution 
of Eqs. (\protect\ref{eq012})-(\protect\ref{eq0}) in the form, 
\begin{eqnarray}
\psi_j=\exp(-i \mu_j t) \left[u_j + \epsilon \left(p_j(x) \exp(\lambda t)
+ q_j^{\star}(x) \exp(\lambda^{\star} t)\right) \right]
\label{linearization}
\end{eqnarray}
where $p_j$ and $q_j$ represent infinitesimal perturbations with
eigenvalues $\lambda \equiv \lambda _{r}+i\lambda _{i}$ (here $\epsilon$ 
is a formal small parameter and the asterisk denotes complex conjugate). 
When its relevant eigenvalues are found to be purely imaginary,
then the corresponding state is linearly stable, while the presence of
eigenvalues with $\lambda_r \neq 0$ is tantamount to instability.
Then, the results of the linear stability analysis are typically presented 
in the spectral plane $(\lambda_r,\lambda_i)$. A typical example is 
provided in Fig. \ref{figNaSim1}, where we show the profiles and the corresponding
spectral planes $(\lambda_r,\lambda_i)$ of the branches S1, AN1, and AS1. 

Coming back to the bifurcation diagram of Fig. \ref{figNa}, 
it is important to highlight that it 
has been constructed by fixing $\mu_{-1}$ and varying only $\mu_{1}$ 
and $\mu_{0}$ (since one of these three quantities is always slaved to the
variation of the other two). For this reason, S2, AN2 and AS2 are horizontal 
lines since $\mu_{-1}$ is fixed in the case studied below 
(i.e., their properties do not change as $(\mu_{1},\mu_{0})$ are varied.)
We will vary $\mu_{0}$ while doing the analysis, with $\mu_{1}$ being determined by 
$\mu_{1}=2\mu_{0}-\mu_{-1}$.


\subsection{Two-component states}
\label{two_comp}

\begin{figure}[tbhp!]
\centering
\includegraphics[width=.3\textwidth]{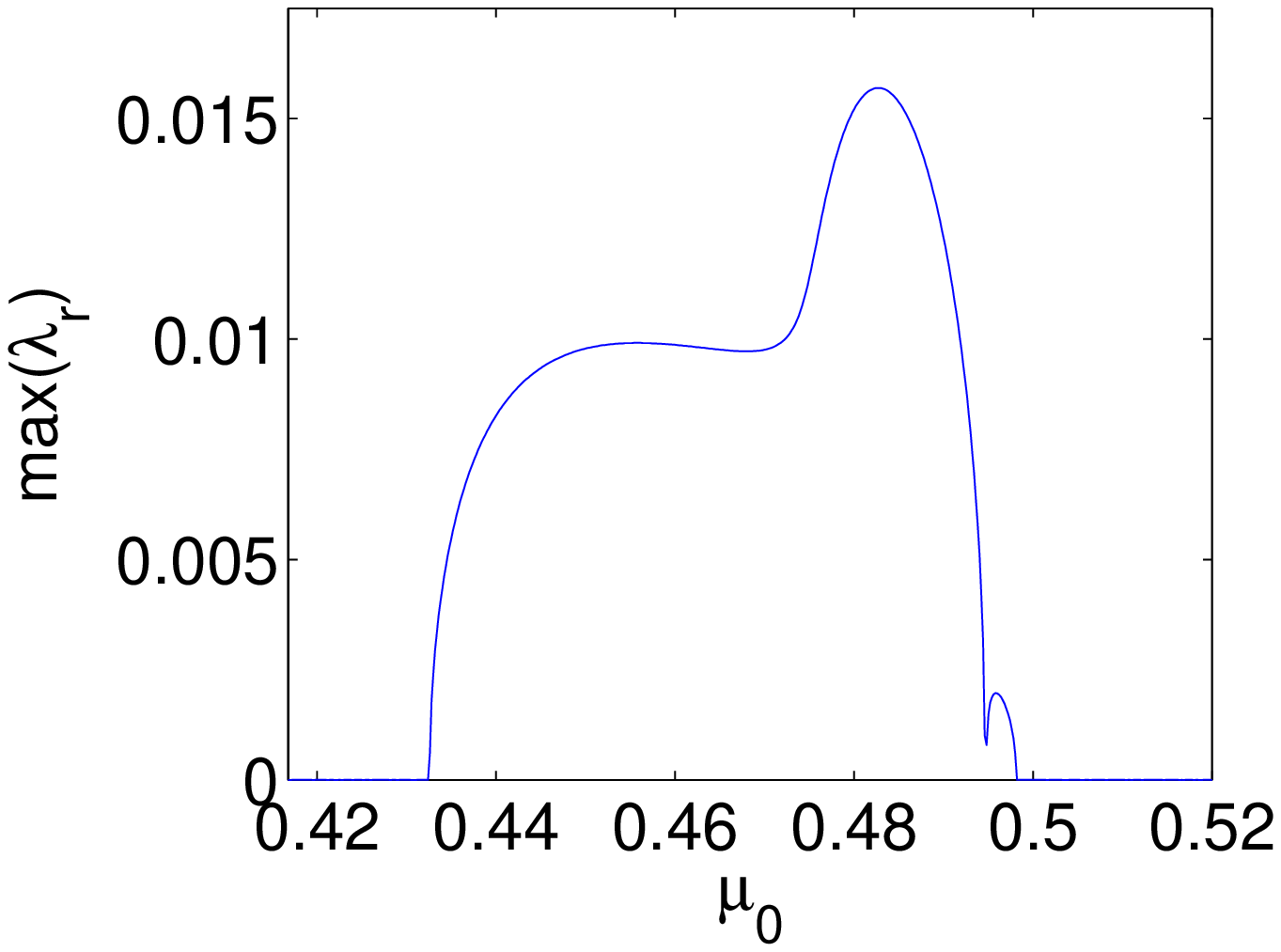}
\includegraphics[width=.3\textwidth]{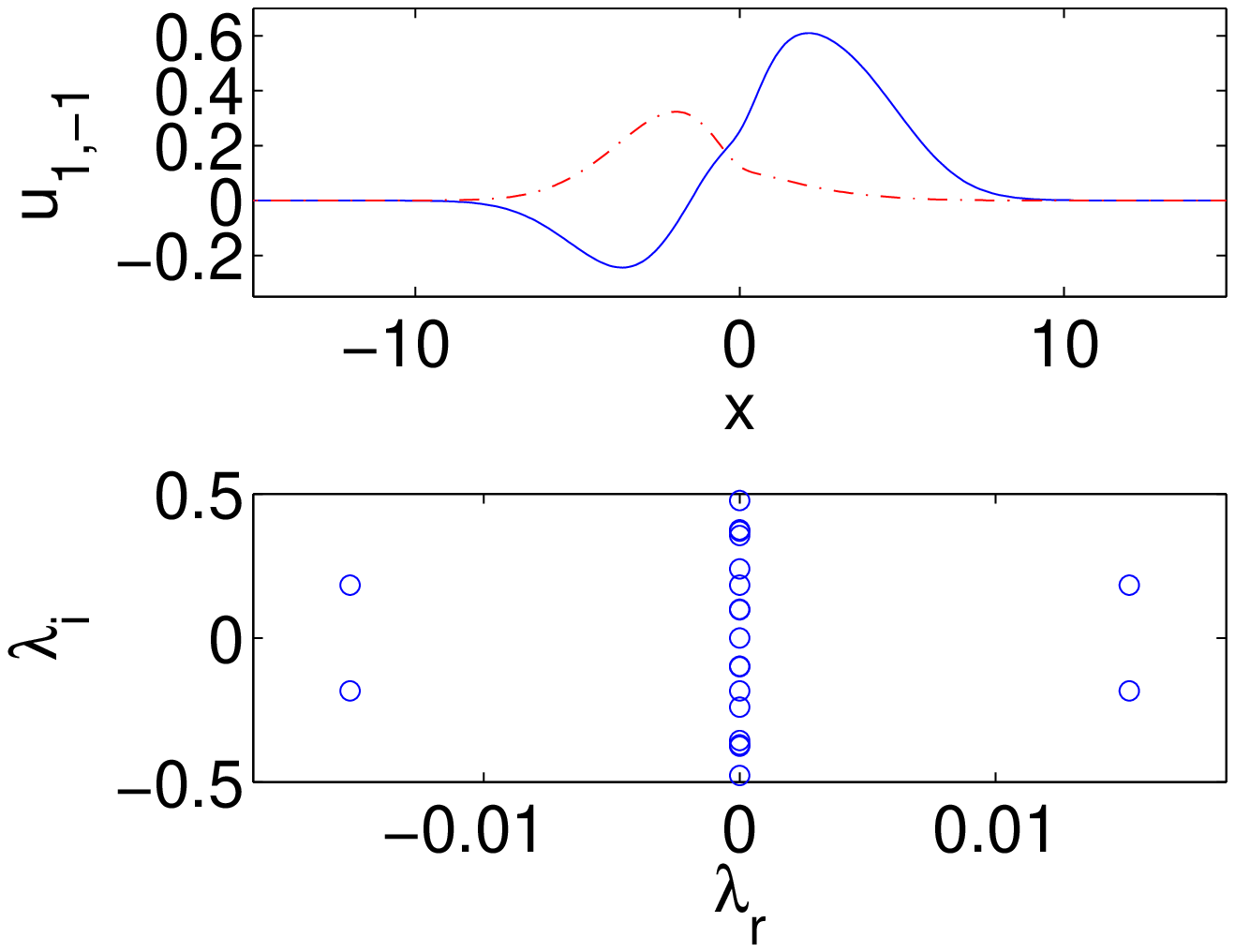} 
\includegraphics[width=.3\textwidth]{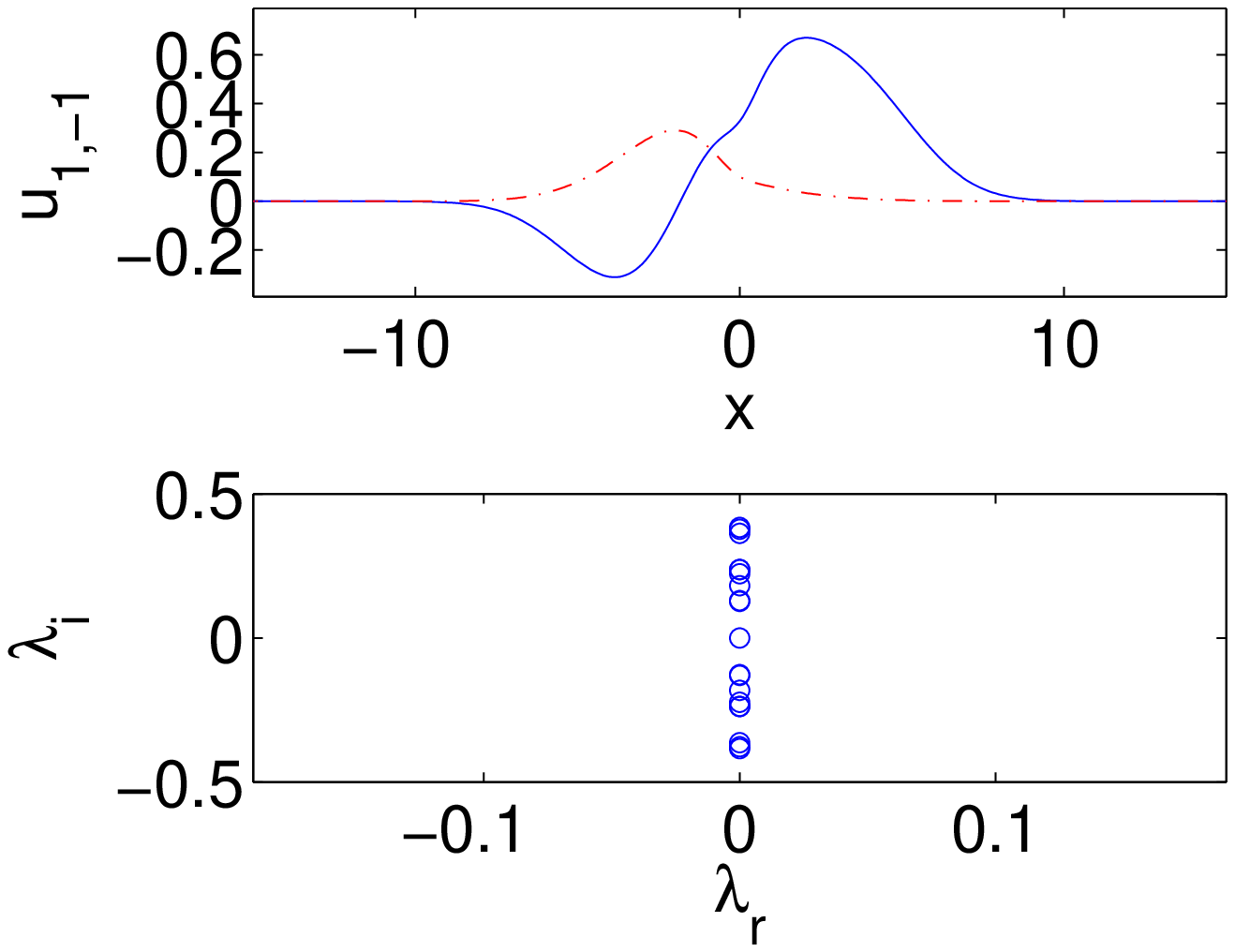}\\
\includegraphics[width=.3\textwidth]{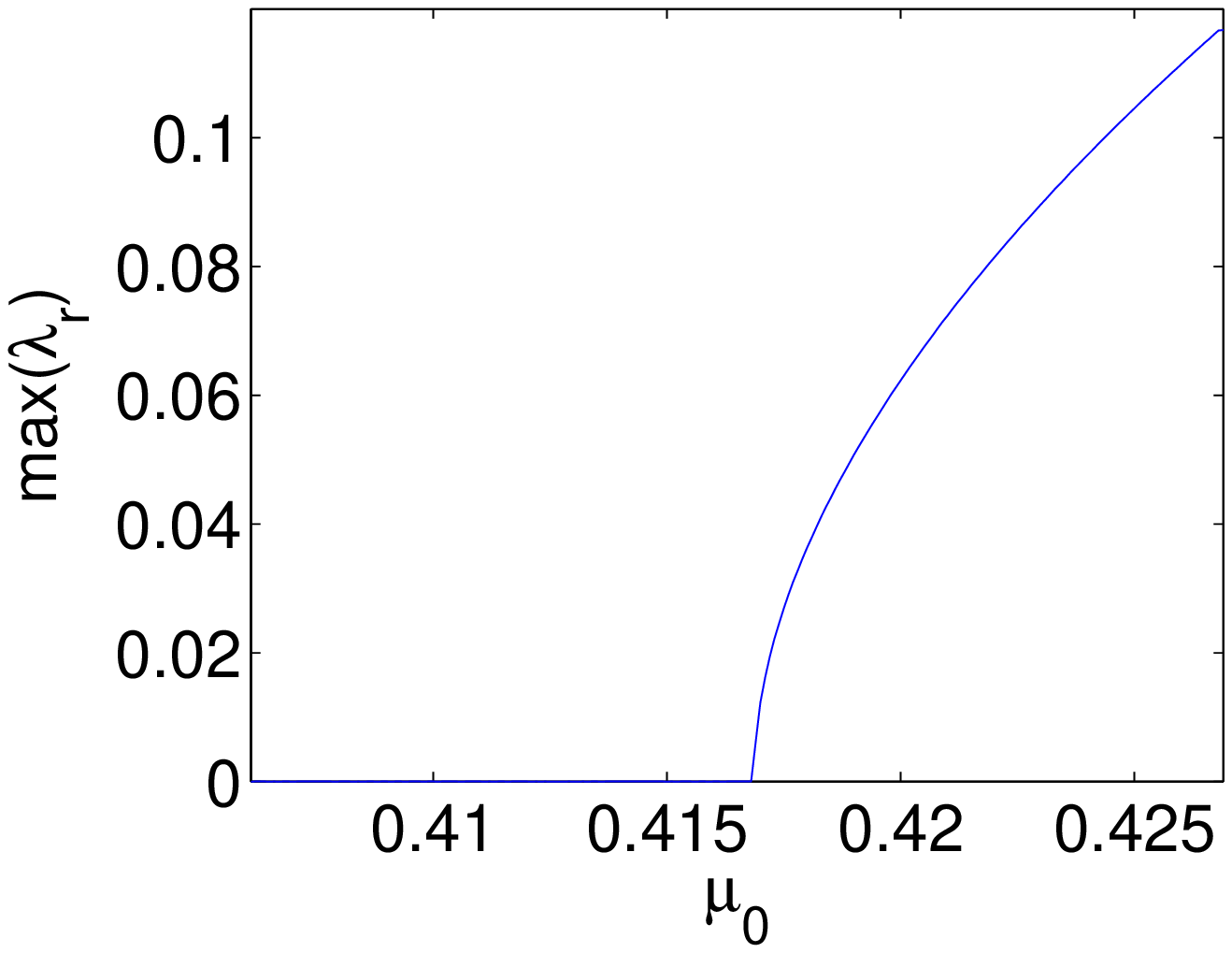}
\includegraphics[width=.3\textwidth]{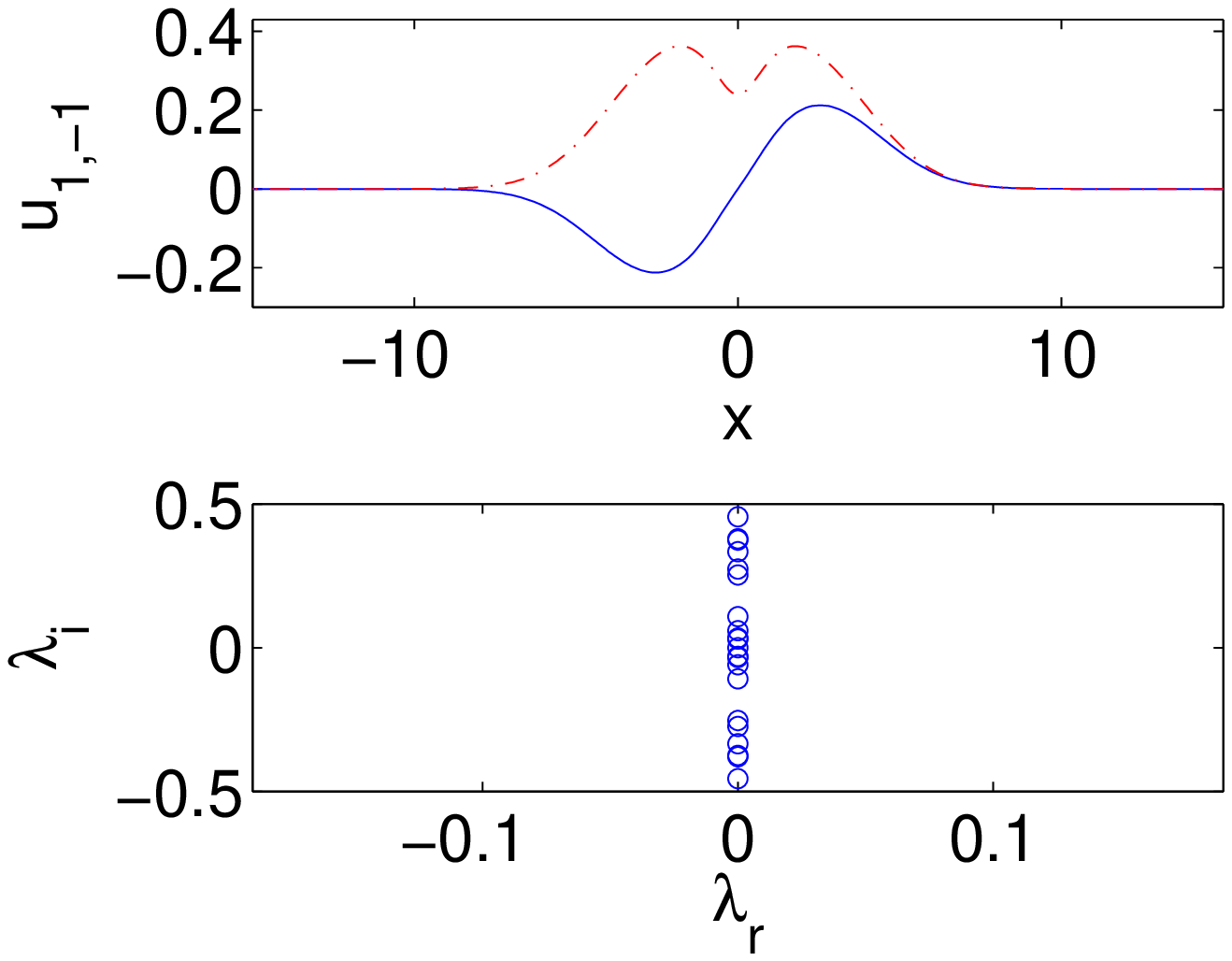}
\includegraphics[width=.3\textwidth]{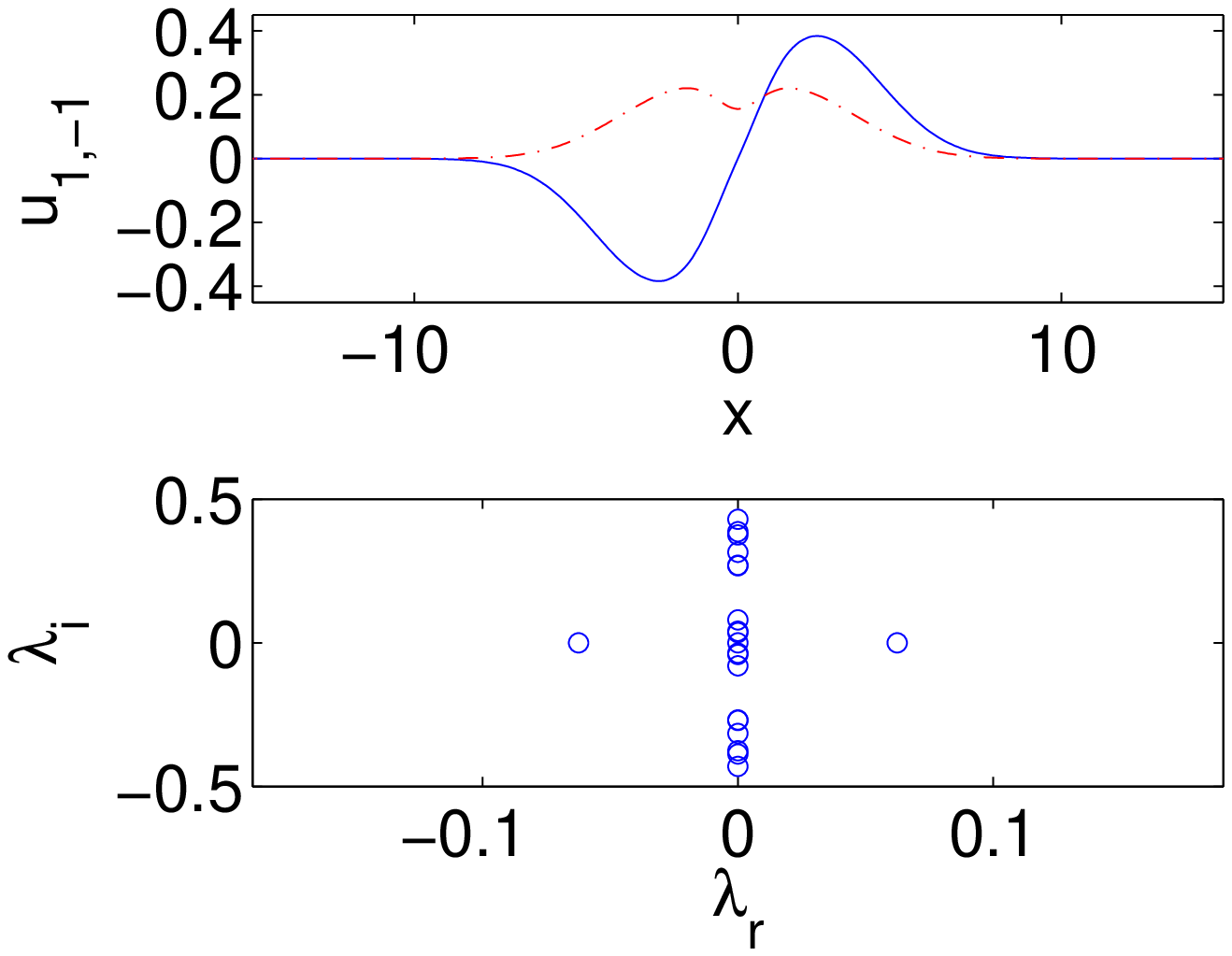}\\
\caption{(Color online) Left panels: the maximum real eigenvalue as a function 
of $\mu_0$ for D3 (top left) and D4 (bottom left) in Fig. \ref{figNa}. 
The middle and right panels 
show the profiles of wave functions (first and third row)
$u_{1}$ and $u_{-1}$ [by solid (blue) and dash-dotted (red) lines, respectively], 
together with the stability eigenvalues corresponding to D3 and D4: 
D3 for $\mu_{0}=0.48$ (top middle), D3 for $\mu_{0}=0.51$ (top right), 
D4 for $\mu_{0}=0.41$ (bottom middle) and D4 for $\mu_{0}=0.42$ (bottom right) 
with $\mu_{-1}=0.38$ in the case of $^{23}$Na condensate.}
\label{figNaD34}
\end{figure}

\begin{figure}[tbhp!]
\centering
\includegraphics[width=.3\textwidth]{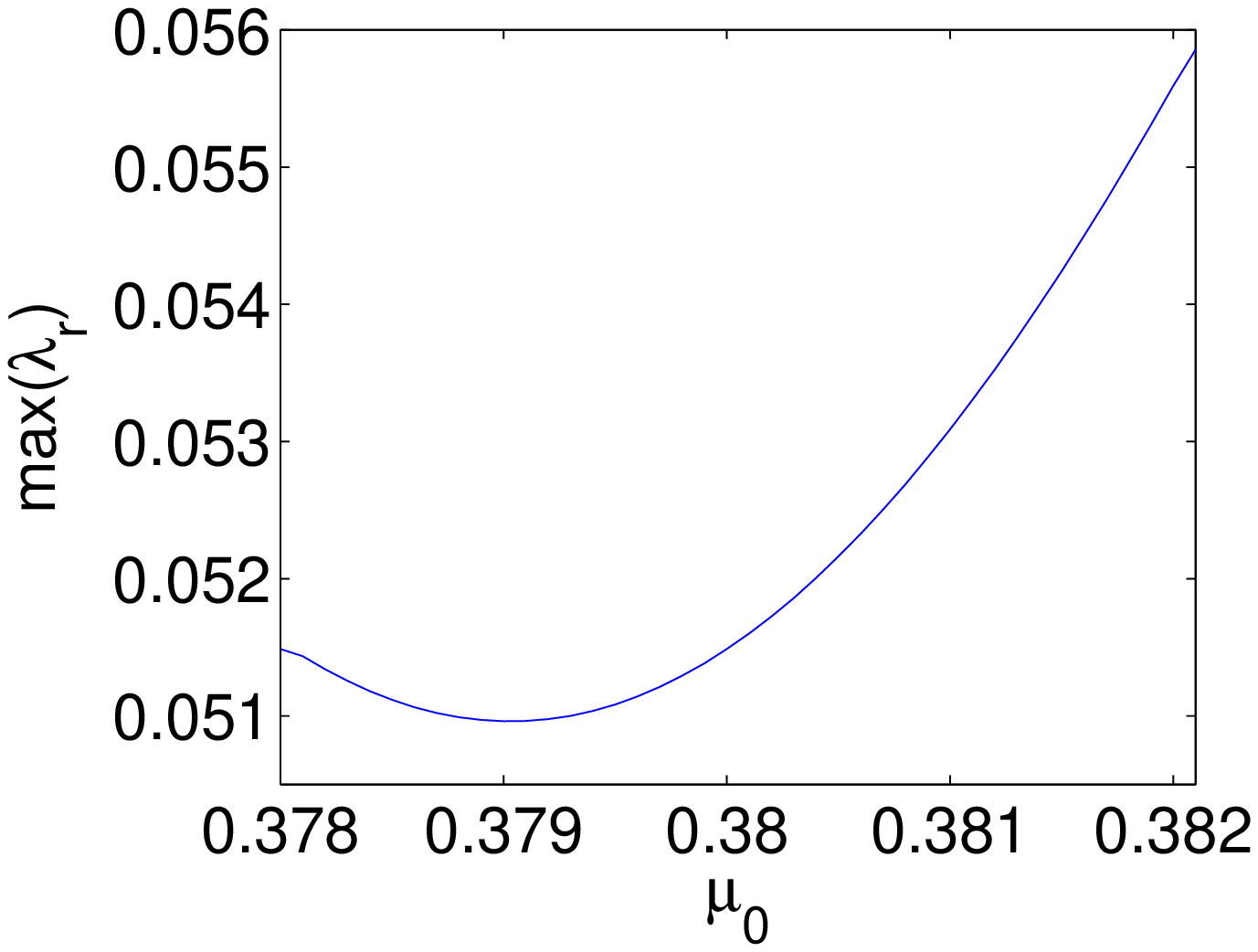}  
\includegraphics[width=.3\textwidth]{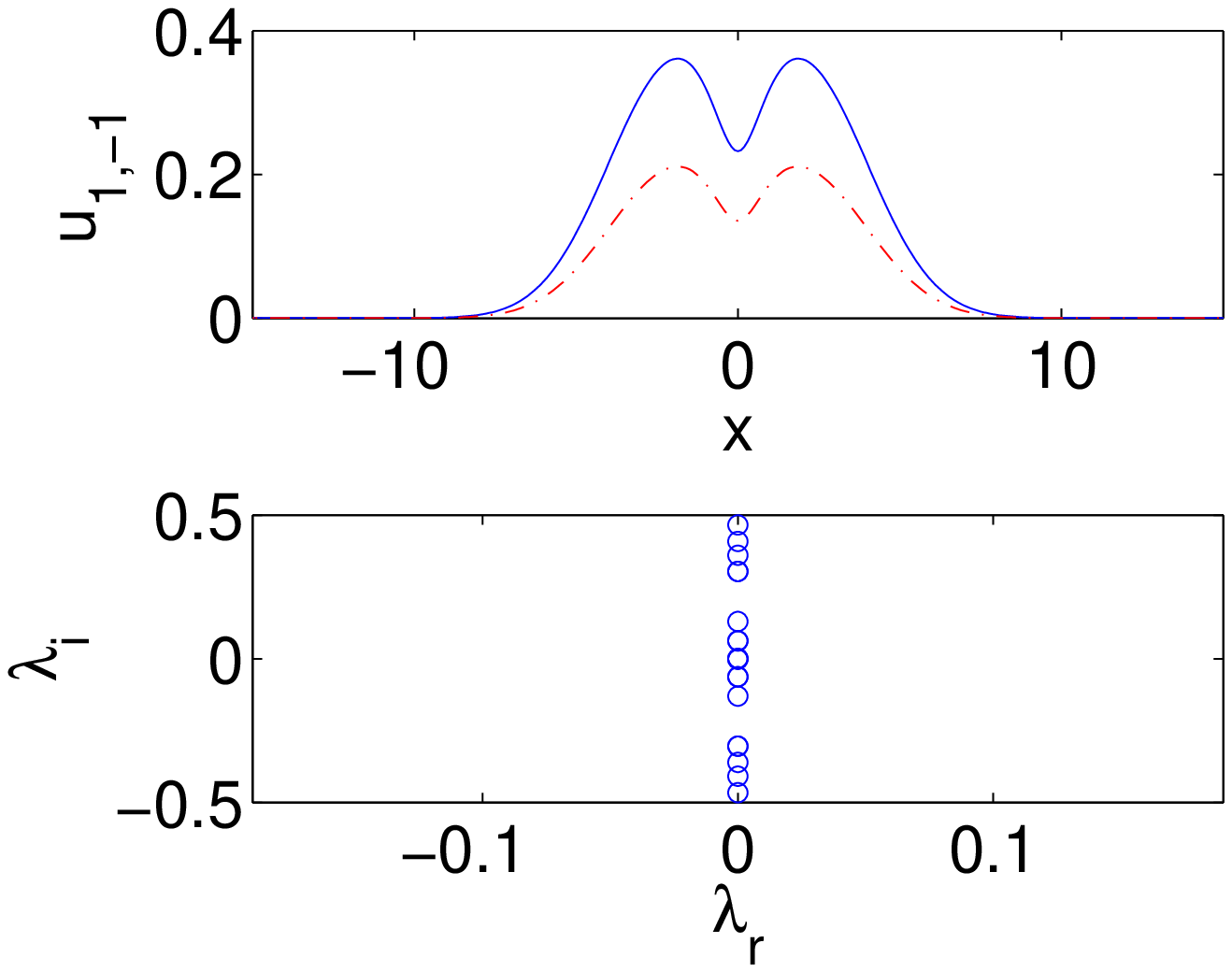}   
\includegraphics[width=.3\textwidth]{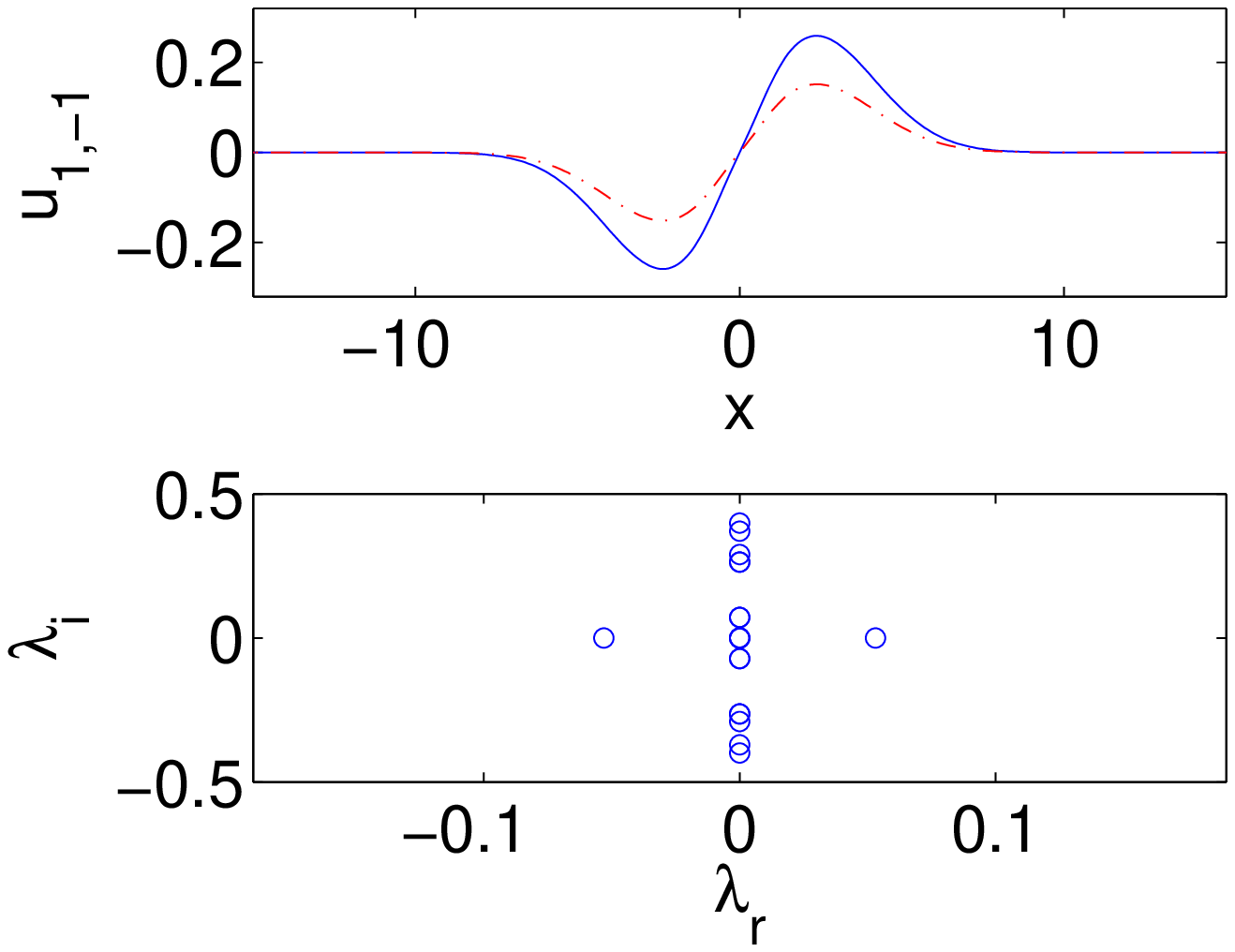}\\
\caption{(Color online) The left panel shows the maximum real eigenvalue as a function of $\mu_0$ 
for D6 in Fig. \ref{figNa}. The middle and right panels 
show the profiles of wave functions $u_{1}$ 
and $u_{-1}$ by solid (blue) and dash-dotted (red) lines, respectively, 
together with stability eigenvalues corresponding to D5 and D6 in Fig. \ref{figNa}: 
D5 for $\mu_{0}=0.382$ (middle), D6 for $\mu_{0}=0.381$ (right).}
\label{figNaD56}
\end{figure}

In addition to 
``pure'' branches populating only one component, 
there also exist additional branches populating two components, 
similarly to the results 
of Ref. \cite{chenyu}.
In Fig. \ref{figNa}, prototypical examples of these branches 
are given by D1-D6 which involve the two components $u_{1}$ and $u_{-1}$. 
More specifically, D2 ``connects'' S1 and AN2. That is to say that 
D2 has a symmetric wavefunction in the component $u_{1}$, while it has 
an antisymmetric one in component $u_{-1}$. 
D1 bifurcates from D2, below a critical number of atoms 
through a pitchfork bifurcation, is asymmetric in both components $u_{1}$ and
$u_{-1}$ and eventually merges into AS2. 
It is relevant to point out here that in the first panel of the
second row of Fig. \ref{figNa}, it appears as if D2 and D1 are
essentially overlapping for the small fraction of the D2 branch
for which they coexist (in this segment, D2 is unstable). However
D2 terminates into AN2, while D1 continues to exist for lower norms, 
down to AS2. In a situation entirely similar to that highlighted above, but now
between S2 and AN1, the branch that ``connects'' them is D4. Out of
that branch with a symmetric wavefunction in $u_{-1}$ and an 
anti-symmetric in $u_{1}$ bifurcates (above a critical norm $N$) the
branch D3 which is again asymmetric in both of these components,
and will eventually merge with AS1. 
Since these two situations (involving D1,D2 and D3,D4, respectively) 
are essentially similar (both of the bifurcation sub-structures are 
shown in Fig. \ref{figNa}), only the latter pair of branches 
is examined in detail in Fig. \ref{figNaD34}. In particular, it can
be seen there that while the D4 branch is stable before D3 appears,
the supercritical pitchfork leading to the emergence of D3 
destabilizes D4. On the other hand, D3 itself may be unstable but
only due to Hamiltonian-Hopf bifurcations \cite{vdm}, associated
with weak complex quartets of eigenvalues. 
Finally, in terms of two-component branches, in addition to 
the above pairs connecting the symmetric nonlinear eigenmodes of 
one component with the anti-symmetric ones of the other, there
also exist the branches D5 and D6 (see again the blowups of
Fig. \ref{figNa}). These connect the symmetric state of $u_1$ 
(branch S1) with the symmetric state of $u_{-1}$ (branch S2) 
and the antisymmetric state of $u_1$ (branch AN1) with the 
antisymmetric state of $u_{-1}$ (branch AN2). The branch D5 is linearly
stable, while D6 is linearly unstable, as indicated in Fig. 
\ref{figNaD56}. Interestingly, the norm of these branches almost
coincides with the norm of branches S0 (for D5) and AN0 (for D6)
although they do not involve a nontrivial component of $u_0$.
This is apparently associated with the fact that branches
such as D5 or D6 can be effectively 
described in the framework of SMA. 
On the other hand,
clearly D1-D4 cannot; thus, they provide a distinct, potentially stable
(and even symmetry-broken) set of states that could be accessible
by the $F=1$ spinor BEC system.

\subsection{Three-component states}
\label{three_comp}

\begin{figure}[tbhp!]
\centering
\includegraphics[width=.3\textwidth]{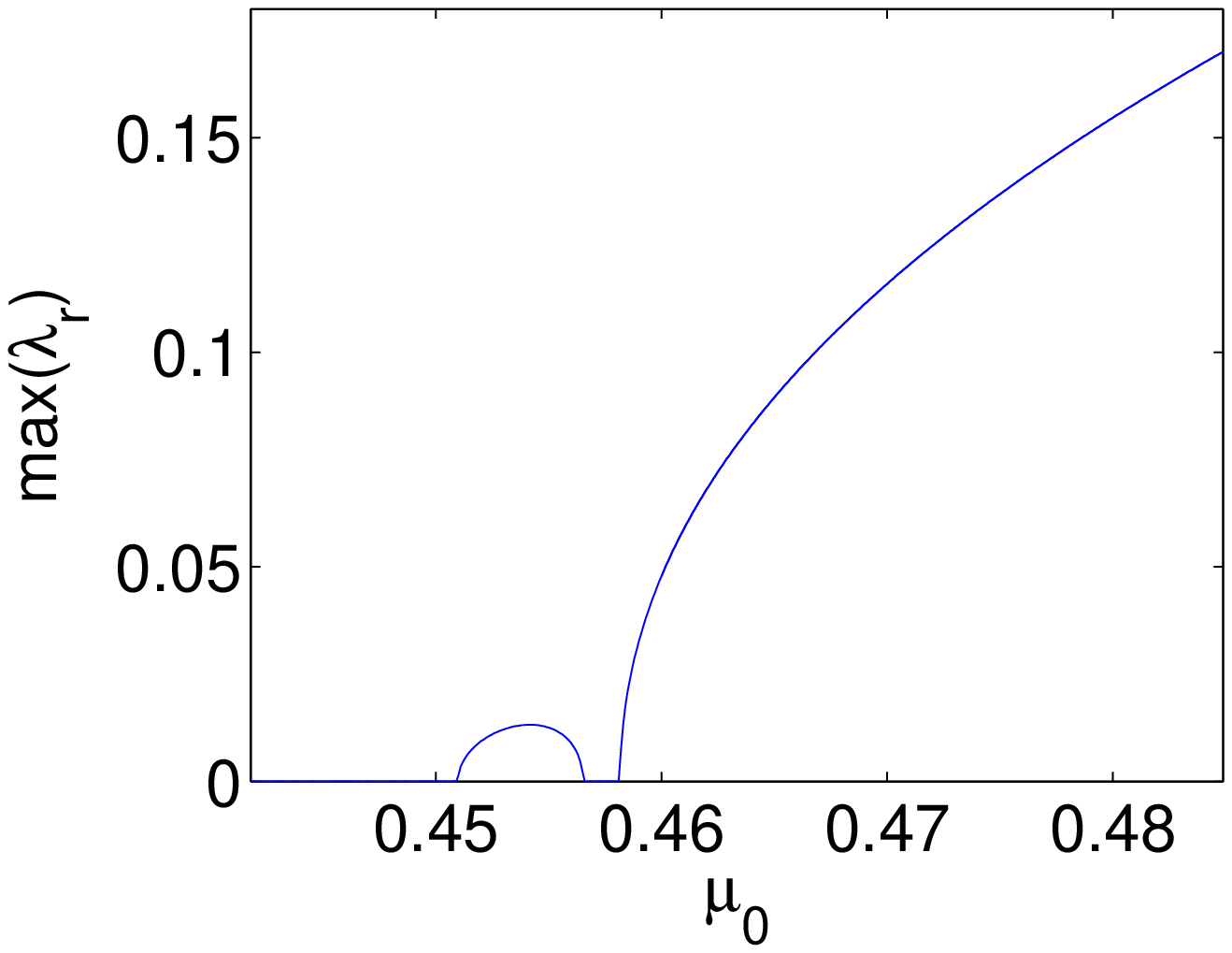}
\includegraphics[width=.3\textwidth]{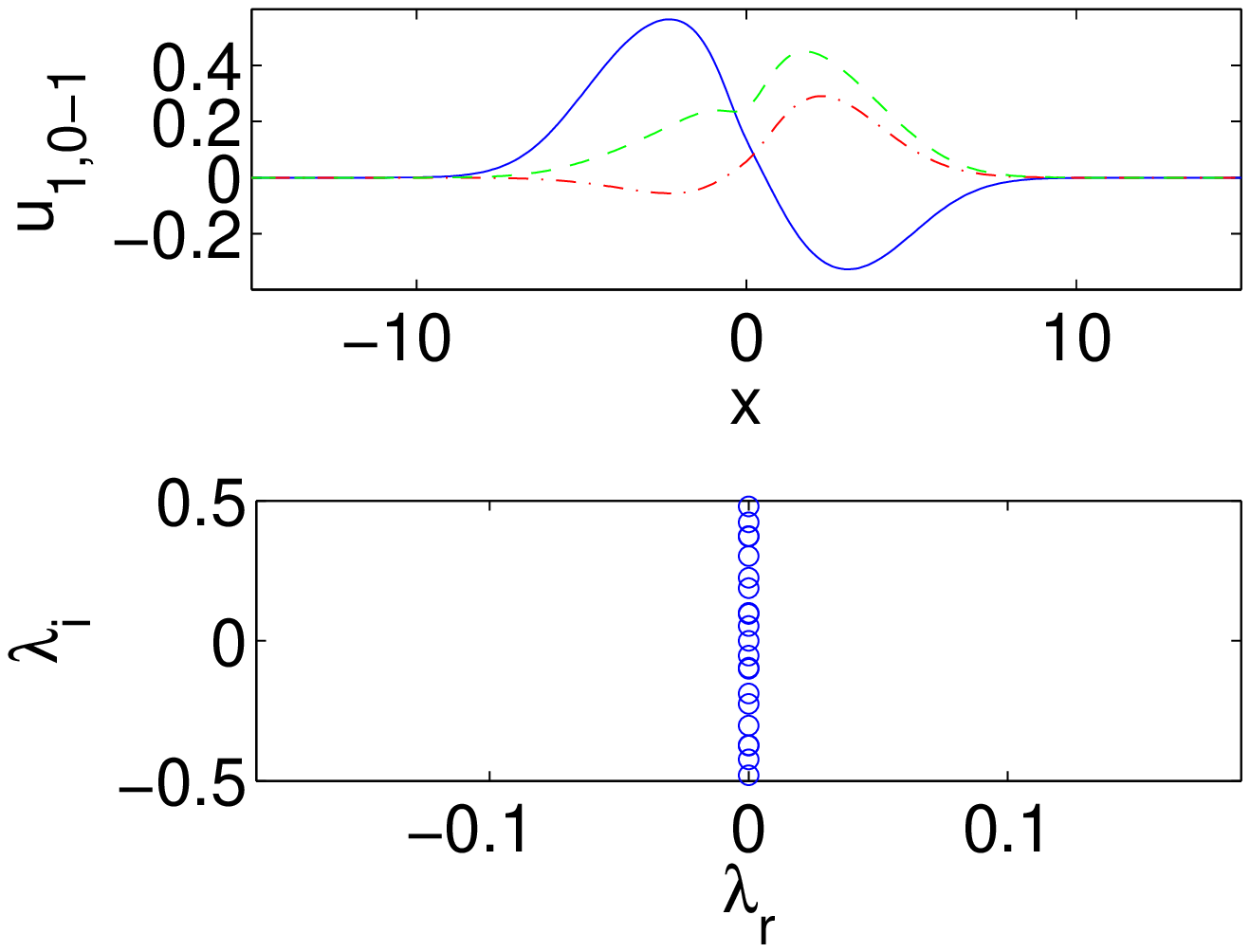}\\  
\includegraphics[width=.3\textwidth]{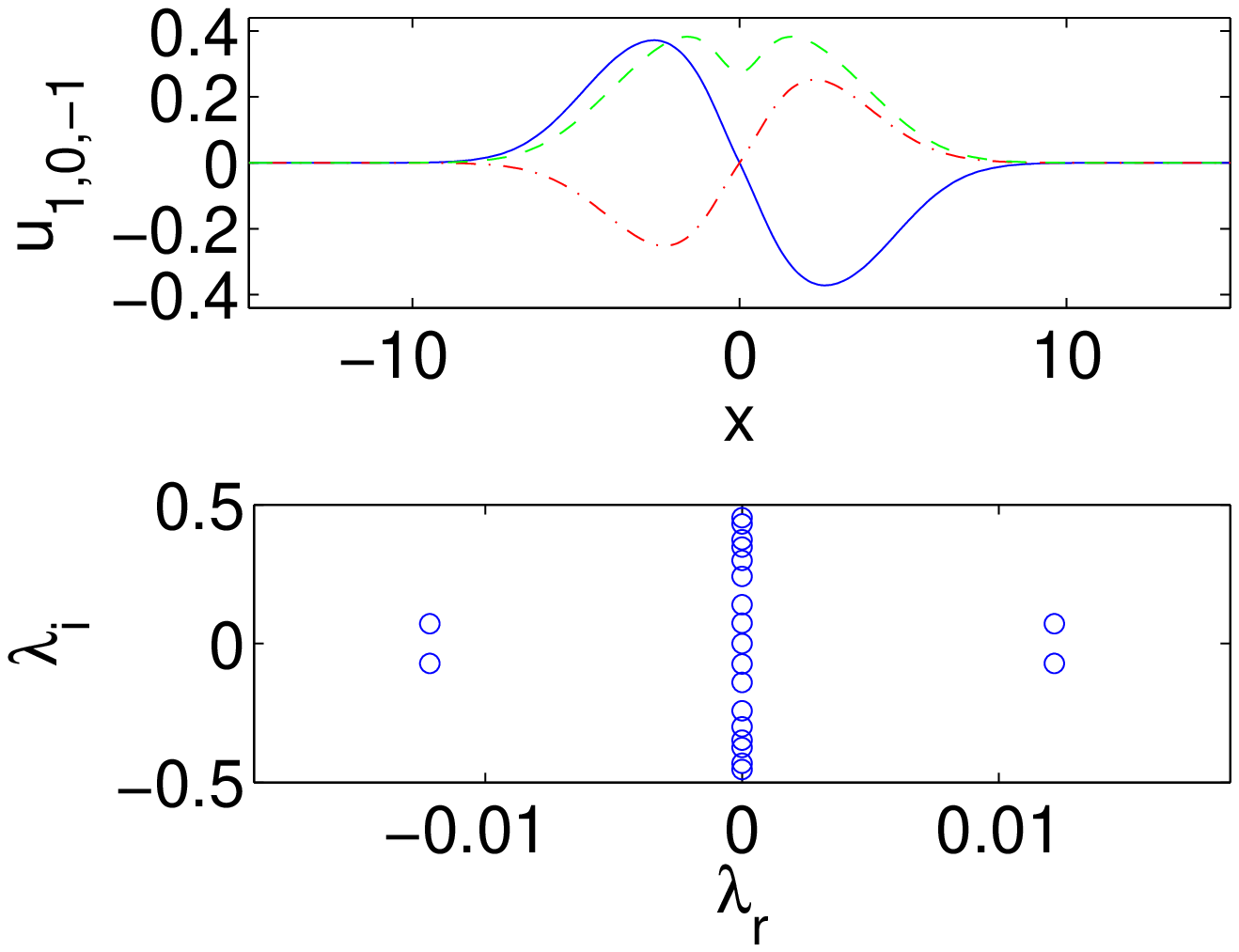}
\includegraphics[width=.3\textwidth]{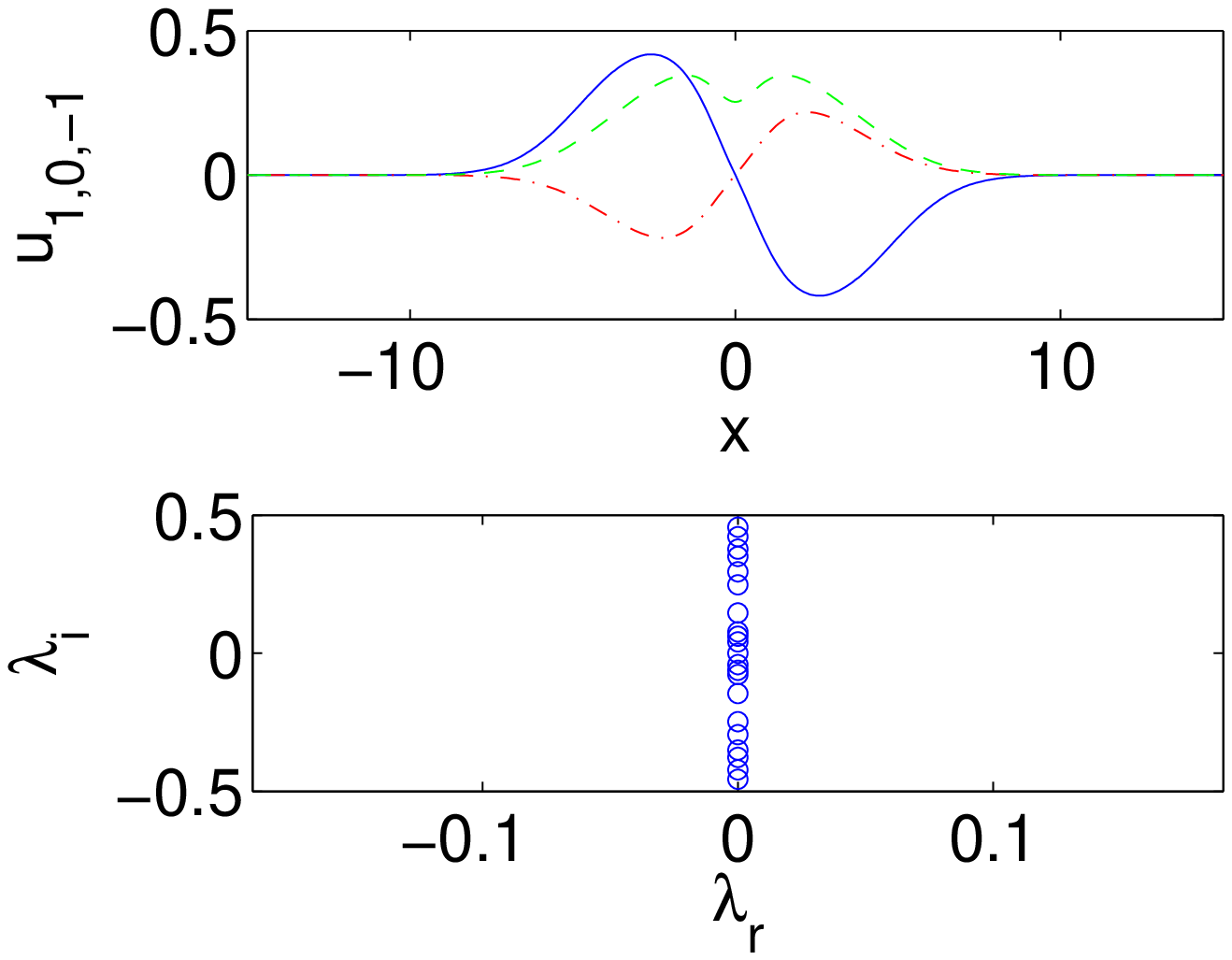}
\includegraphics[width=.3\textwidth]{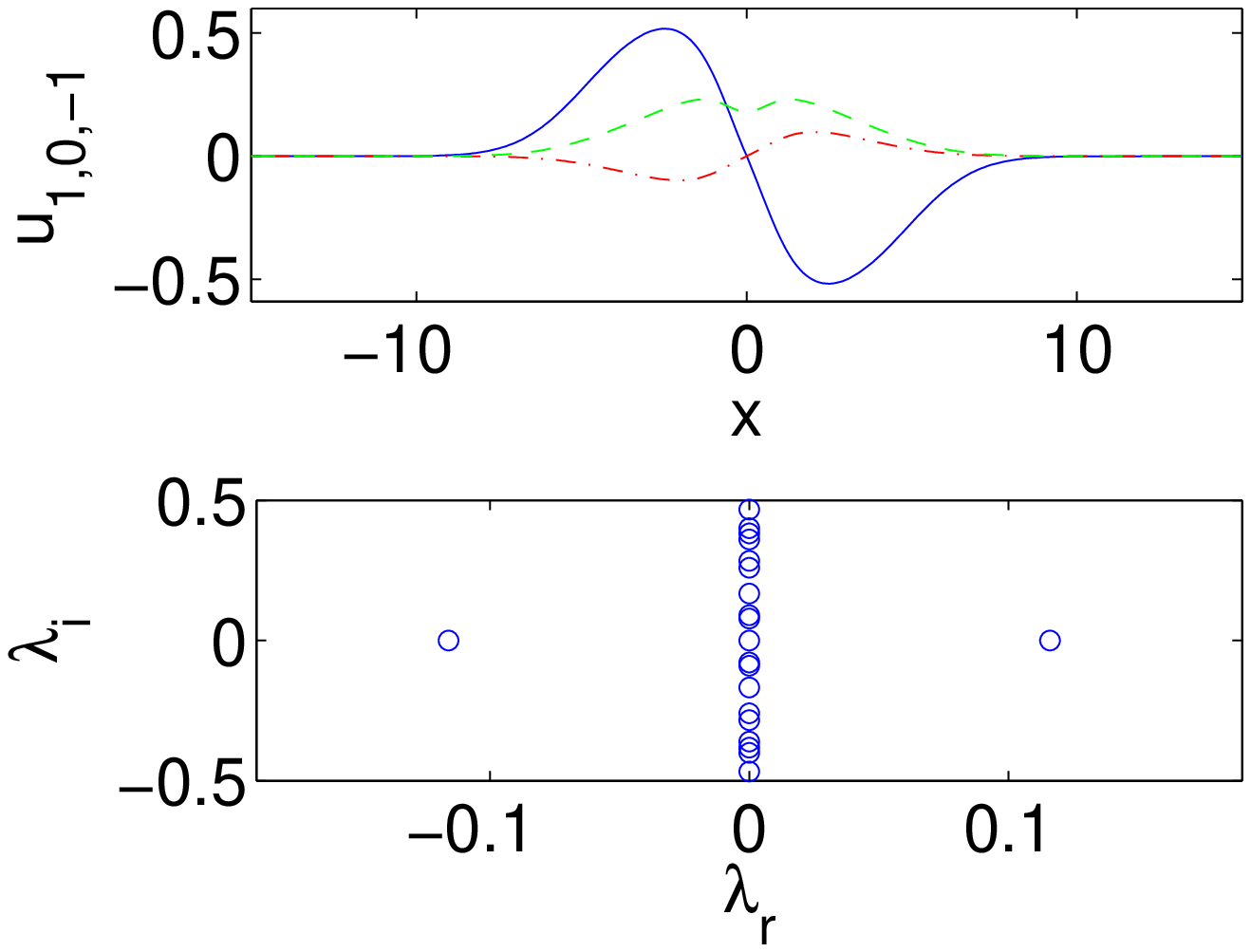}\\
\caption{(Color online) The top left panel shows the maximum real 
eigenvalue as a function of $\mu_0$ for C2 in Fig. \ref{figNa}. 
The other panels 
show the profiles of wave functions 
$u_{1}$, $u_{0}$ and $u_{-1}$ by solid (blue), dashed (green) and dash-dotted (red) lines, 
respectively, and stability eigenvalues corresponding to branches C1 and C2 in Fig. \ref{figNa}: 
C1 for $\mu_{0}=0.48$ (top right), C2 for $\mu_{0}=0.453$ (bottom left), 
C2 for $\mu_{0}=0.457$ (bottom middle) and C2 for $\mu_{0}=0.47$ (bottom right) 
with $\mu_{-1}=0.38$. The profiles of $u_{-1}$ 
[dash-dotted (red) lines] are multiplied by 30 to be more visible.}
\label{figNaC12}
\end{figure}

\begin{figure}[tbhp!]
\centering
\includegraphics[width=.3\textwidth]{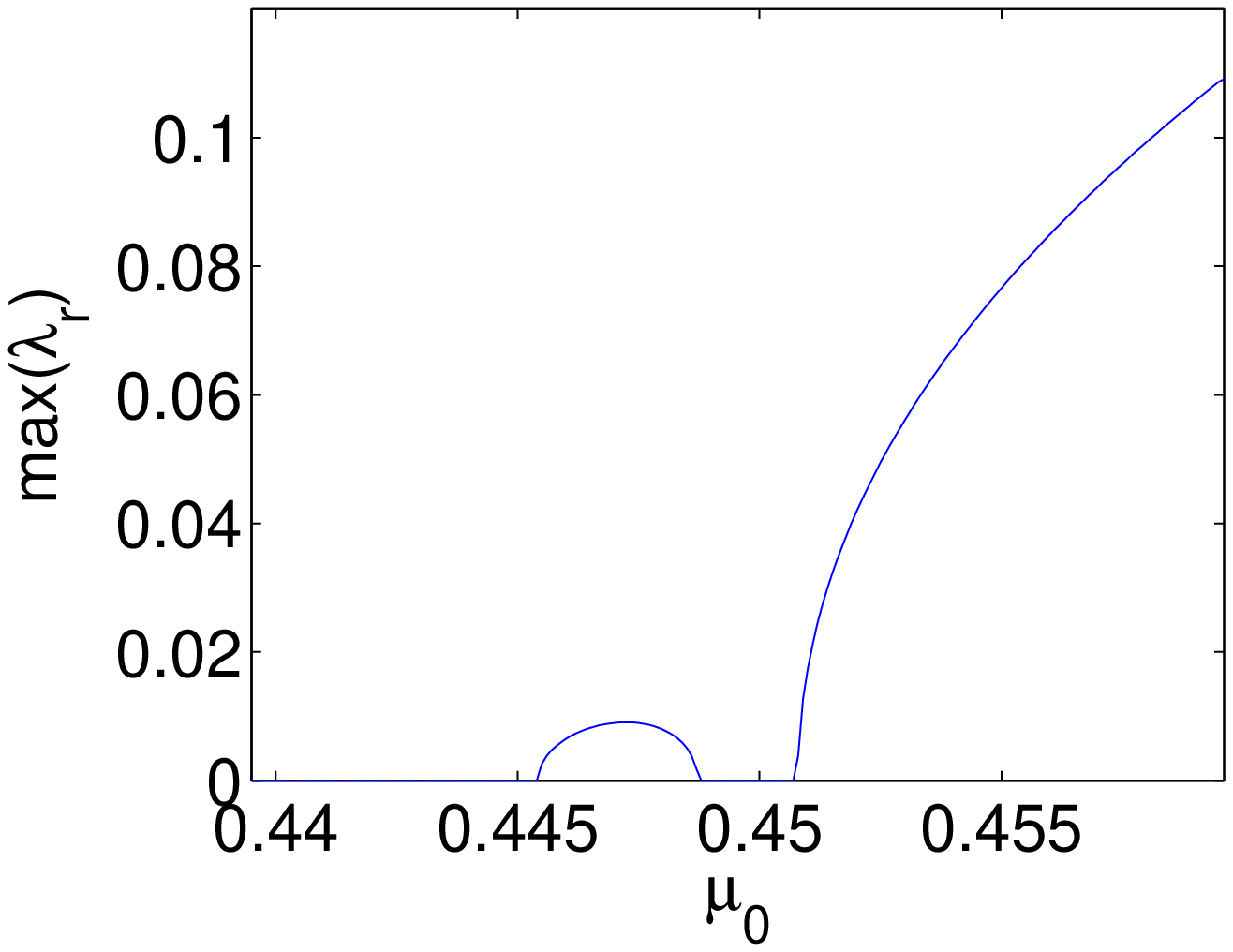}
\includegraphics[width=.3\textwidth]{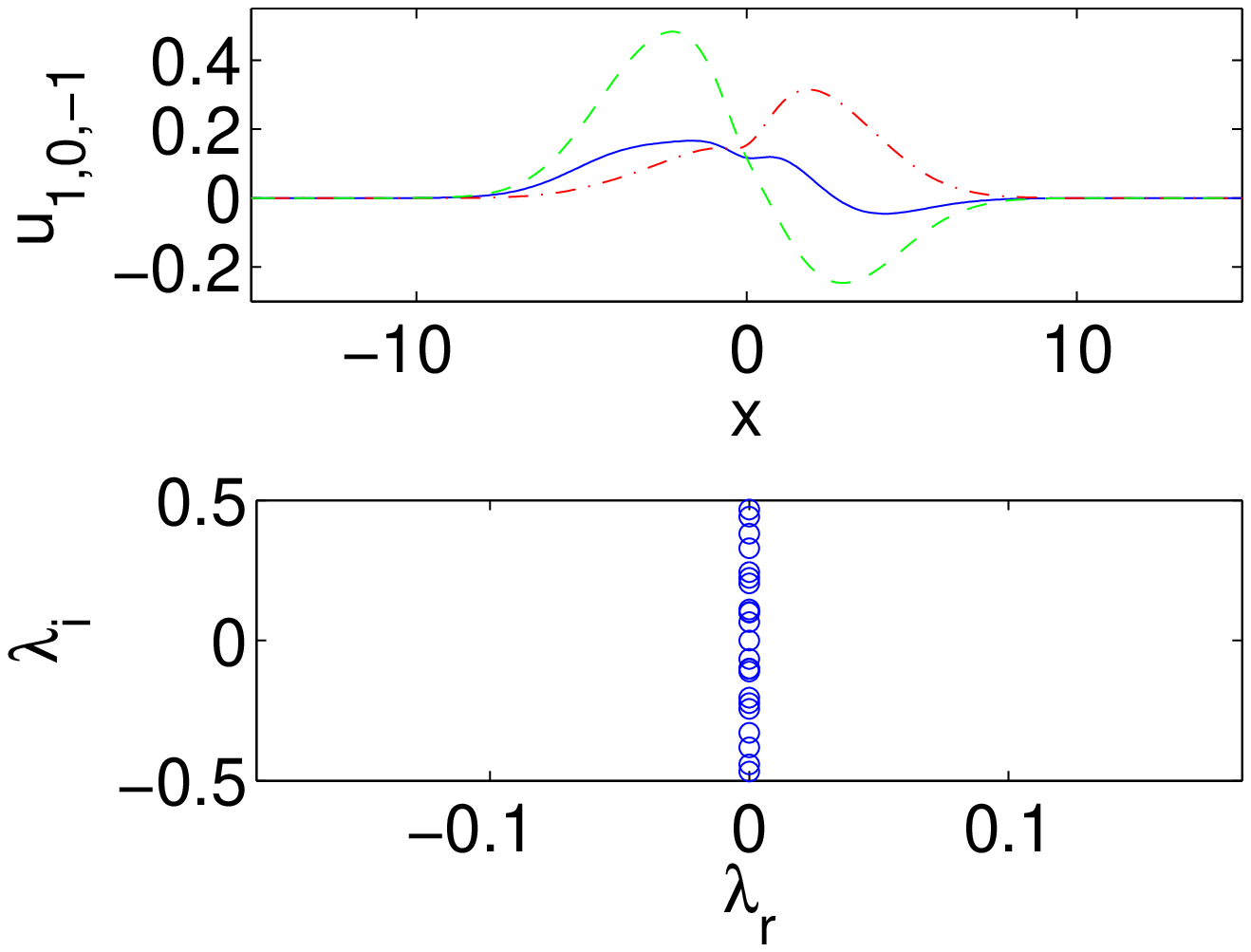}\\  
\includegraphics[width=.3\textwidth]{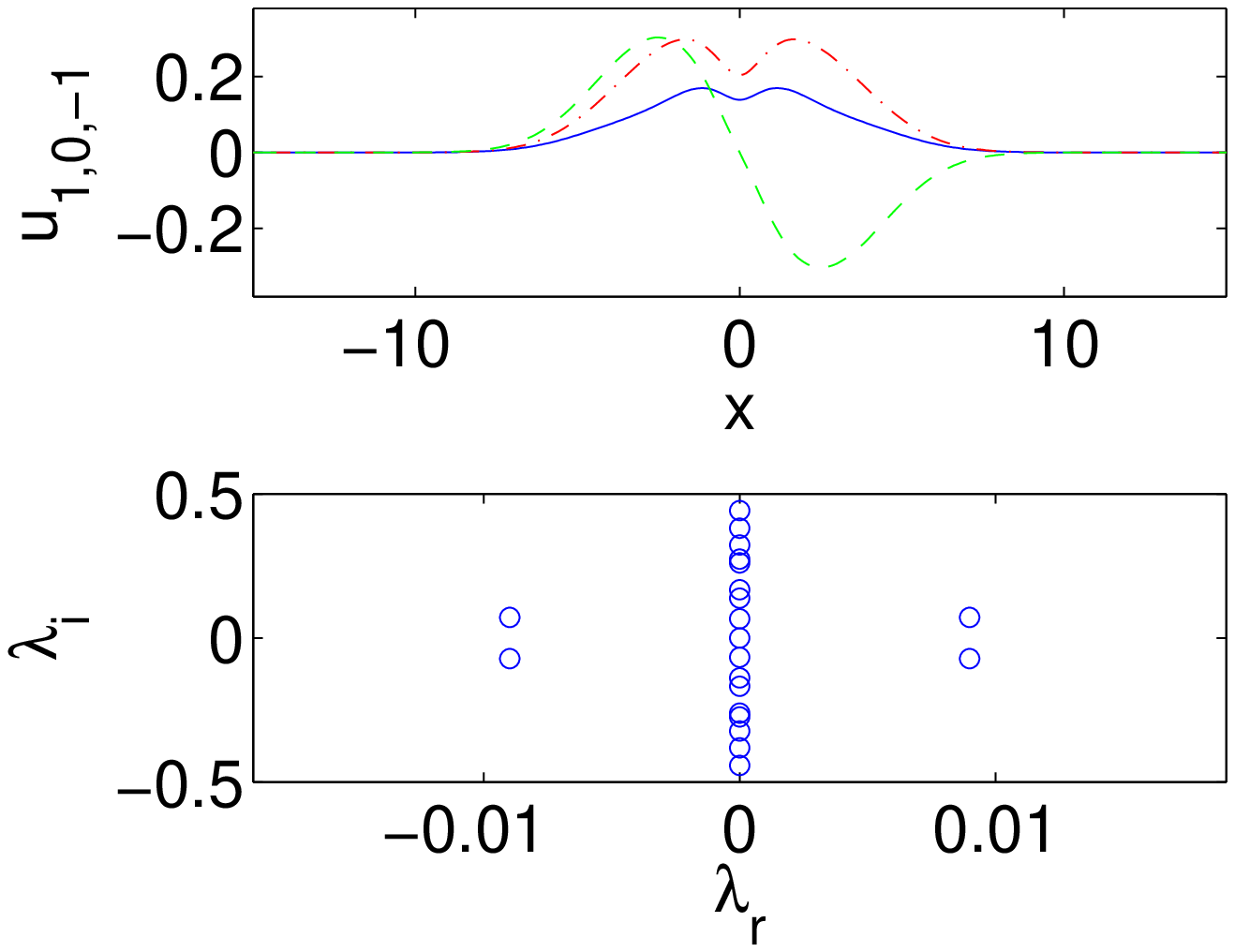}  
\includegraphics[width=.3\textwidth]{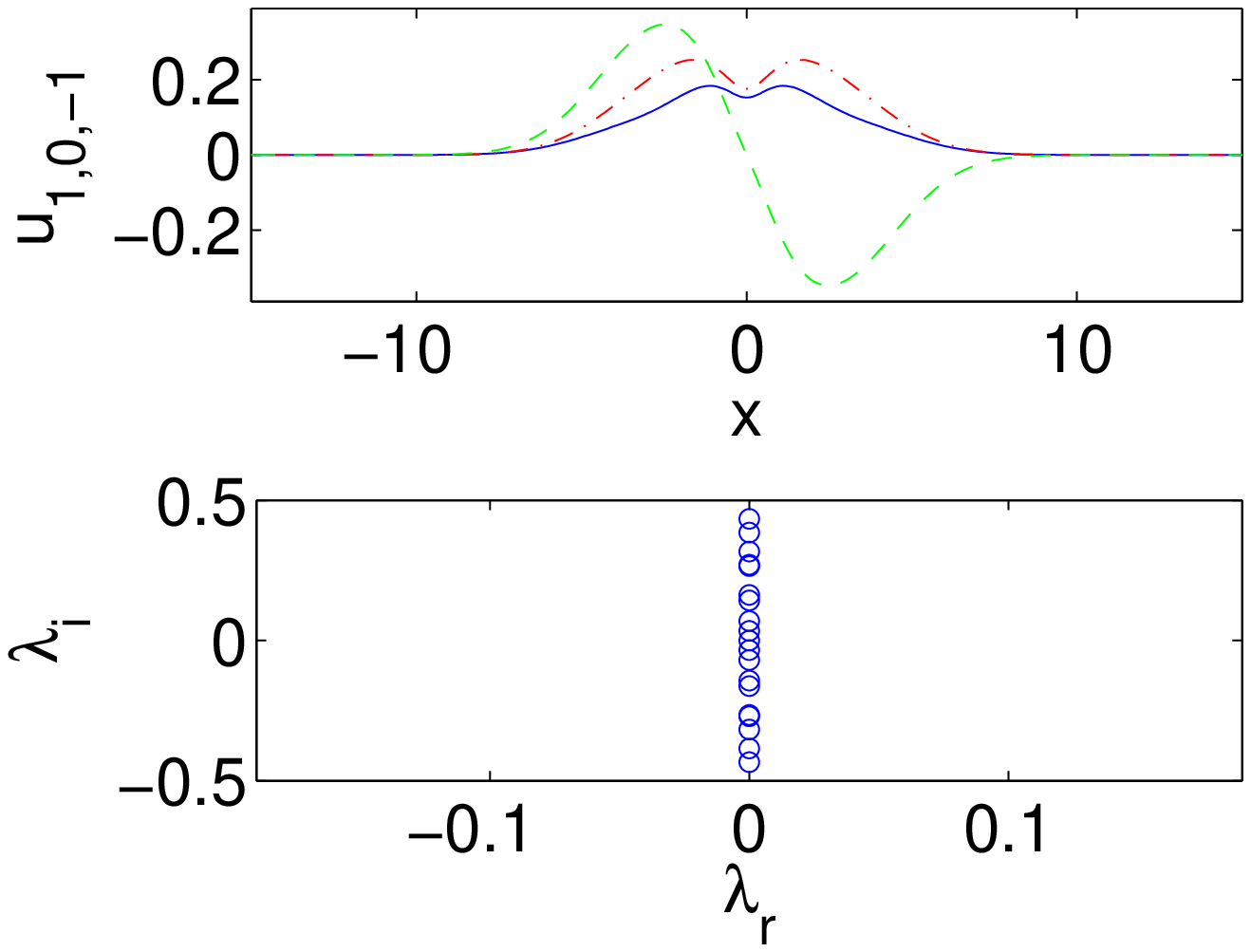}
\includegraphics[width=.3\textwidth]{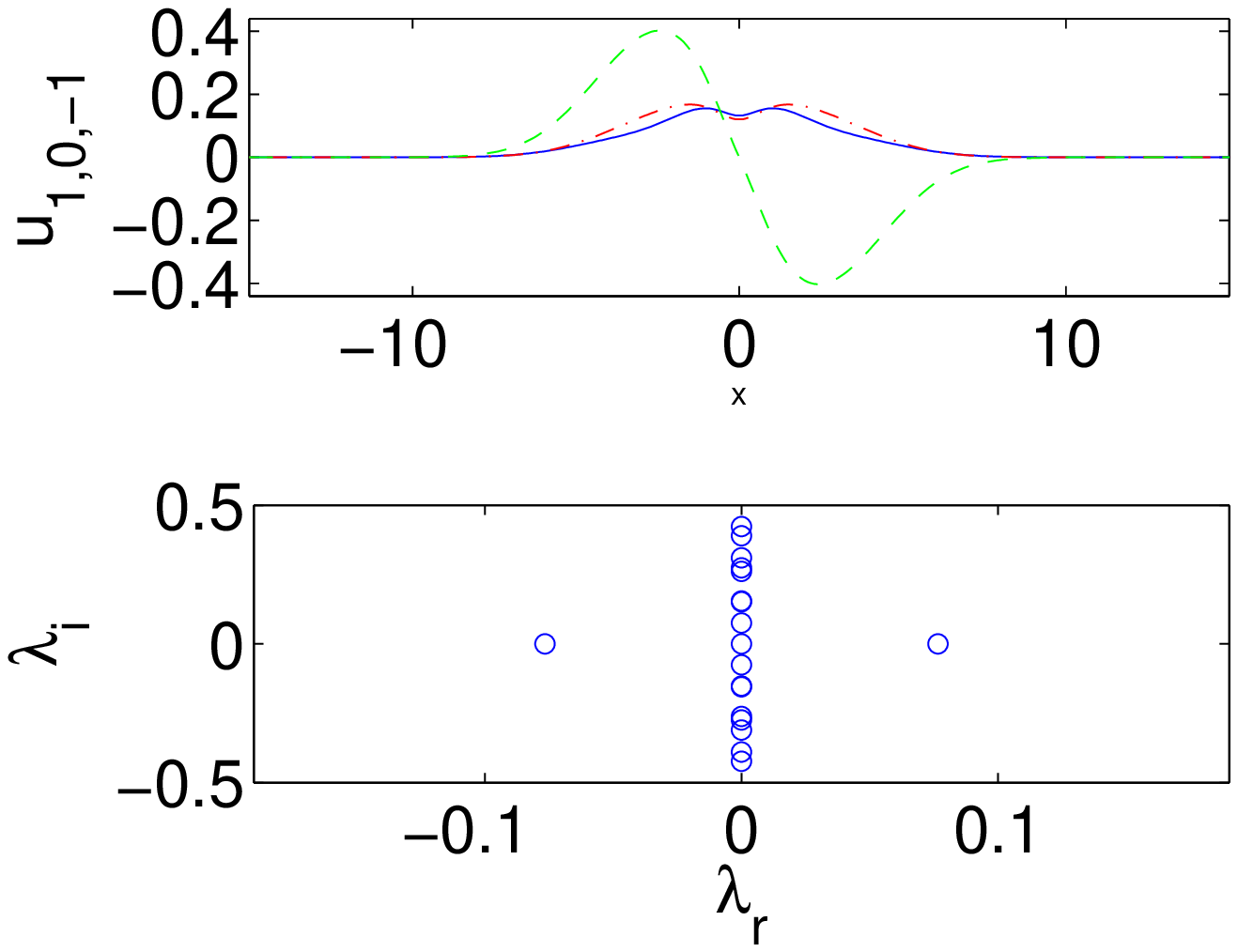}\\
\caption{(Color online) The top left panel shows the maximum real eigenvalue 
as a function of $\mu_0$ for C6 in Fig. \ref{figNa}. The other panels show 
the profiles of wave functions $u_{1}$, $u_{0}$, $u_{-1}$ [by solid (blue), 
dashed (green) and dash-dotted (red) lines, respectively], and stability 
eigenvalues corresponding to branches C5 and C6 in Fig. \ref{figNa}: C5 for $\mu_{0}=0.48$ 
(top right), C6 for $\mu_{0}=0.447$ (bottom left), C6 for $\mu_{0}=0.45$ 
(bottom middle) and C6 for $\mu_{0}=0.455$ (bottom right) with $\mu_{-1}=0.38$. 
The profiles of $u_{1}$ [solid (blue) lines] are multiplied by 40 to be more visible.}
\label{figNaC56}
\end{figure}

We now turn to states involving all three components, such as the
ones analyzed in Figs. \ref{figNaC12} and \ref{figNaC56}.
These branches resemble the ones that we 
analyzed above regarding
states D1-D4; in particular, they connect S0 with AN1 (or with AN2)
or AN0 with S1 (or with S2). However, there is a subtle difference: 
in the case of the branches D1-D4 the coupling of the
components $u_1$ and $u_{-1}$ did not involve any contribution from
(or to) component $u_0$, since in this case the last term in 
the system of Eqs. (\ref{eq012})-(\ref{eq0}) is inactive. 
{\it However}, in the presence
of a nonvanishing component $u_0$, this last (four-wave-mixing type) 
term becomes activated and even though the branch connects, say, S0 with 
AN1 (as is the case for the branch C2), this waveform introduces a 
contribution in the component $u_{-1}$, rendering the solution nonzero 
in all three components. Furthermore, the nature of the profile 
of $u_0$ and $u_1$ combined with the functional form of this term 
($\propto u_0^2 u_1$) determine the parity of the $u_{-1}$ component 
[in the case of C2 it should be anti-symmetric as is indeed shown 
in Fig. \ref{figNaC12}]. Out of this genuinely three-component
solution C2, bifurcates an asymmetric variant thereof (again through
a supercritical pitchfork) destabilizing the solution and the resulting
new branch, C1, is asymmetric and again involves all three components of
the spinor condensate. [It is worthwhile to note in passing that as shown in
 Fig. \ref{figNaC12}, the solution branch C2 could even be unstable
prior to the critical point of the symmetry breaking pitchfork bifurcation
due to a complex quartet and an associated oscillatory instability]. 
Branches C4 and C3 are the analogs of branches C2 and C1 but involve predominantly
the component $u_{-1}$ (and weakly generate the component $u_1$).
Branches C3 and C1 eventually merge with AS2 and AS1 respectively.
Then, branch C6 involves a similar coupling which joins S2 and AN0,
as is shown in Fig. \ref{figNa} and Fig. \ref{figNaC56} (instead of S0 and AN2);
out of that bifurcates the stable, three-component asymmetric branch
C5 also shown in Fig. \ref{figNaC56}. The analog of C6 and C5 involving
a coupling between components $u_{1}$ and $u_{0}$ (instead of $u_{-1}$ and $u_{0}$)
is given by the branches C8 and C7 which have also been identified in
Fig. \ref{figNa} (but for  $\mu_{-1}=0.48$). 
Branches C7 and C5 both merge with AS0 eventually. 
Notice that in the rightmost panel of the third row of Fig. \ref{figNa},
the branch C8 joining S1 and AN0 and the stable asymmetric 
branch C7 bifurcating off of that again appear to almost coincide (norm-wise).
Importantly, it should be noted that none of these 
solutions can be captured in any way by the SMA. 

On the other hand, as is evident in Fig. \ref{figNa}, by the comparison
of the top right panel of the two-mode Galerkin-type 
approximation with the full results of the original system, 
the two-mode approximation performs very well in providing a semi-analytical prediction 
(obtained through the solution of a few simple nonlinear algebraic
equations) of the bifurcation diagram of the  full system.
It is clear that quantitatively the two-mode approximation 
represents the relevant branches less accurately for higher values
of the number of atoms $N$; e.g., the branch C1 reaches values
of $N \approx 3$ as $\mu \rightarrow 0.52$ (at the right end of the
Fig. \ref{figNa}), while 
from the two-mode approximation it only reaches 
$N \approx 2.5$. Nevertheless, the method traces in a systematic
manner {\it all} the branches that can also be found in the full numerical
results and their corresponding bifurcations. 

From the above, it can be inferred that the full bifurcation diagram
is extremely complex in the spinor BEC case involving not only
``pure'' one-component states, but also additional two-component
and in fact even fully three-component ``spinorial'' states. The new
emergent states are always a form of intermediary between original
pure states and also possess symmetry breaking bifurcations of their
own giving rise to genuinely spinorial, spatially symmetry broken states.
Despite the fact that many of these states (including the asymmetric ones
or even the symmetric/antisymmetric ones prior to their pitchfork bifurcation points)
are genuinely stable dynamical states of the $F=1$ spinor BEC system, they are 
{\it not} captured in the framework of the SMA. 


\begin{figure}[tbph]
\centering
\includegraphics[width=.35\textwidth]{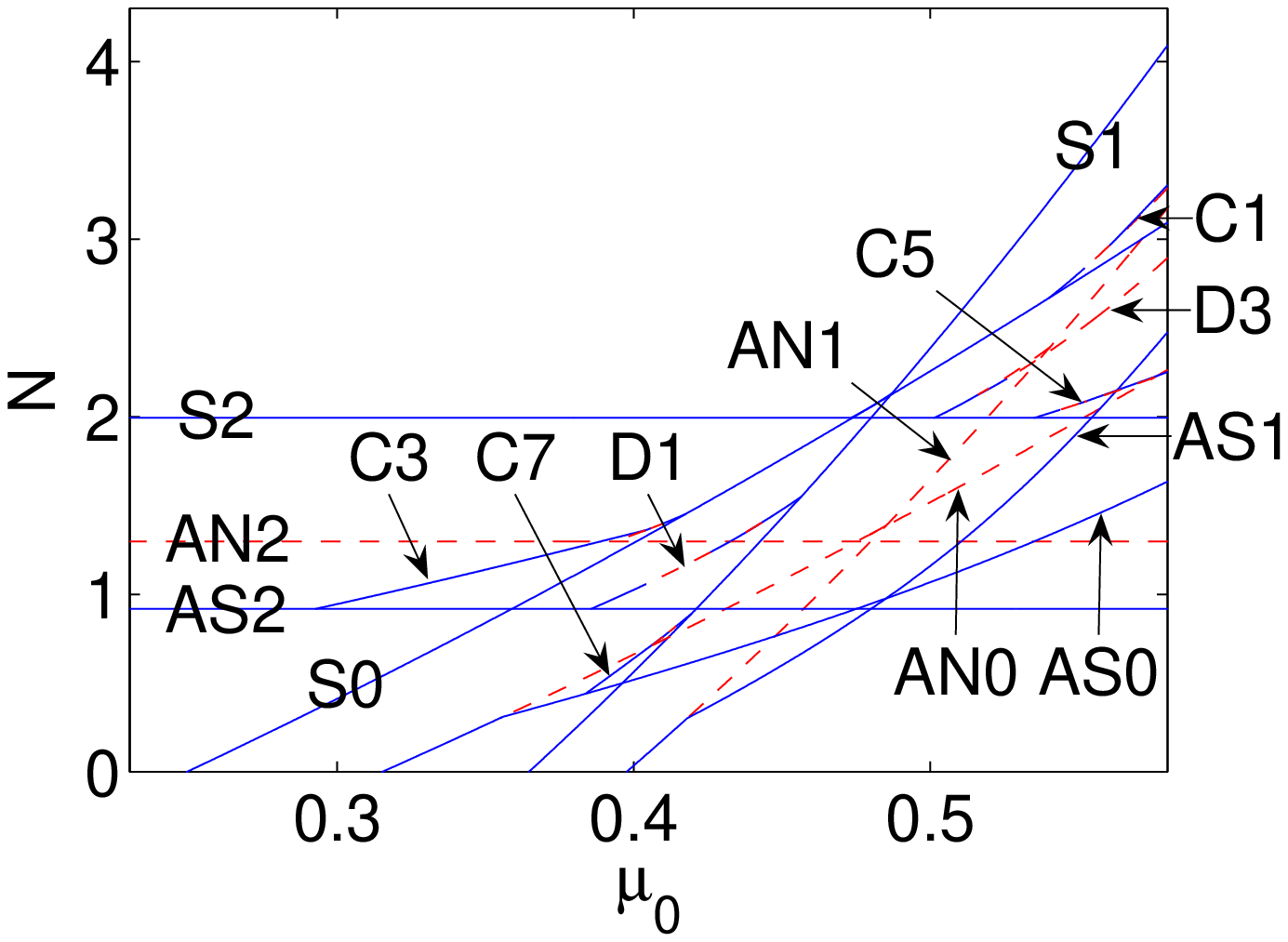} 
\includegraphics[width=.35\textwidth]{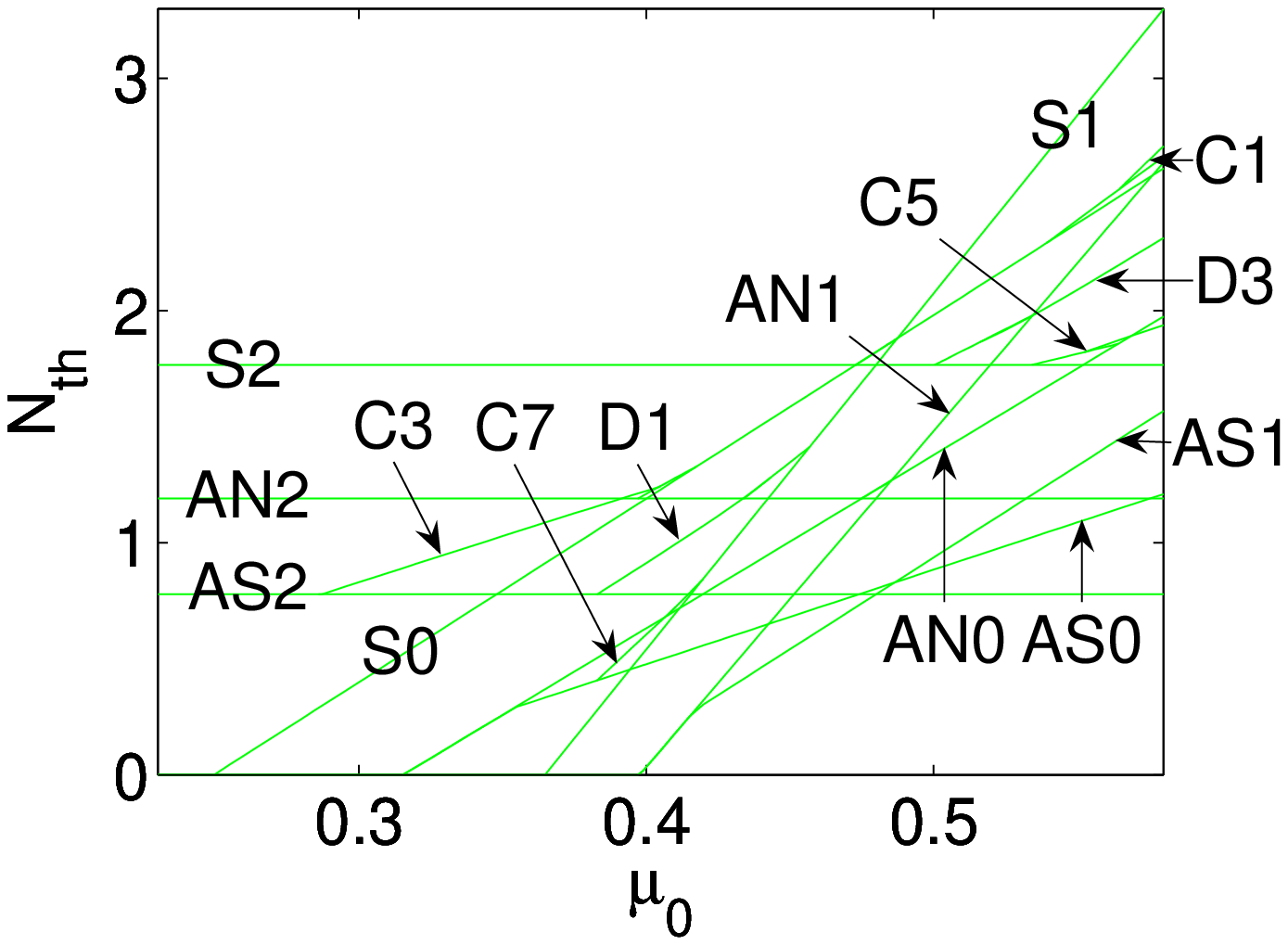}\\
\includegraphics[width=.35\textwidth]{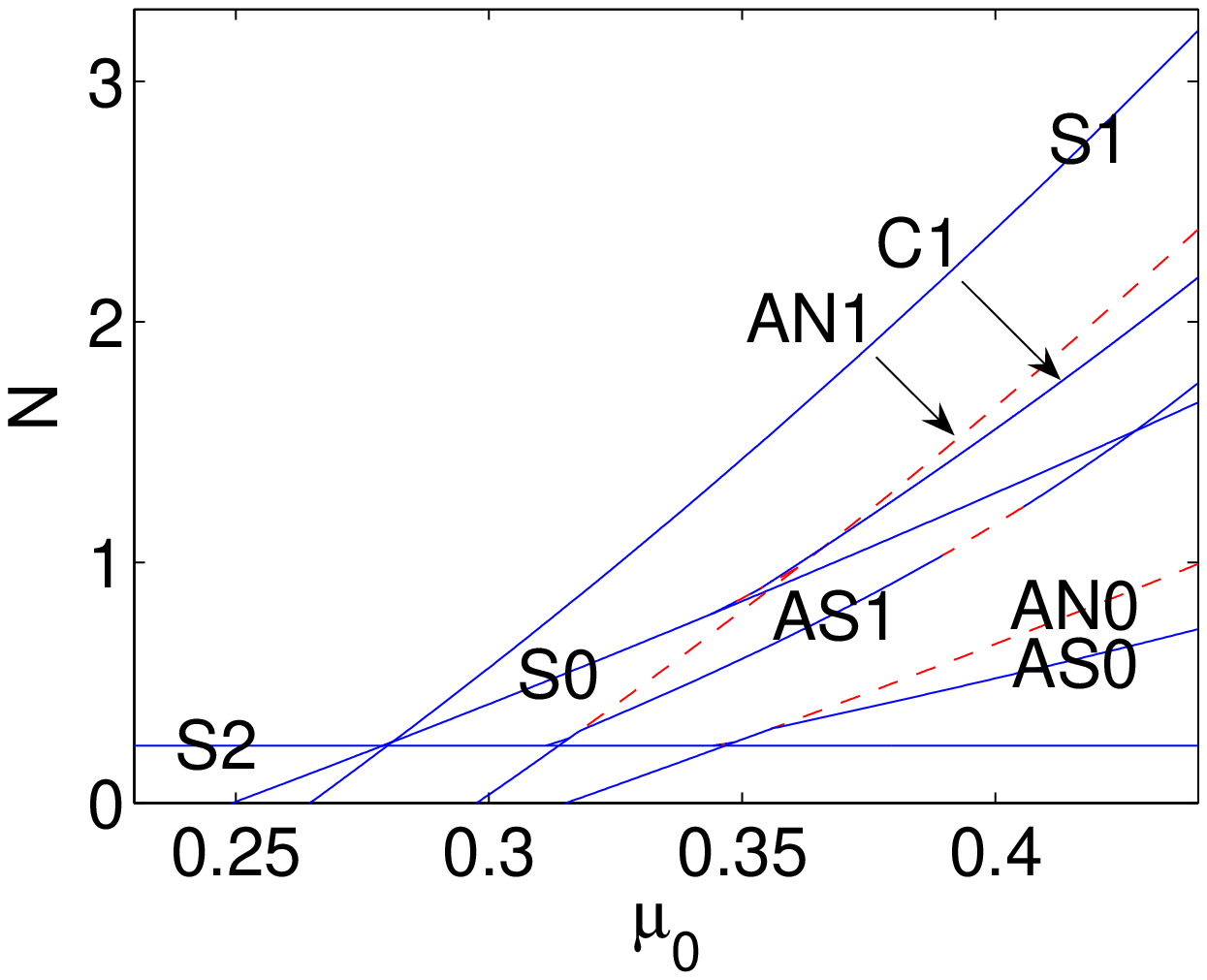} 
\includegraphics[width=.35\textwidth]{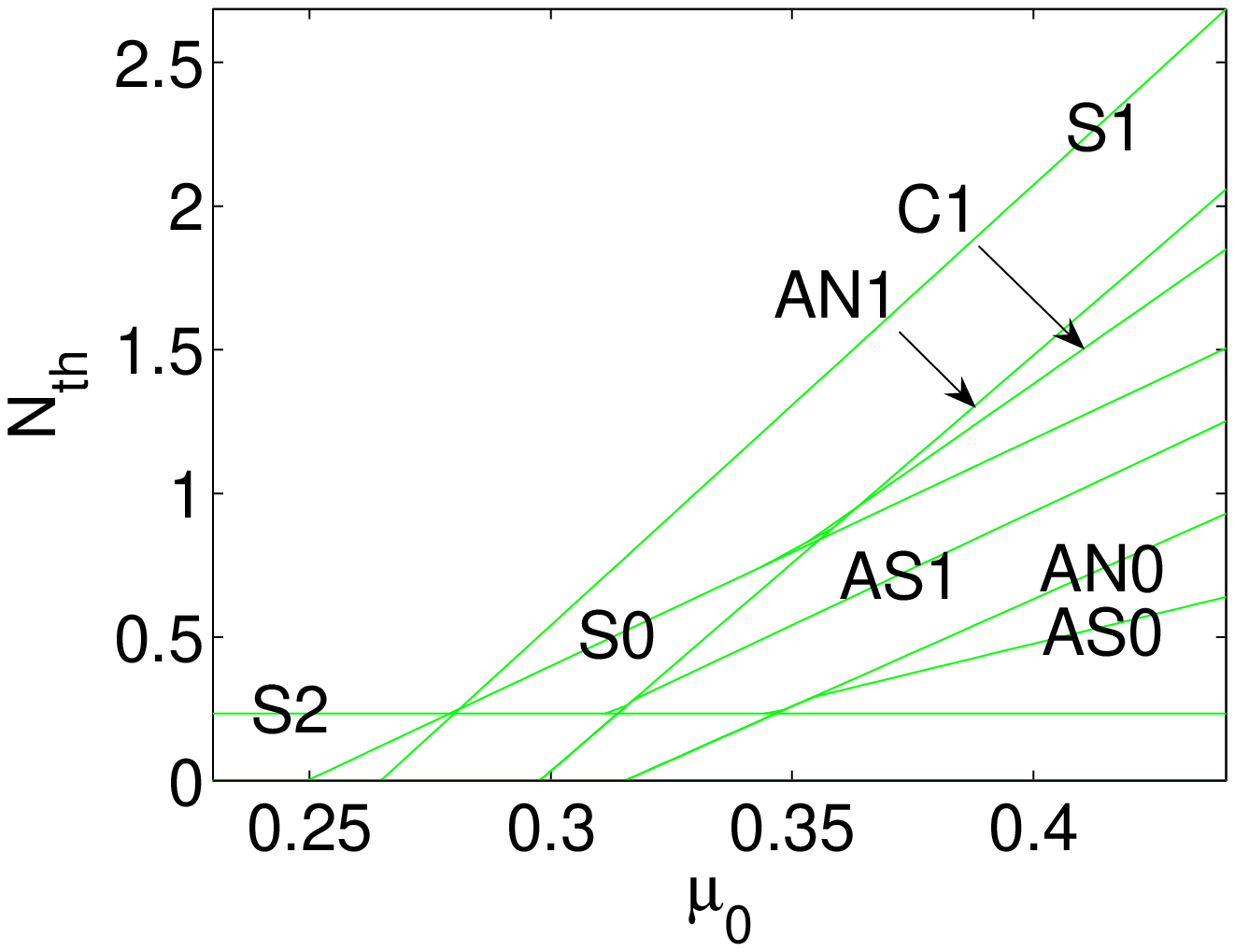}\\
\includegraphics[width=.35\textwidth]{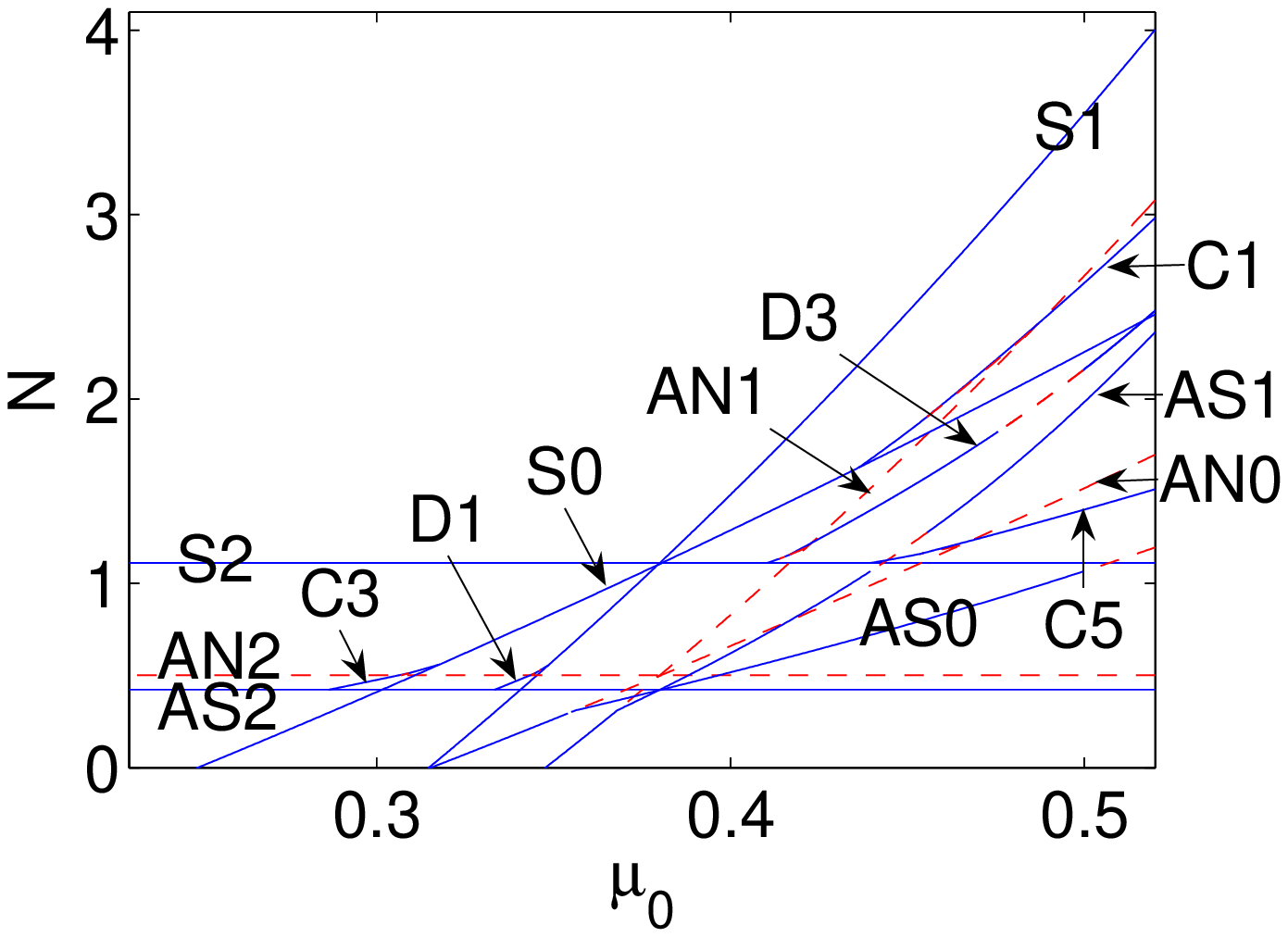} 
\includegraphics[width=.35\textwidth]{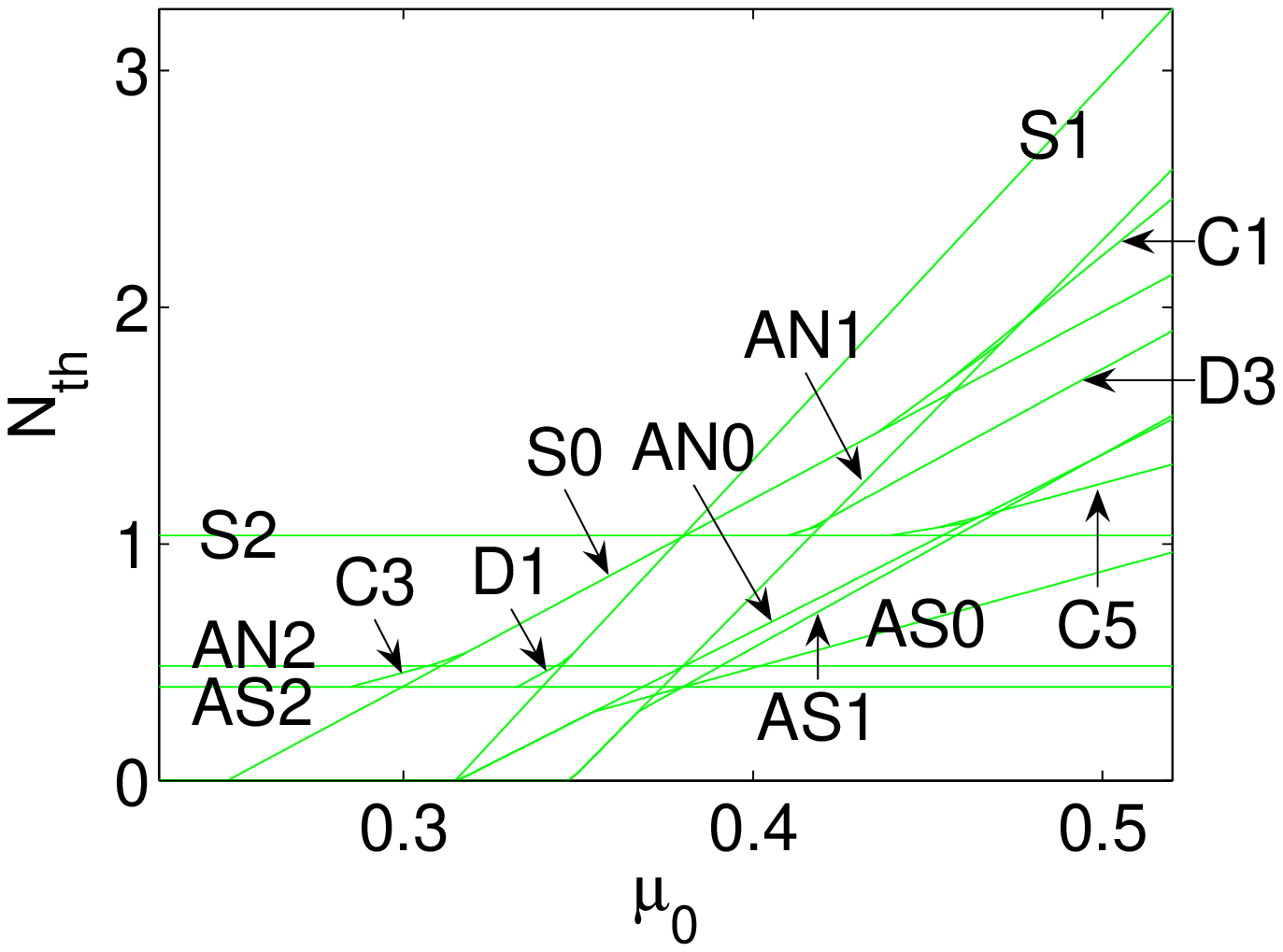}\\
\caption{(Color online) The norm of the numerically found (left) and the approximate two-mode (right)
solutions of Eqs. (\protect\ref{eq012})-(\protect\ref{eq0}) as a 
function of $\mu_{0}$ for $\mu_{-1}=0.48$ (top) and $\mu_{-1}=0.28$ (middle) 
in the case of $^{23}$Na spinor BEC, and $\mu_{-1}=0.38$ in the case of $^{87}$Rb spinor BEC (bottom). 
The notation is the same as in Fig. \protect\ref{figNa}.}
\label{figEx}
\end{figure}

Since Fig. \ref{figNa} was constructed for a particular value of 
the chemical potential $\mu_{-1}$, it is worth commenting on 
how the relevant features may change upon variation of $\mu_{-1}$. 
For this reason, Fig. \ref{figEx} examines both higher and lower 
values of this chemical potential (both in the full system, as well 
as in the reduced two-mode approximation of the six algebraic equations 
--left and right panels, respectively of the top two rows of the figure--).
What is observed is that while a slight increase of the parameter does
not substantially alter the phenomenology, a slight decrease thereof
may result in a drastic decrease in the number of branches observed.
This can be qualitatively understood as follows. When $\mu_{-1}$ 
is decreased, initially AS2, later AN2, and finally even S2 will 
disappear. For $\mu_{-1}=0.28$ shown in Fig. \ref{figEx}, 
already branches AS2 and AN2 have disappeared and only S2 persists among 
the pure $u_{-1}$ branches. As a result, all the branches that 
would ``connect'' to AN2 are also forced to disappear, including, e.g., 
D2 and D1, D3, D6 or C4, C3 and C5. Hence, as a rule of thumb,
decreasing the chemical potentials (in this case of repulsive
interactions) generally reduces the number of available states
that can exist. A guide for potentially identifying such more complex
combined states is the corresponding presence of ``pure'' states
in the system. Notice that in all our considerations herein, we
have focused on chemical potentials $\mu_{-1}<0.61$ such that
nonlinear modes associated with the second excited state do not become
``activated''. For higher values of the chemical potential, obviously
such higher excited states will come into play and the two-mode Galerkin-type 
approximation will no longer be sufficient to address their existence.
Nevertheless, in considering such higher excited states, we have 
typically observed them to be unstable and hence do not pursue them
further herein.

We should note that we also considered along the same
vein the bifurcation diagram for the case of the \textit{ferromagnetic} $^{87}$Rb spinor condensate 
(recall that in this case $\delta = \nu_a/\nu_s = -4.66\times 10^{-3} < 0$).
While this difference in the sign of $\delta$ plays a key role in
important dynamical phenomena such as the modulational 
instability \cite{ofy2,boris,ourbd}, nevertheless the structure and
main features of the resulting state diagram are observed to 
essentially be the same between the two cases, perhaps due to the strong
confinement considered here within the realm of the double well potential.

\subsection{Dynamics}

\begin{figure}[tbhp!]
\centering
\includegraphics[width=.3\textwidth]{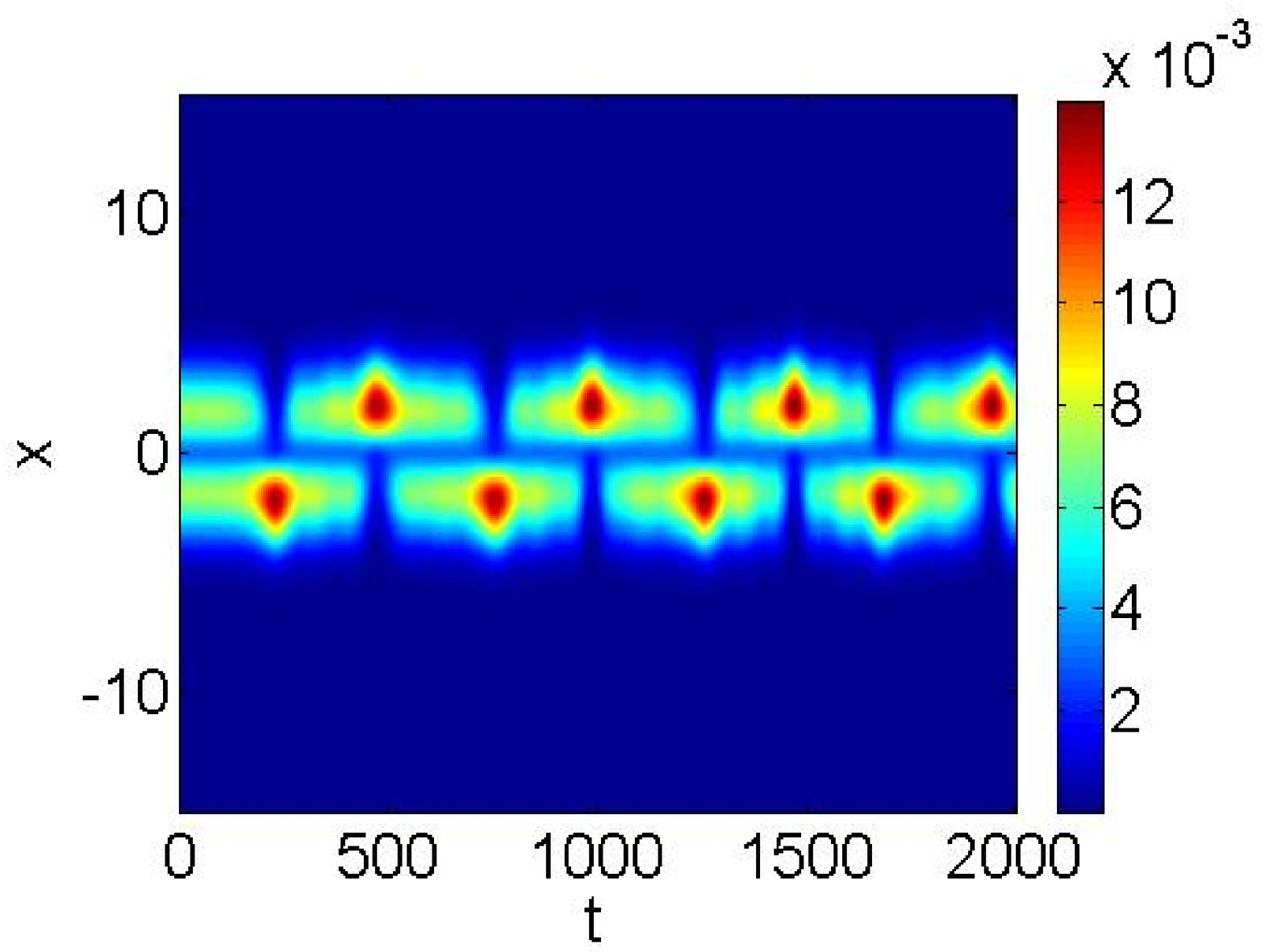}
\includegraphics[width=.3\textwidth]{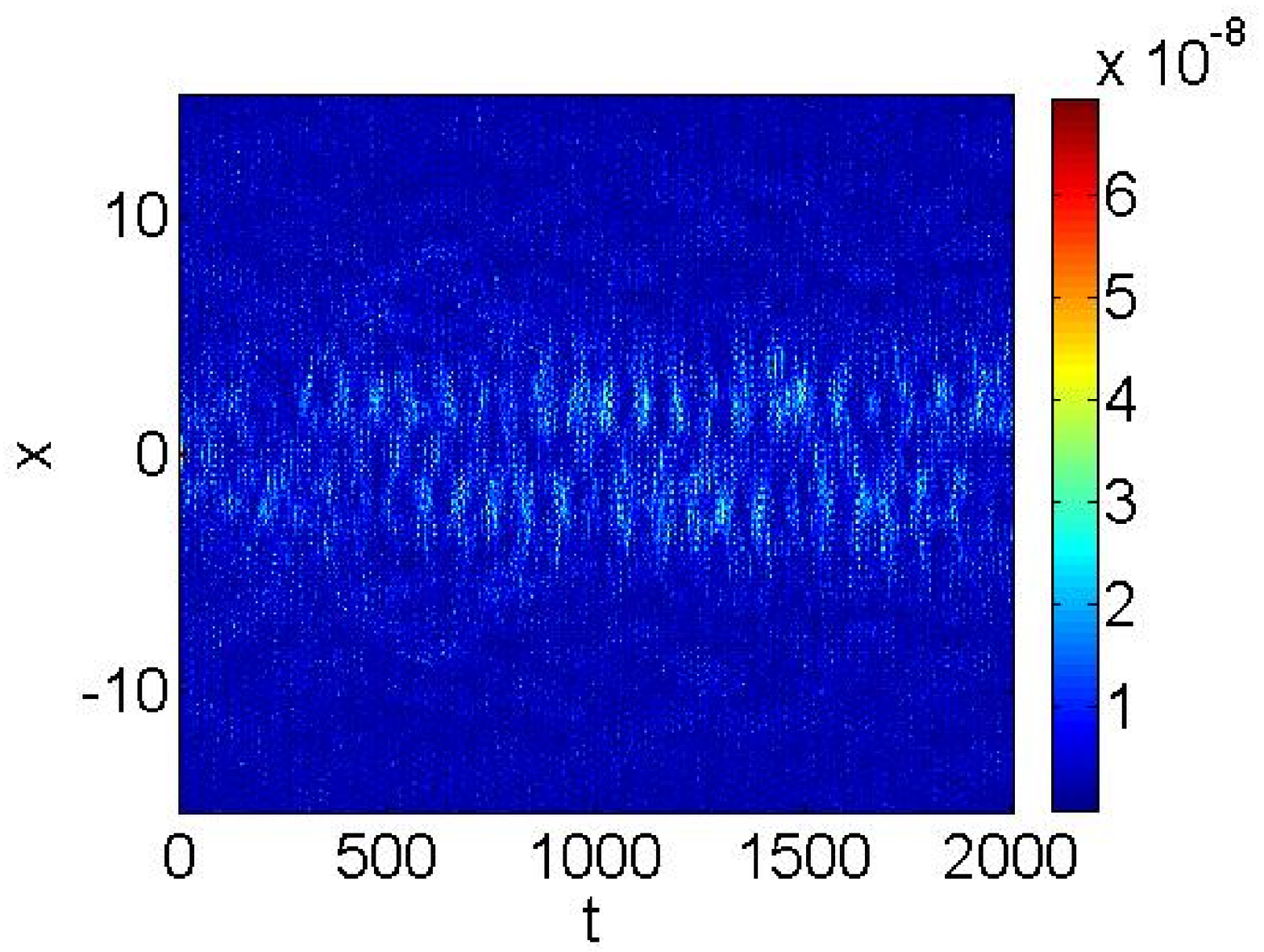} 
\includegraphics[width=.3\textwidth]{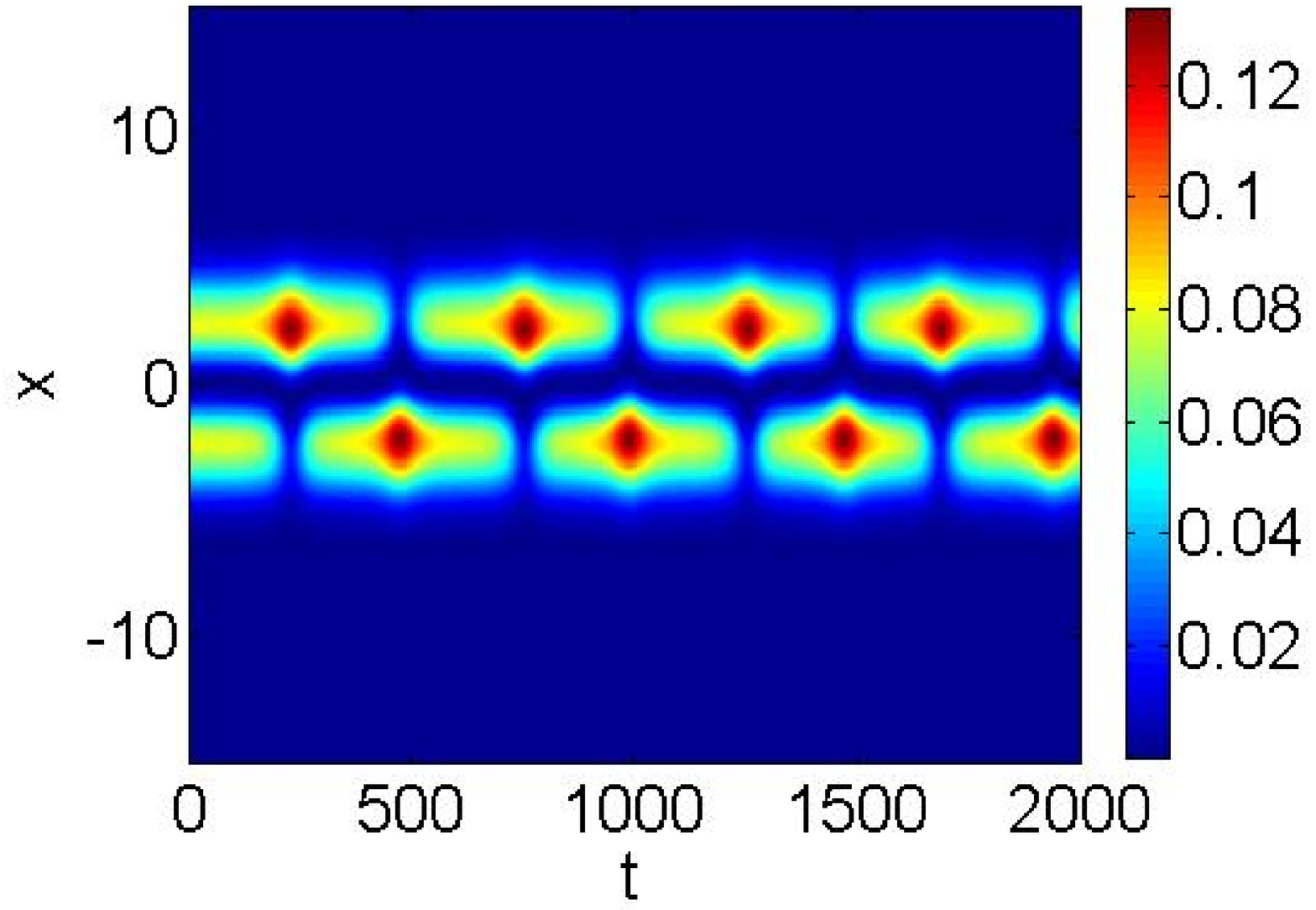}\\
\includegraphics[width=.3\textwidth]{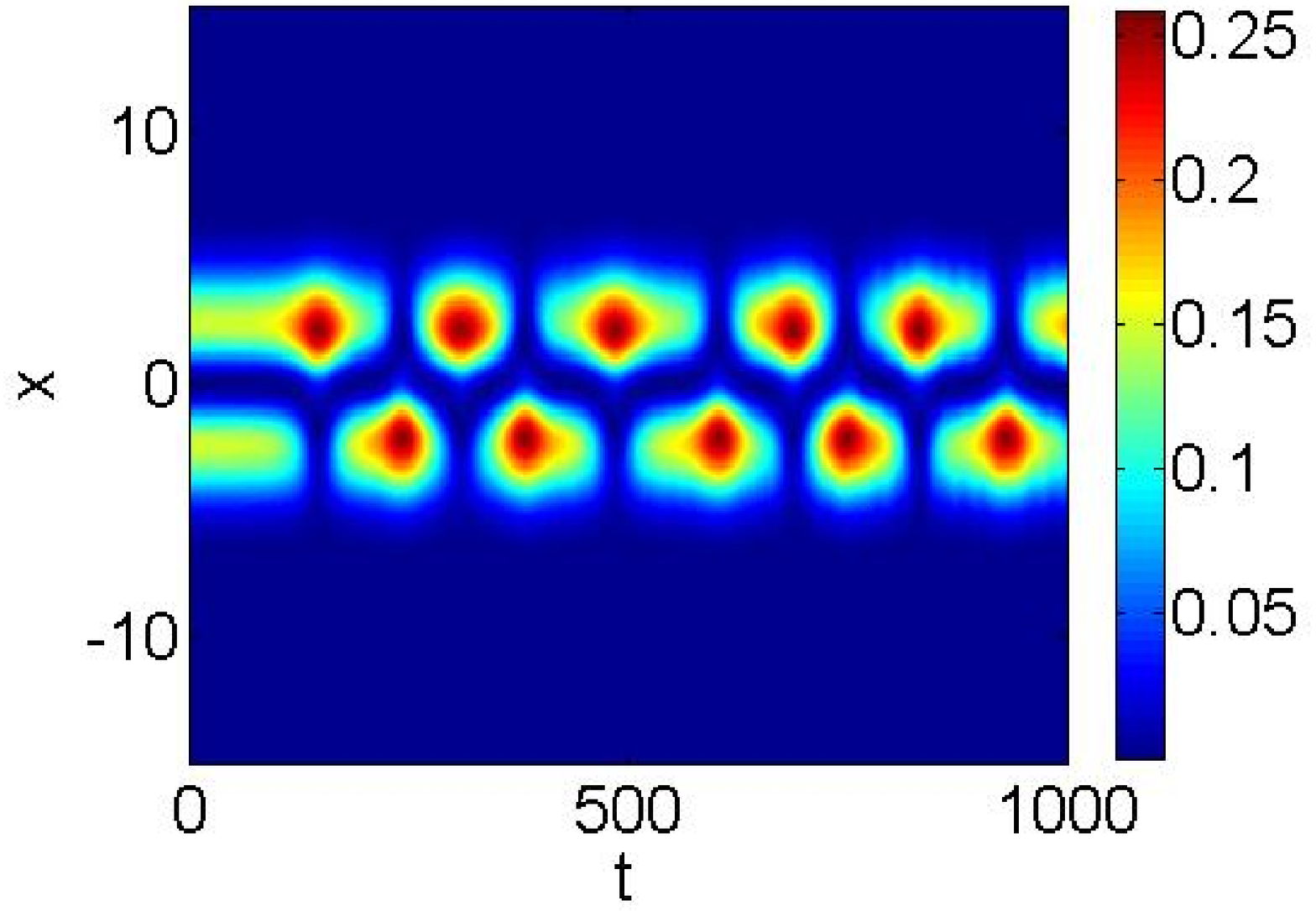}
\includegraphics[width=.3\textwidth]{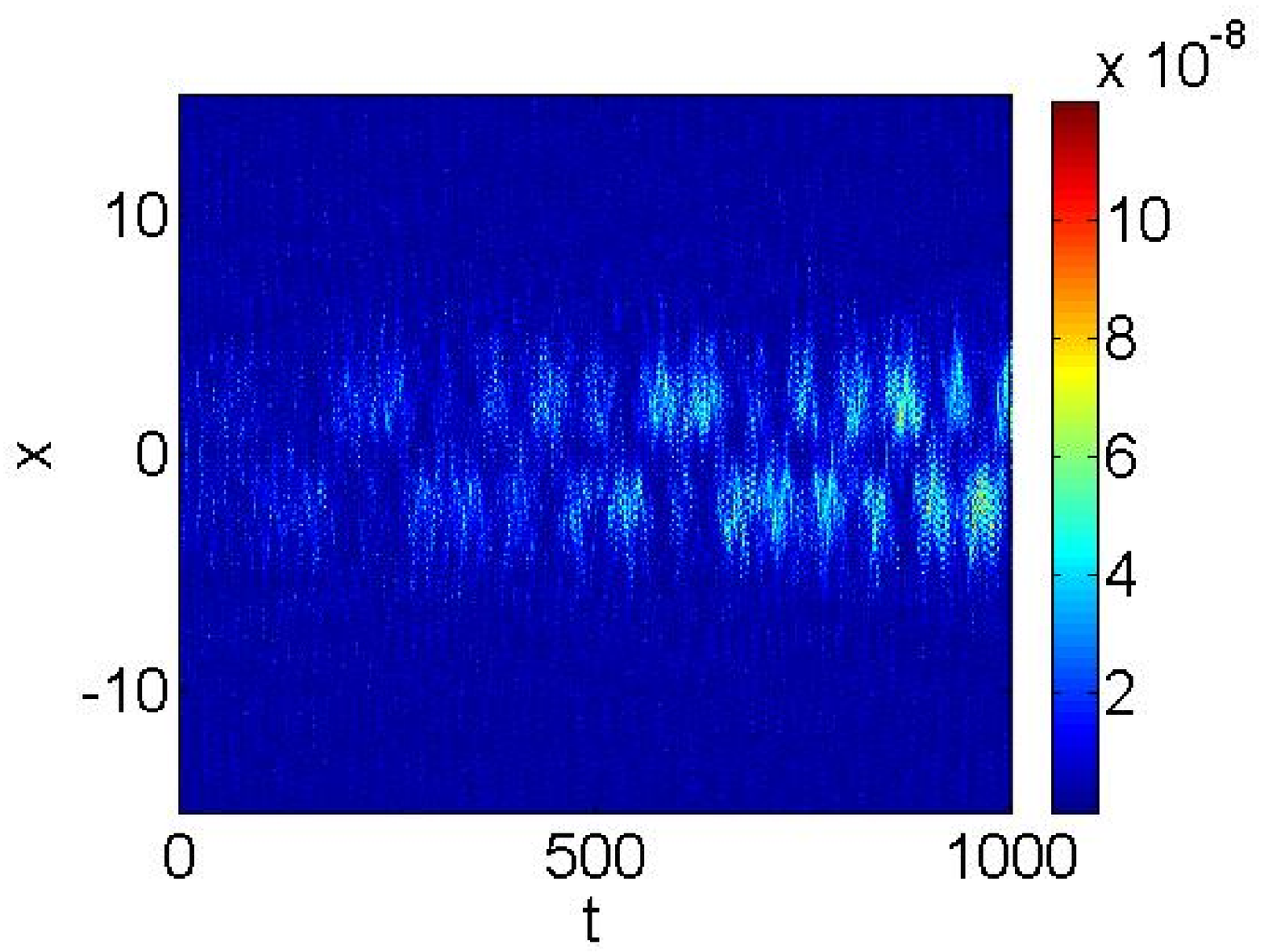}
\includegraphics[width=.3\textwidth]{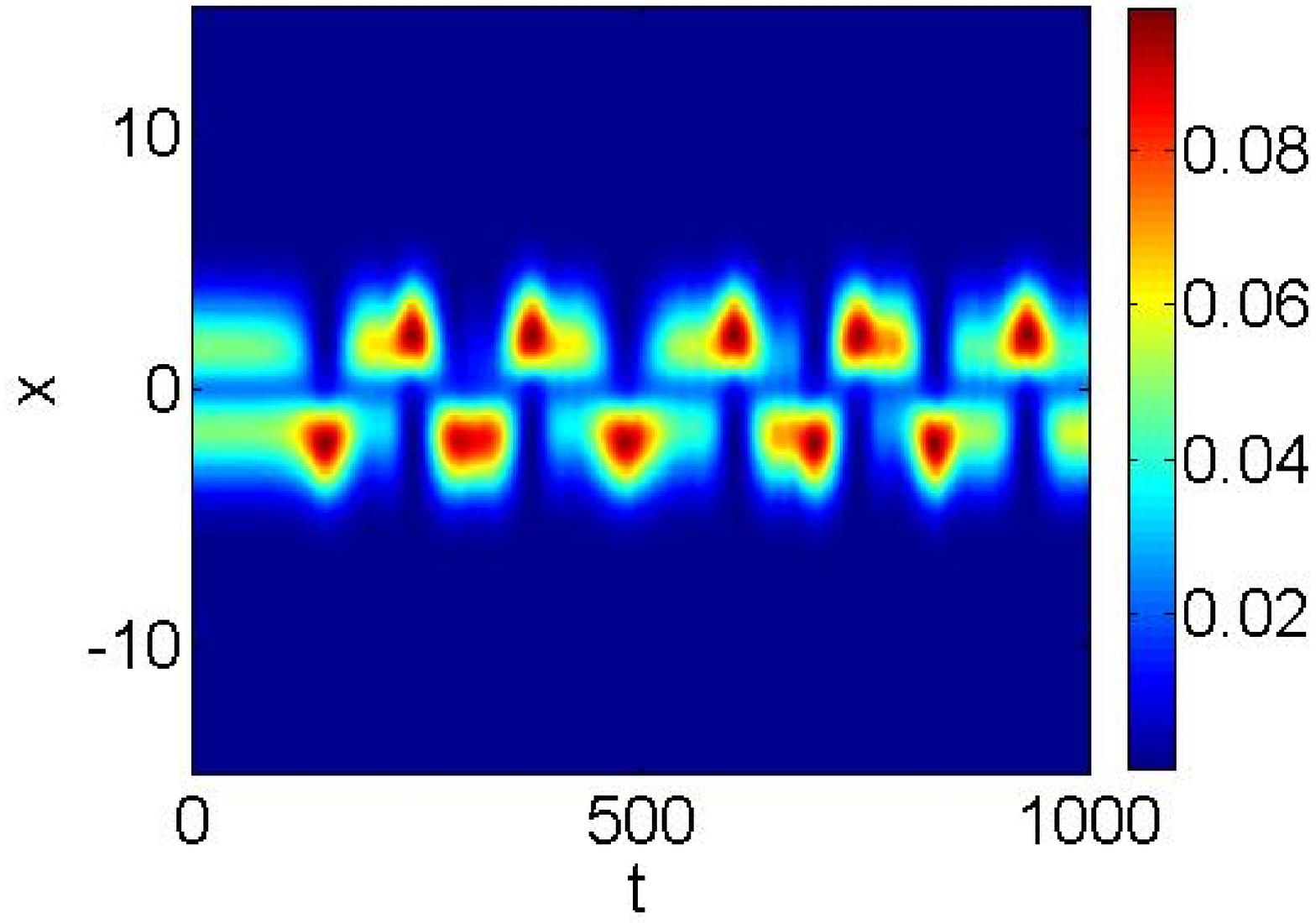}\\
\includegraphics[width=.3\textwidth]{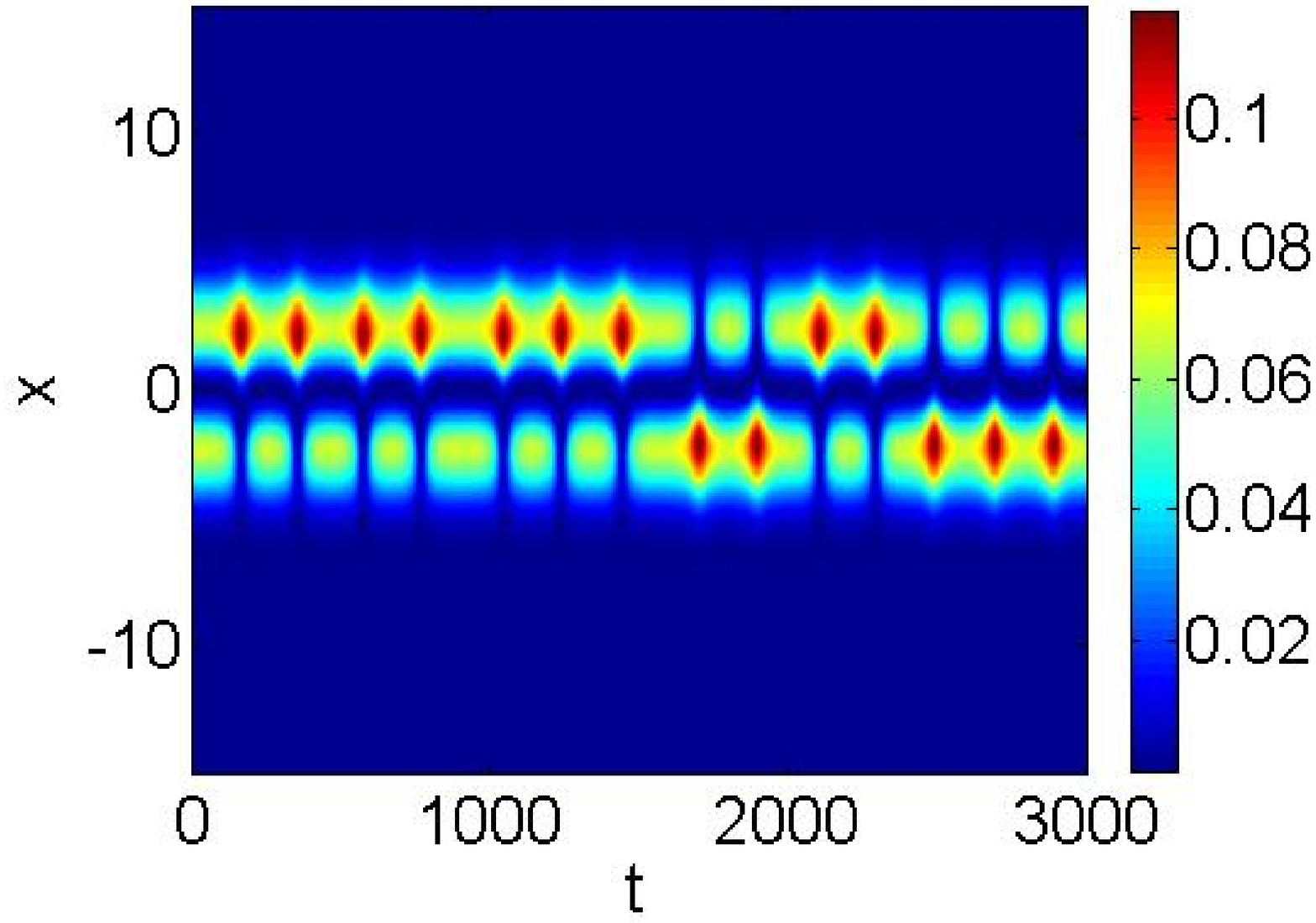}
\includegraphics[width=.3\textwidth]{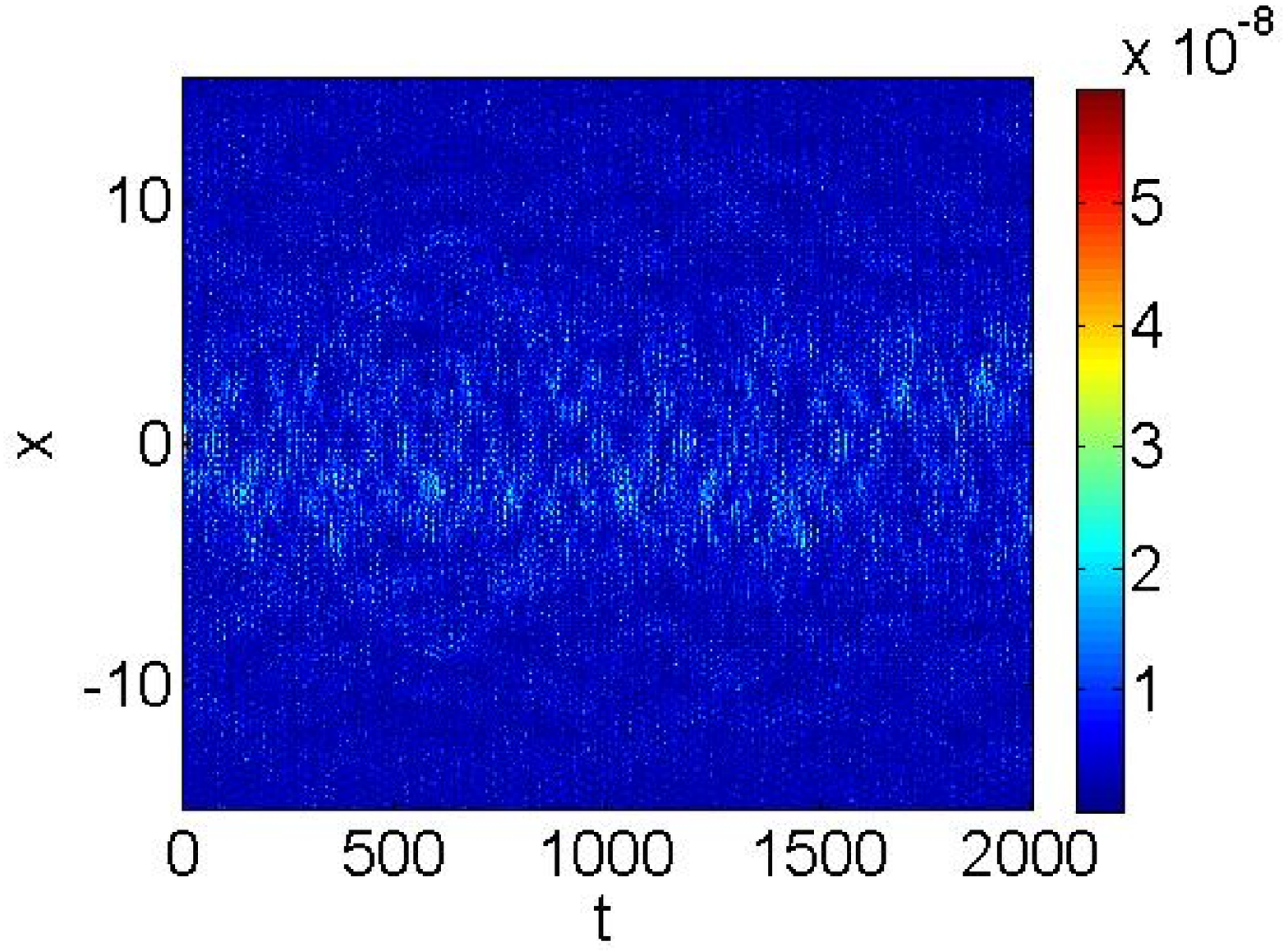}
\includegraphics[width=.3\textwidth]{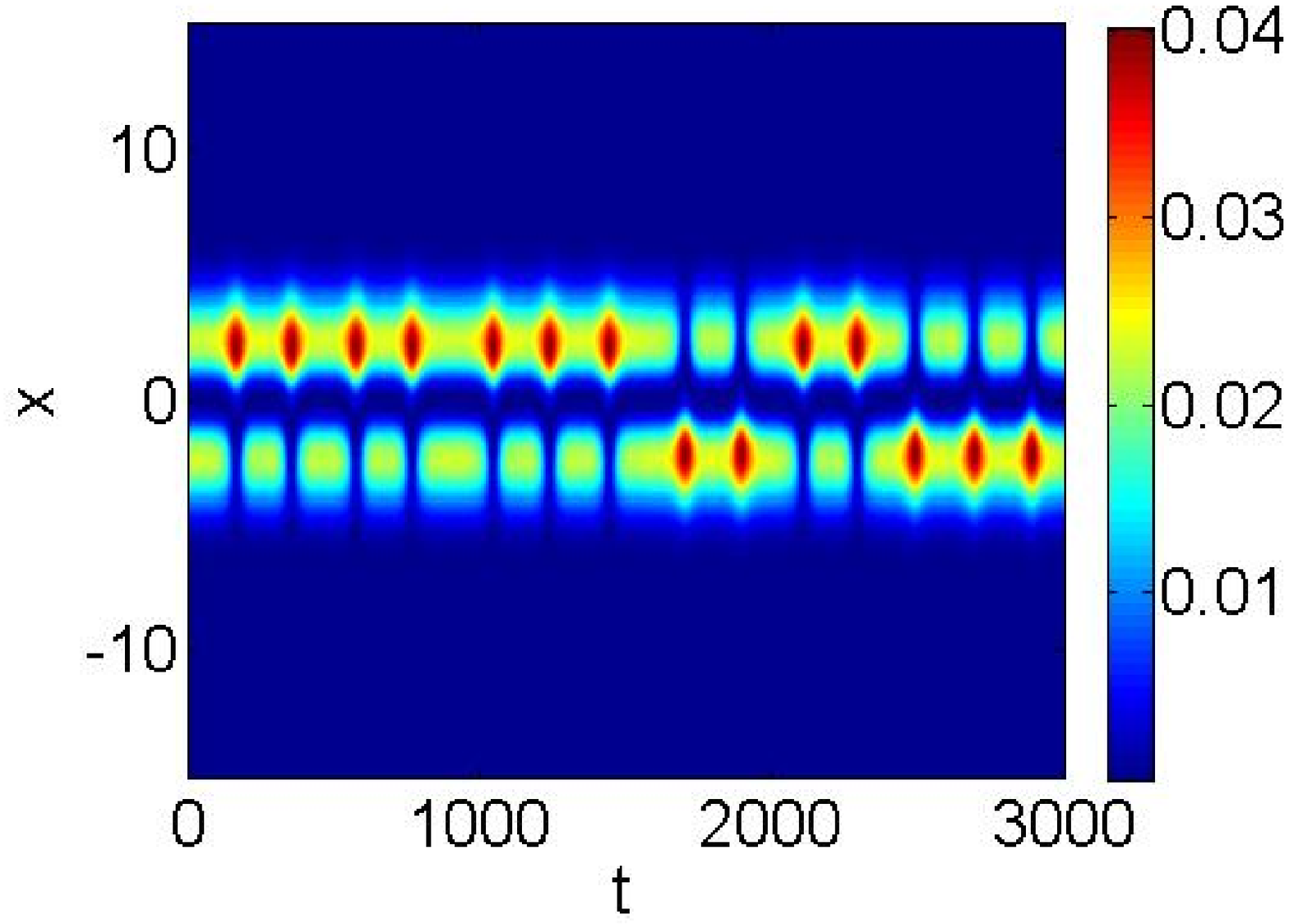}\\
\caption{(Color online) Spatio-temporal contour plots of the densities of 
unstable combined two-component solutions in the case of $^{23}$Na spinor BEC. 
The panels show the simulated evolution of wave functions $\psi_{1}$ (left), 
$\psi_{0}$ (middle) and $\psi_{-1}$ (right) in unstable solutions of D2 (top), 
D4 (middle) and D6 (bottom) from Fig. \ref{figNa}, respectively.}  
\label{figEvolD246}
\end{figure}

We now 
provide some representative examples of dynamics of the various
branches. We commence by considering the branches D2, D4 and D6 respectively
in the top, middle and bottom row of Fig. \ref{figEvolD246}. We can see
that the dynamics and instability of these branches does not strongly
couple to the 0-th component of the spinorial state (and it only does
because of the initial small amplitude noise seeded in that state,
a seeding that takes place in all our simulations in the initial conditions).
In that sense, the dynamical evolution of these states closely resembles
the observations of Ref. \cite{chenyu}. In particular, we observe a nearly
periodic breaking of the symmetry leading to amplification of one
of the two components in one of the wells, while typically the other 
component is amplified in the other well [a notable exception to that
is the case of D6 where the wavefunctions have the same parity, both
being anti-symmetric].

\begin{figure}[tbhp!]
\centering
\includegraphics[width=.3\textwidth]{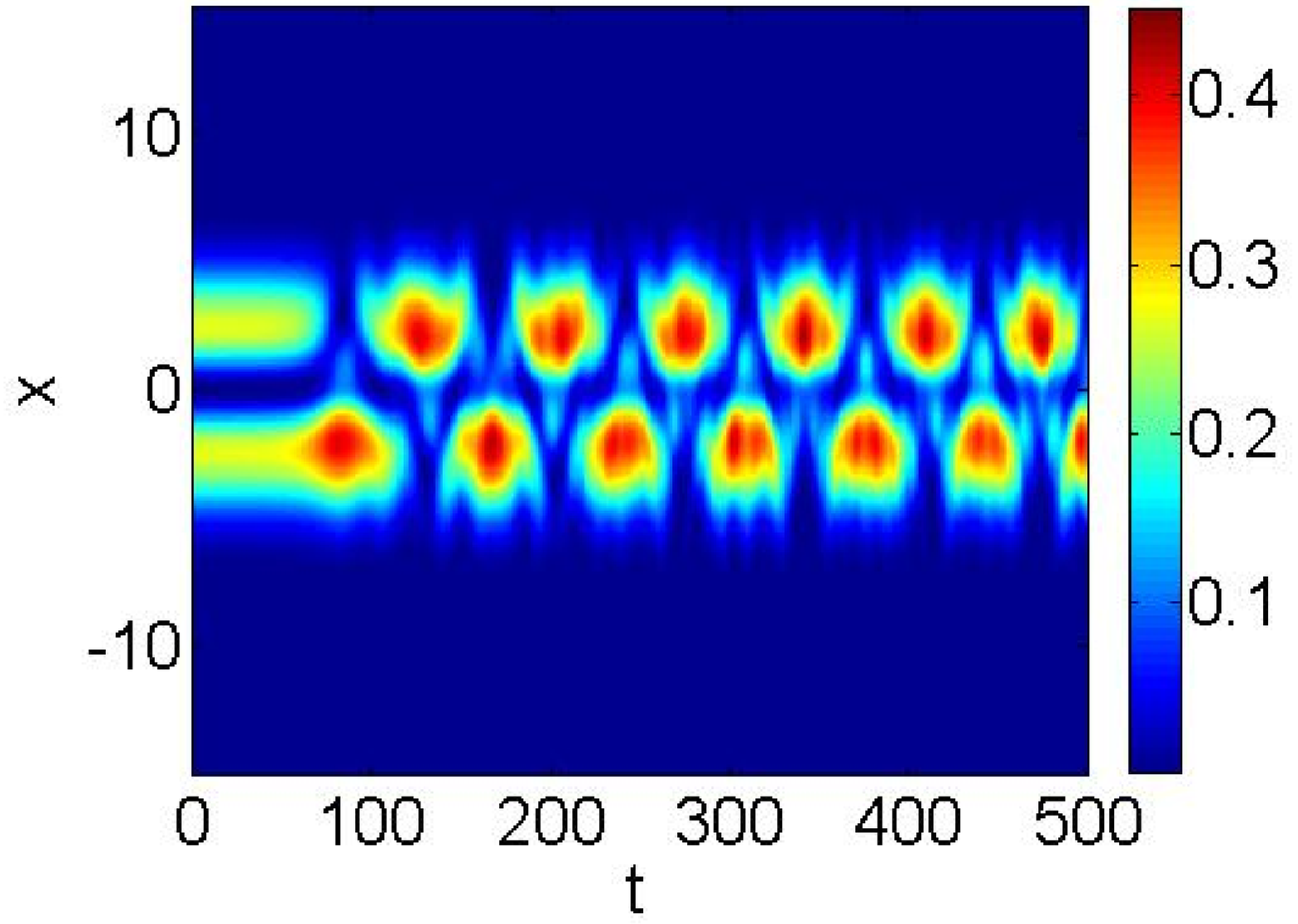}
\includegraphics[width=.3\textwidth]{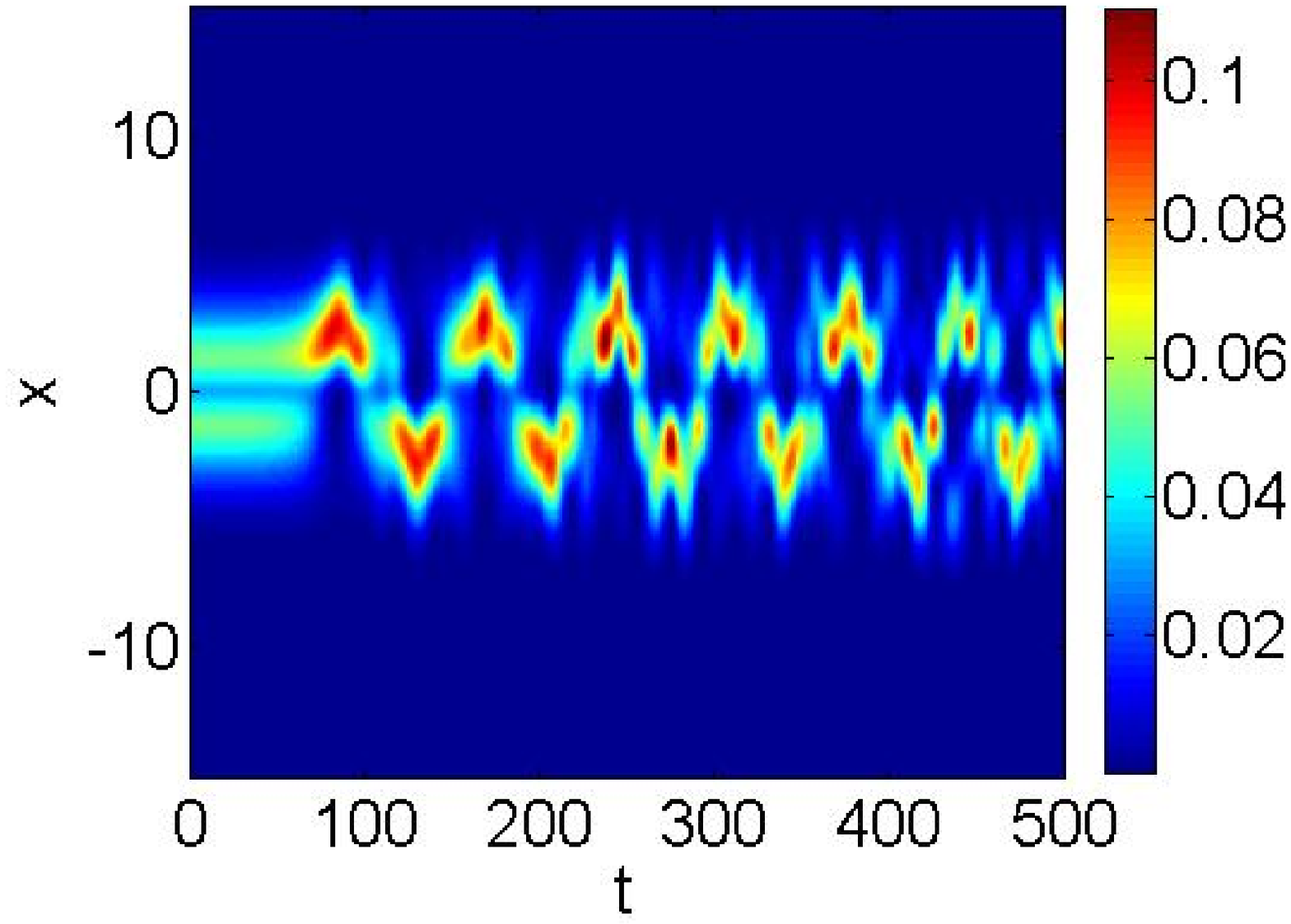} 
\includegraphics[width=.3\textwidth]{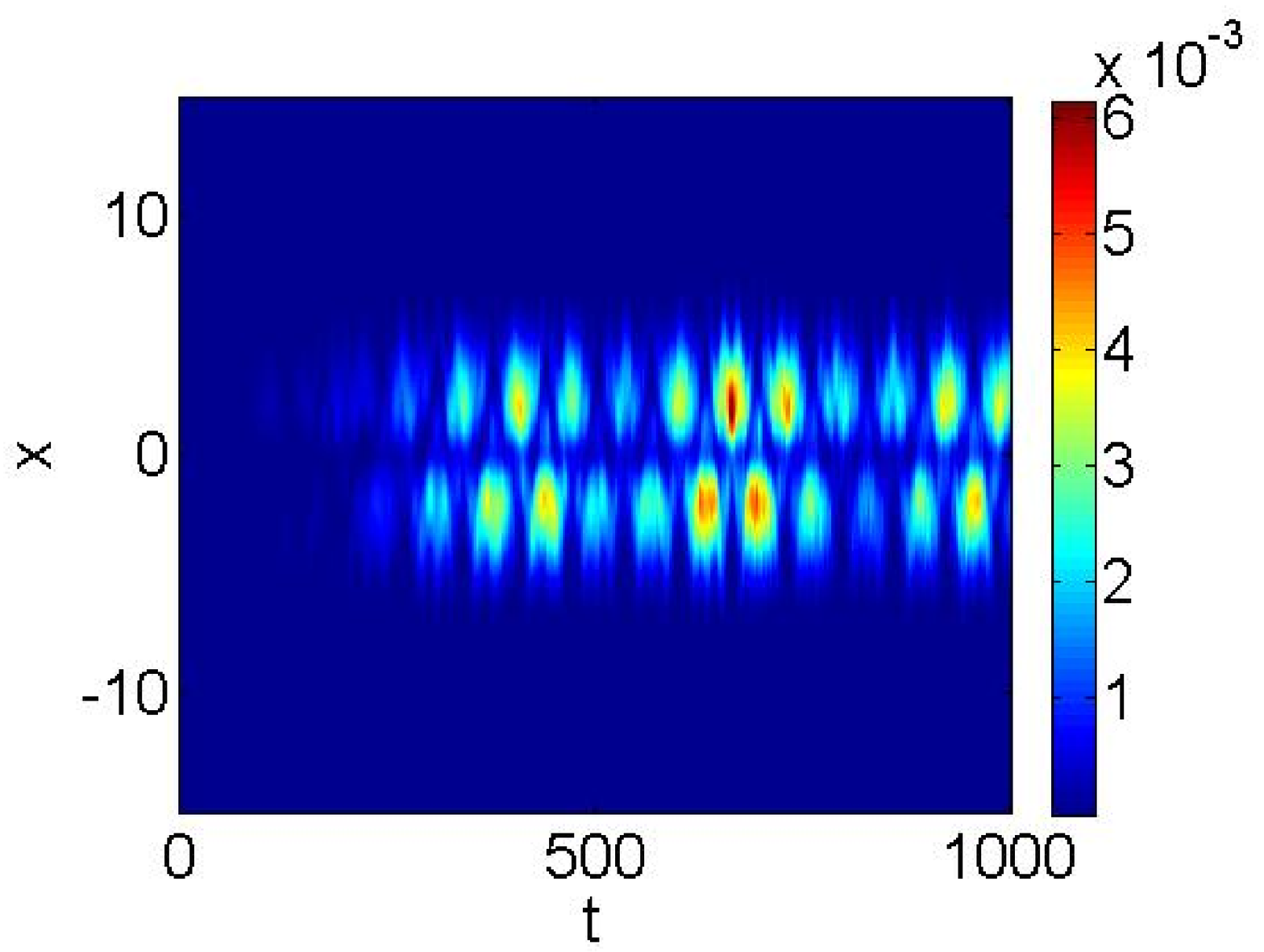}\\
\includegraphics[width=.3\textwidth]{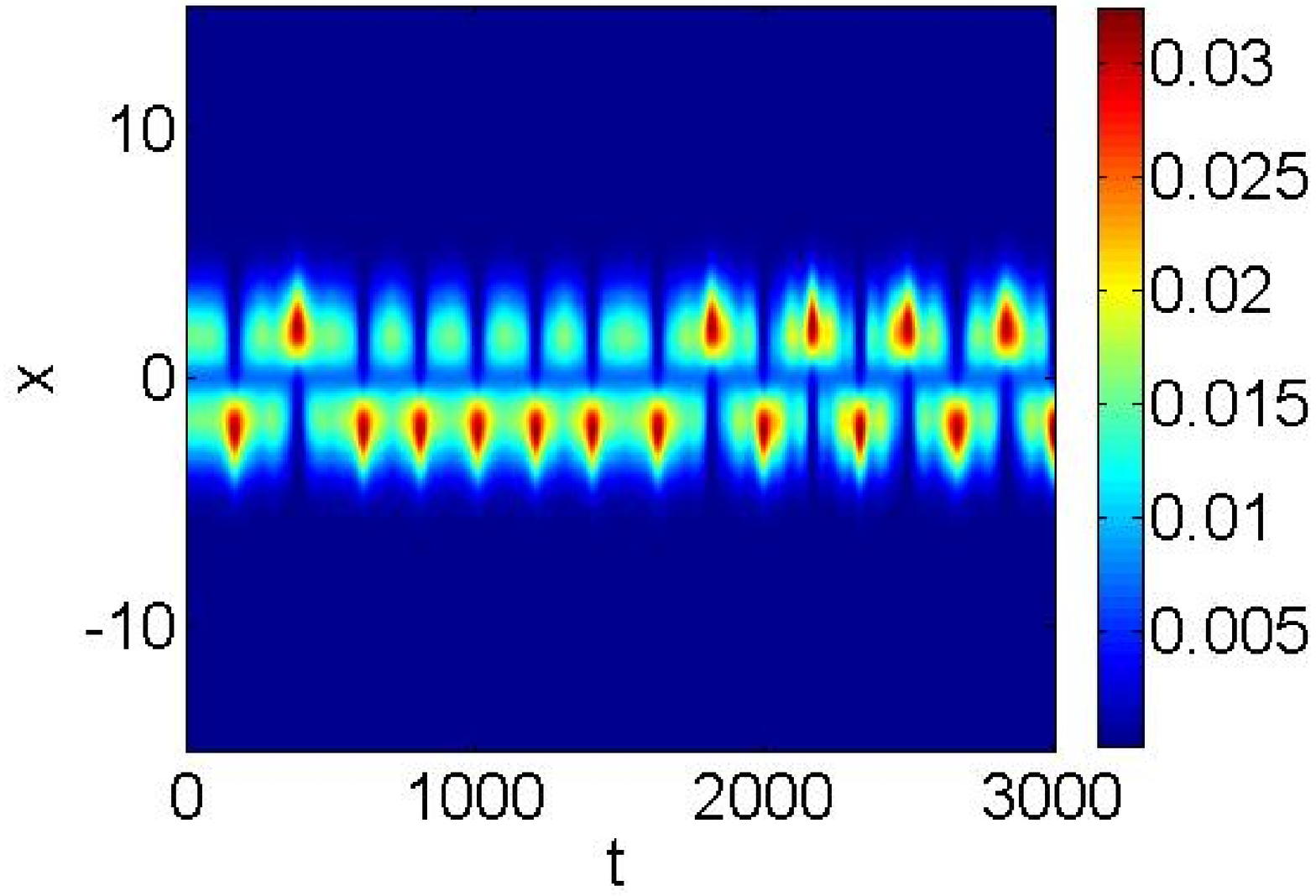}
\includegraphics[width=.3\textwidth]{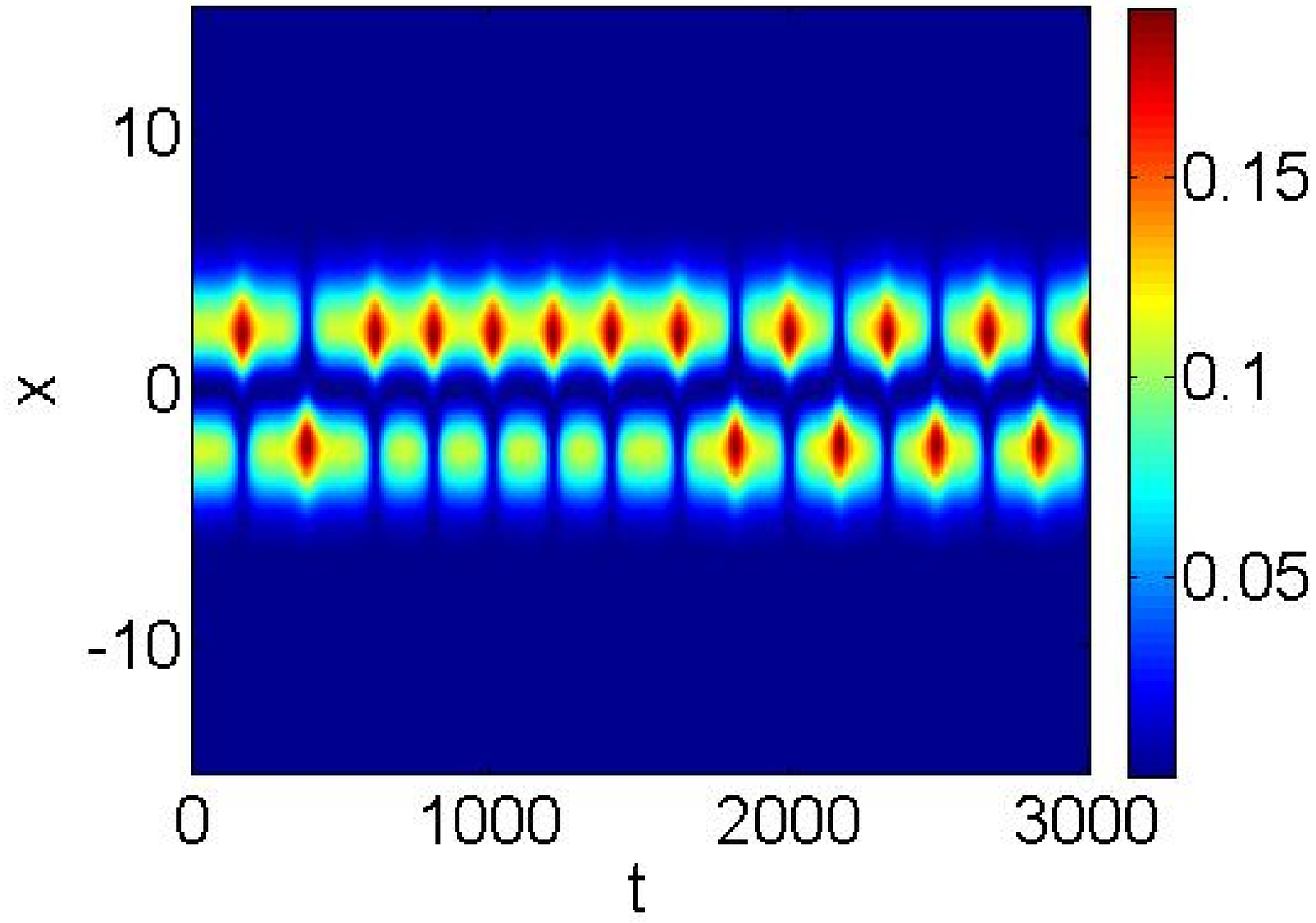}
\includegraphics[width=.3\textwidth]{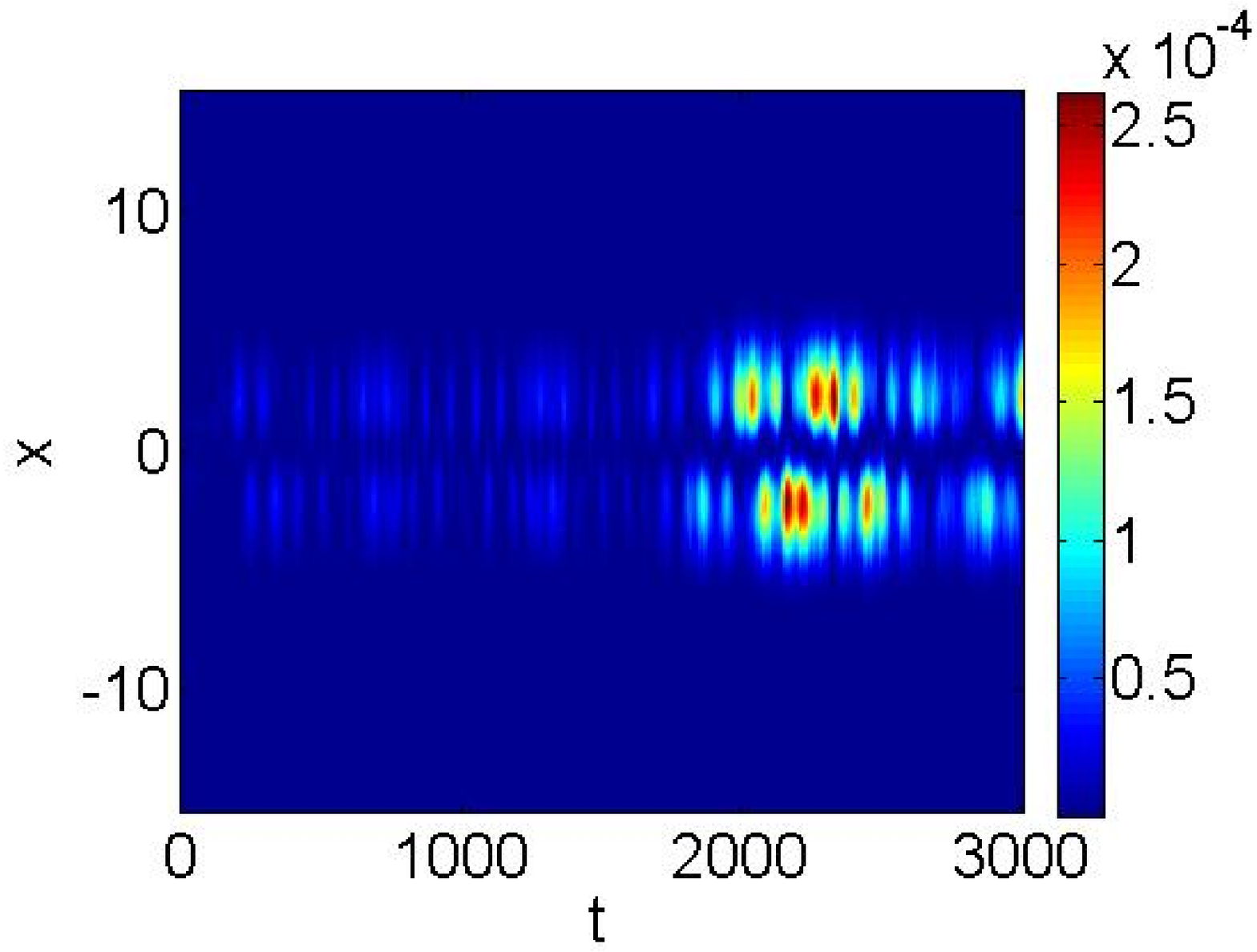}\\
\caption{(Color online) Spatio-temporal contour plots of the densities, $|u_{1}|^{2}$, $|u_{0}|^{2}$ 
and $|u_{-1}|^{2}$, of unstable combined three-component solutions in the case of $^{23}$Na spinor BEC. 
The top and bottom panels show the simulated evolution of wave functions 
$\psi_{1}$ (left), $\psi_{0}$ (middle) and $\psi_{-1}$ (right) in unstable 
solutions of C2 (top) and C8 (bottom) from Fig. \ref{figNa}, respectively.}  
\label{figEvolC28}
\end{figure}

\begin{figure}[tbhp!]
\centering
\includegraphics[width=.3\textwidth]{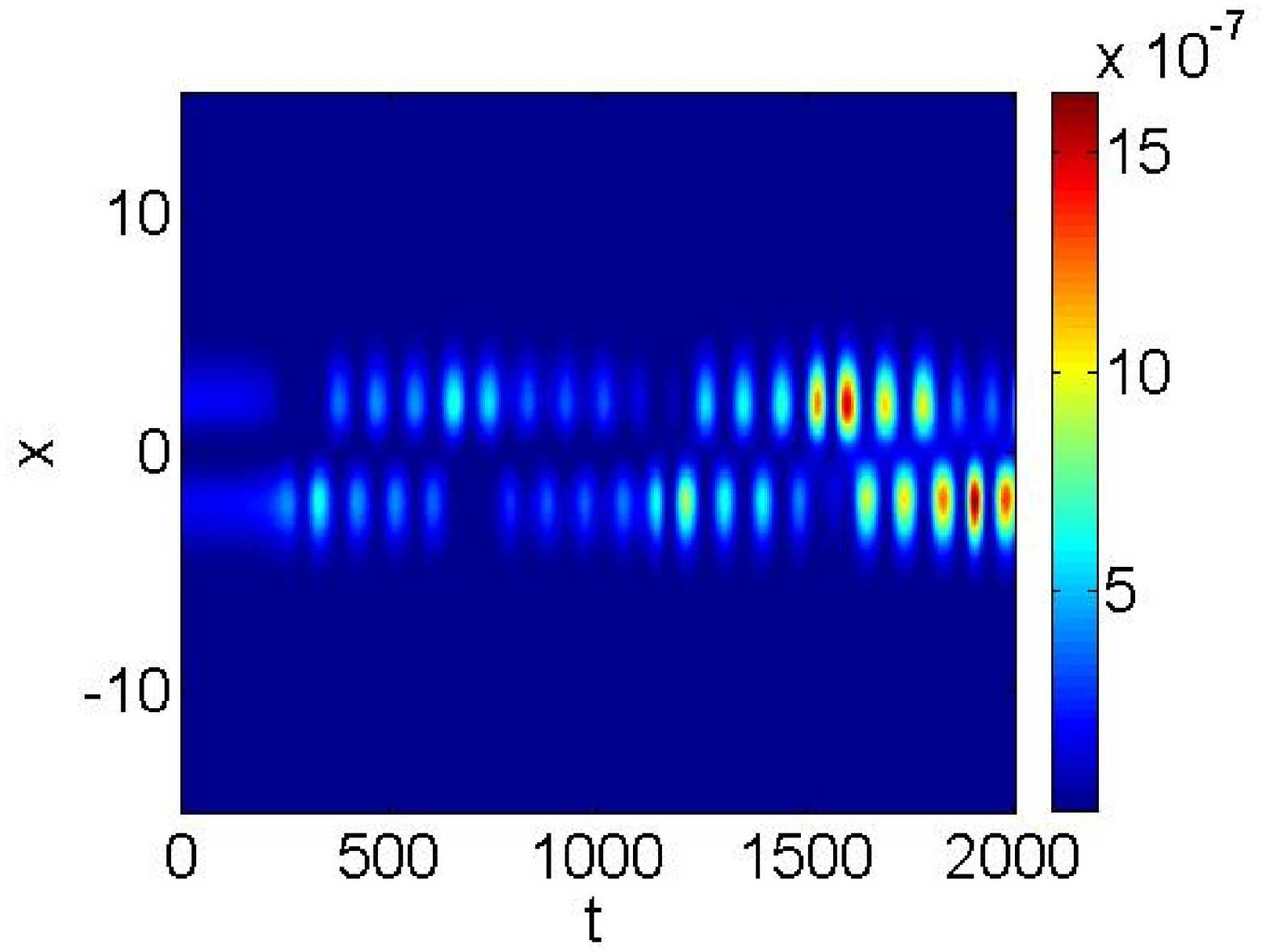}
\includegraphics[width=.3\textwidth]{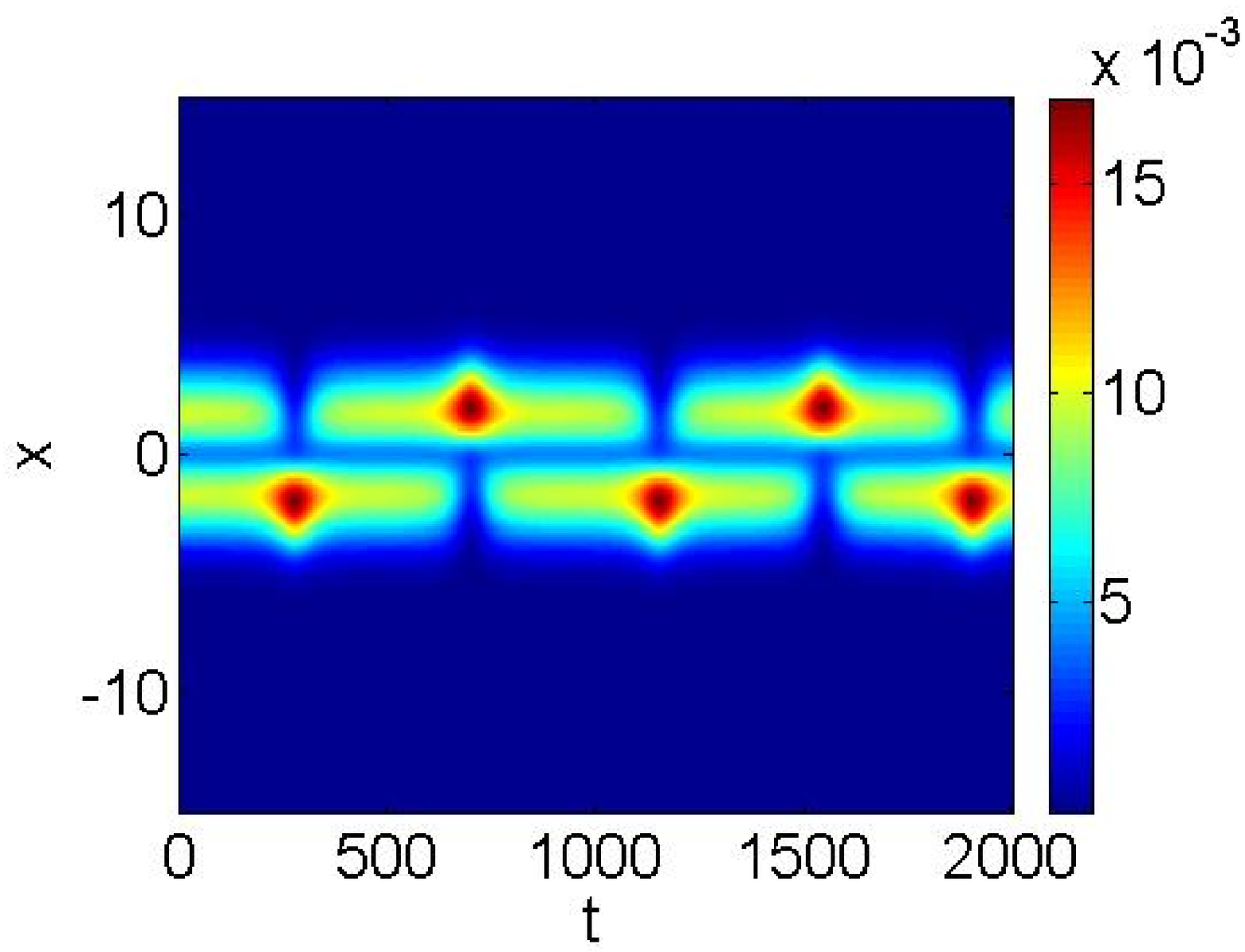} 
\includegraphics[width=.3\textwidth]{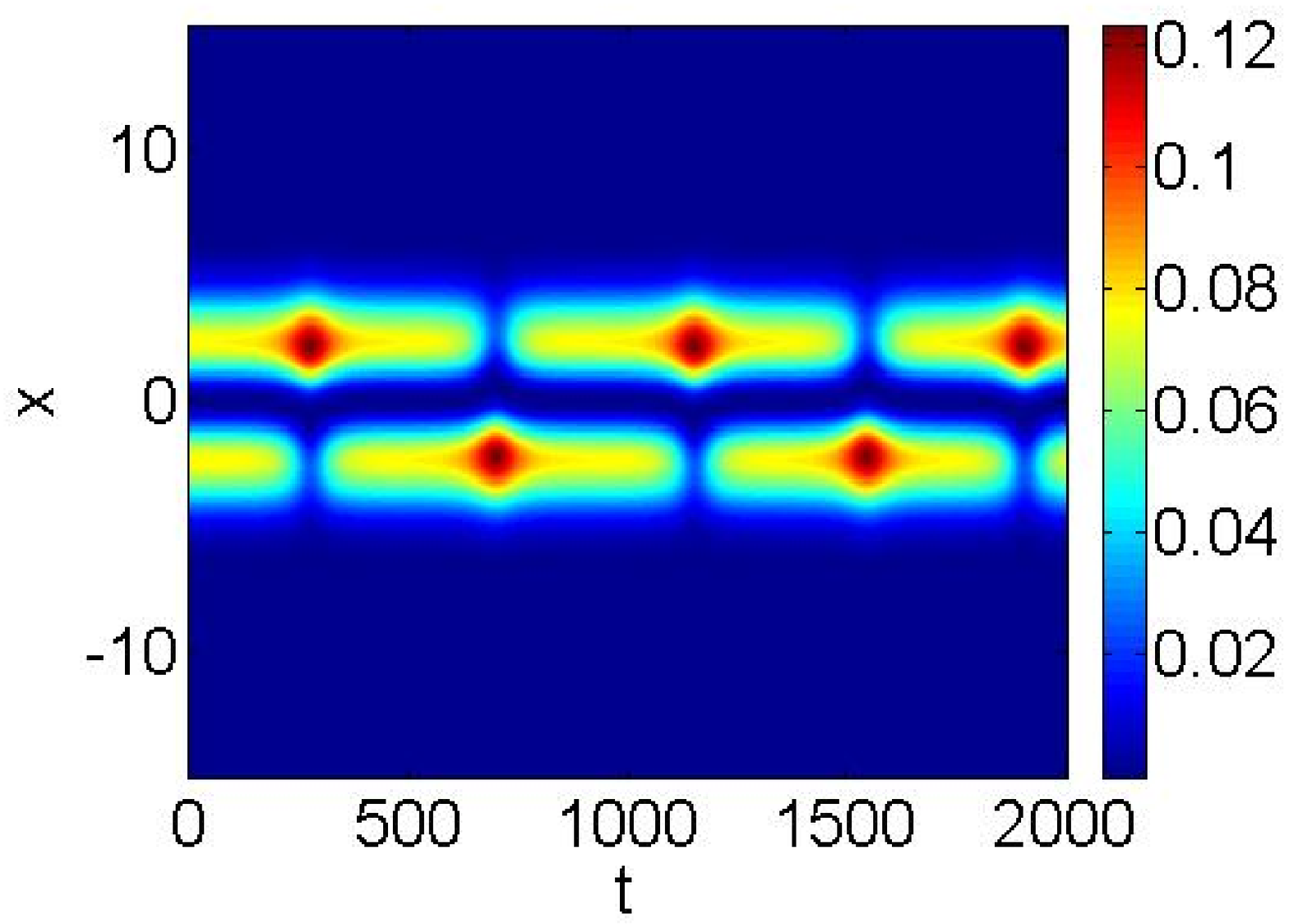}\\
\includegraphics[width=.3\textwidth]{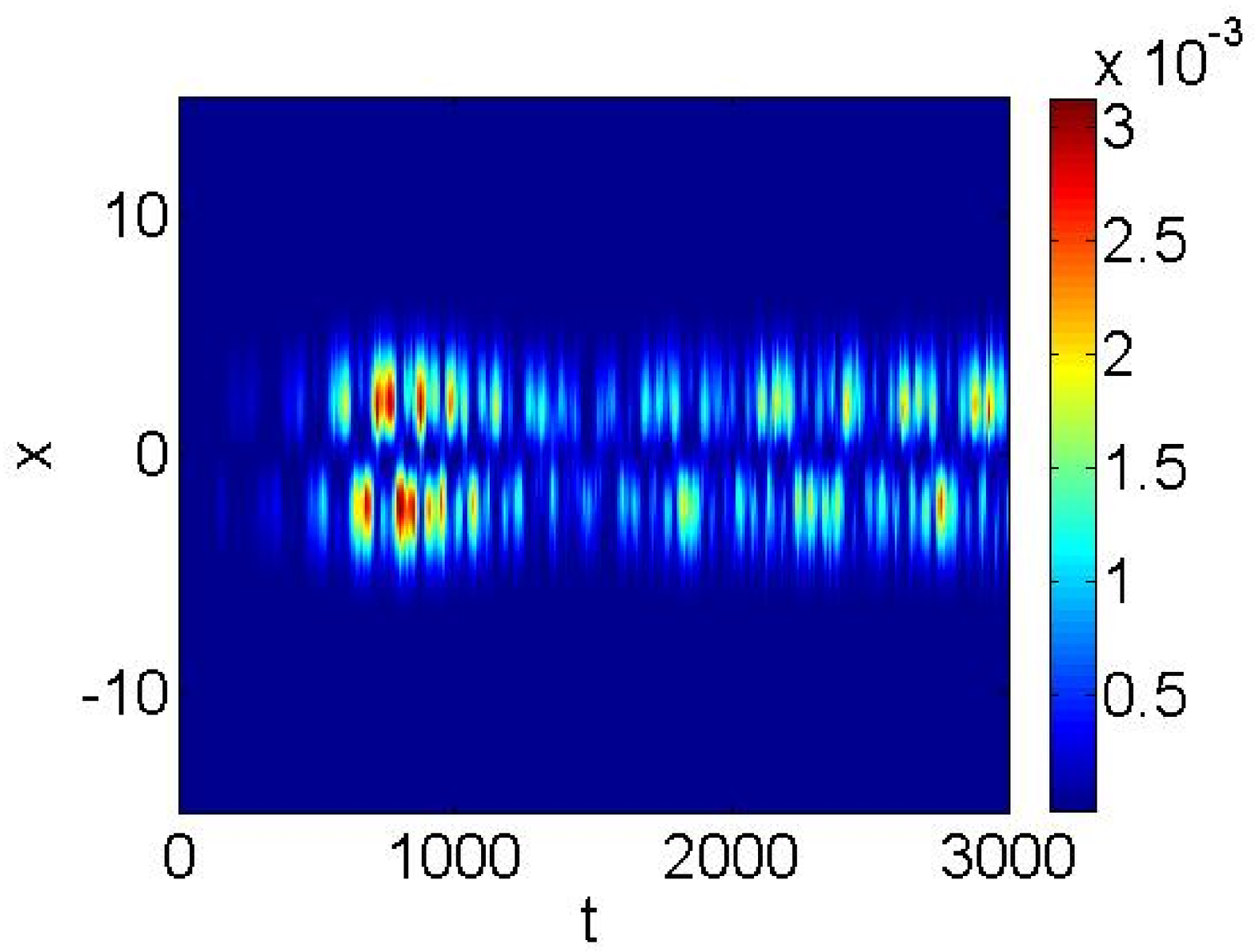}
\includegraphics[width=.3\textwidth]{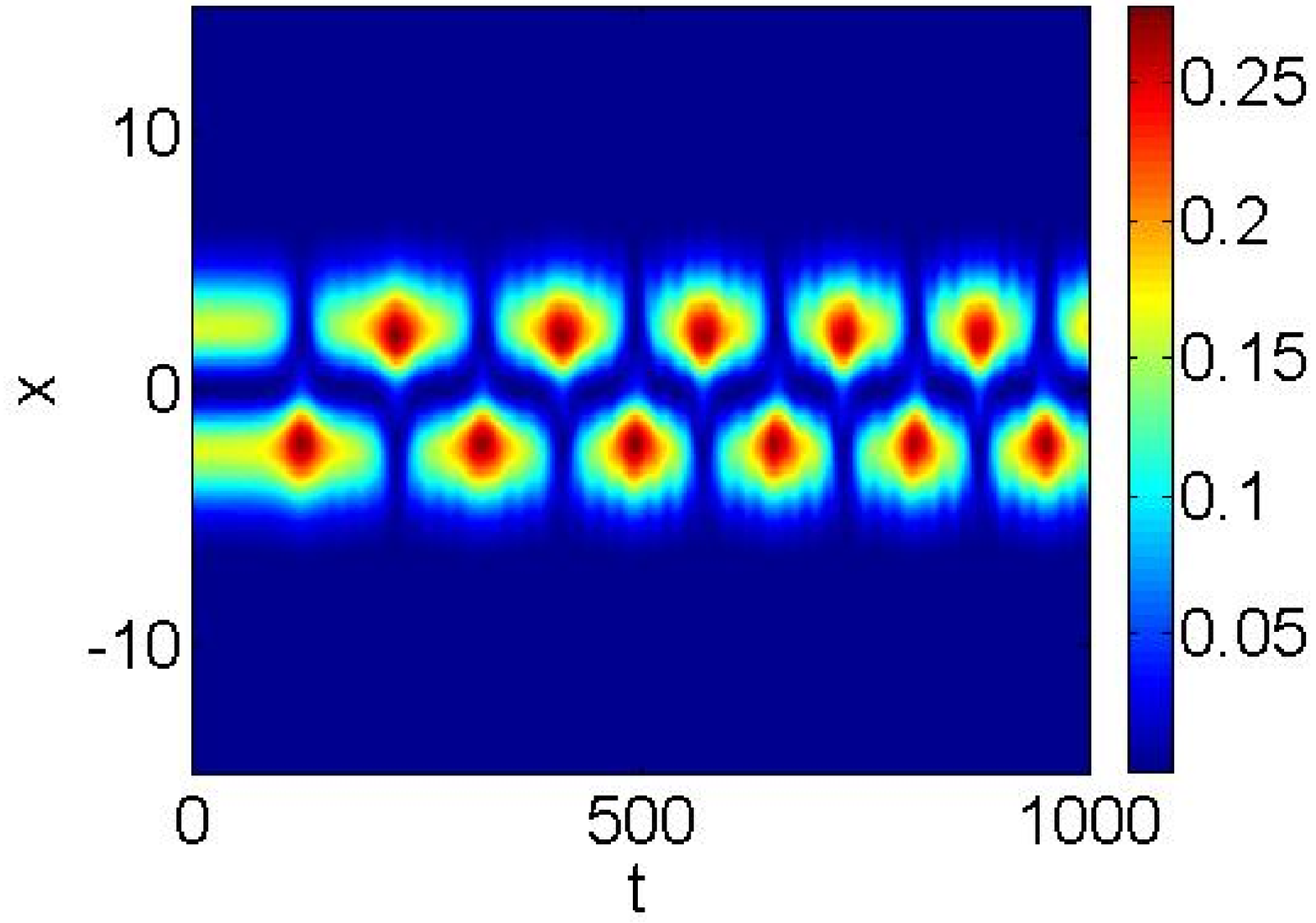}
\includegraphics[width=.3\textwidth]{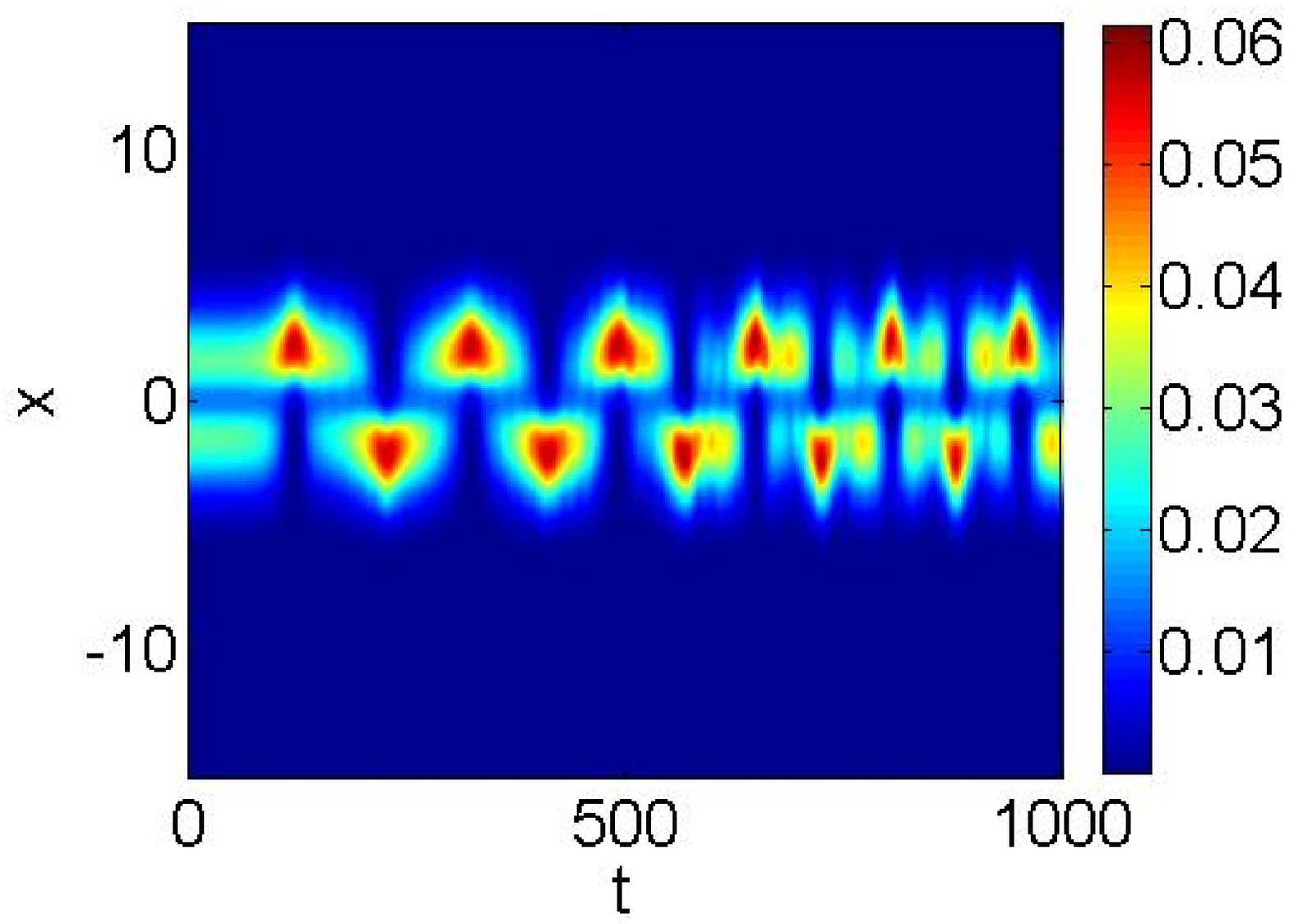}\\
\caption{(Color online) Same as in Fig. \ref{figEvolC28}. 
The top and bottom panels depict the simulated evolution of 
the densities of components $1$ (left), $0$ (middle) and $-1$ (right) 
in 
unstable solutions of C4 (top) and C6 (bottom) as shown in 
Fig. \ref{figNa}, respectively.}  
\label{figEvolC46}
\end{figure}

Roughly similar behavior can be observed in the dynamics of the combined
branches involving all three spinorial components, such as the ones
for C2 and C8 in Fig. \ref{figEvolC28}, as well as the ones for C4 and
C6 in Fig. \ref{figEvolC46}. In Fig. \ref{figEvolC28}, the solution is predominantly
supported in the components $u_{1}$ and $u_{0}$ (respectively $u_{-1}$ and $u_{0}$ in 
Fig.  \ref{figEvolC46}) and has a small amplitude in the remaining
component. The two predominant components exhibit similar recurrent
behavior, whereby at roughly periodic intervals the solution becomes
asymmetric with stronger support for one component in the one well,
while stronger support for the other component is in the second well.
The weaker component is eventually more strongly amplified by the
instability, but still does not appear to play a critical role in
affecting the dynamics of the two dominant components. Nevertheless,
we clearly see the symmetry-breaking manifestation of the relevant
instability and the recurrent emergence of the ensuing asymmetric 
waveforms.

In all of the above examples, we have seeded the instability by a 
random (uniformly distributed) perturbation imposed to the original
stationary solutions with a small amplitude ($10^{-4}$), essentially 
emulating the presence of a background of experimental ``noise'' in a potential
experiment. While this was not based on a physical model of quantum or
thermal noise present in an actual experiment, the main thing
that matters (at least in as far as the deterministic evolution of the
mean field spinor models considered herein is concerned) 
is the projection of the relevant
perturbation on the unstable eigendirection (i.e., eigenvector) 
of the linearization around
the solution. Our results (for different amplitudes,  as well as  
different realizations of the noise) clearly illustrate this fact 
(see below).

\begin{figure}[tbhp!]
\centering
\includegraphics[width=.3\textwidth]{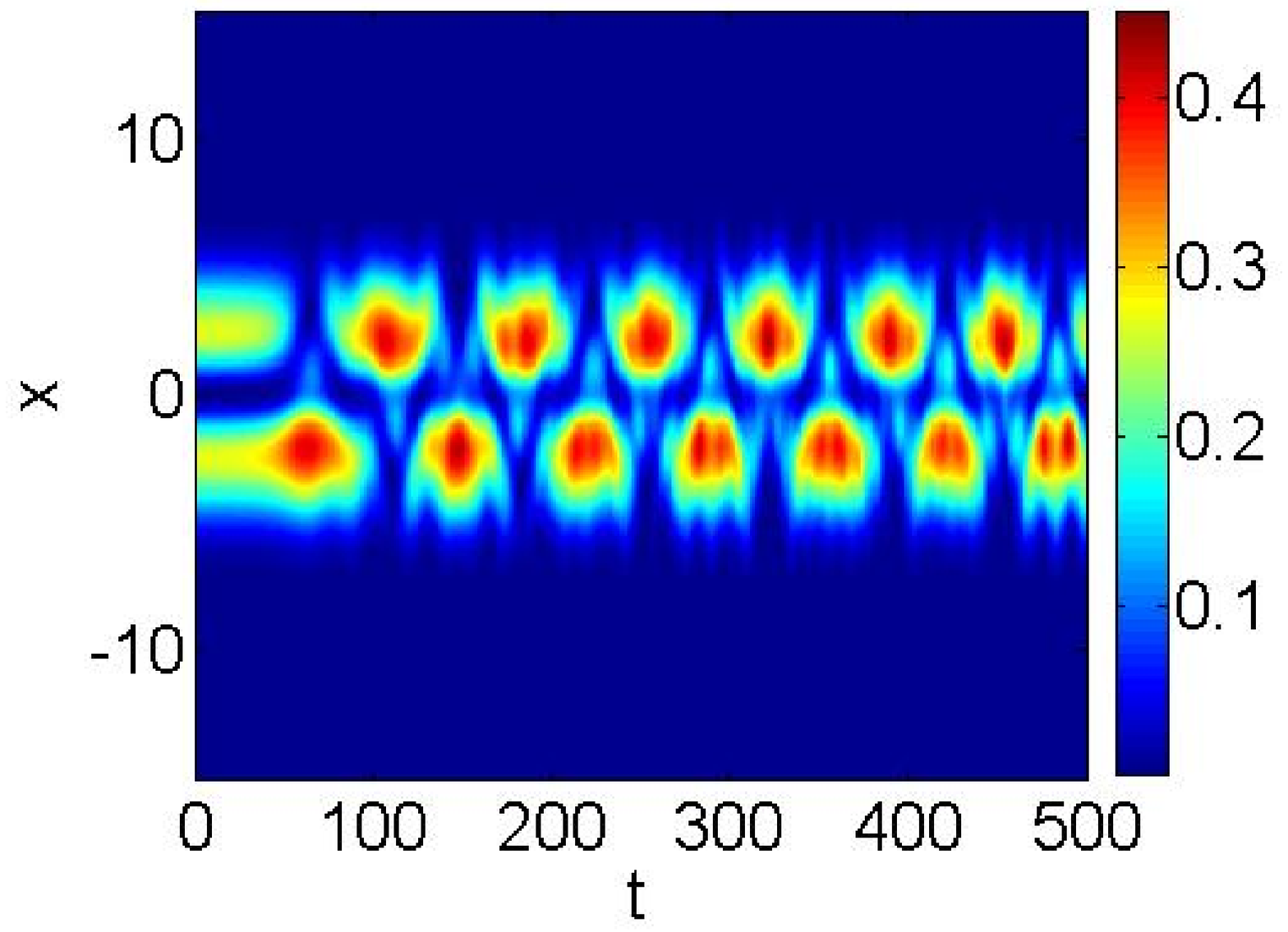}
\includegraphics[width=.3\textwidth]{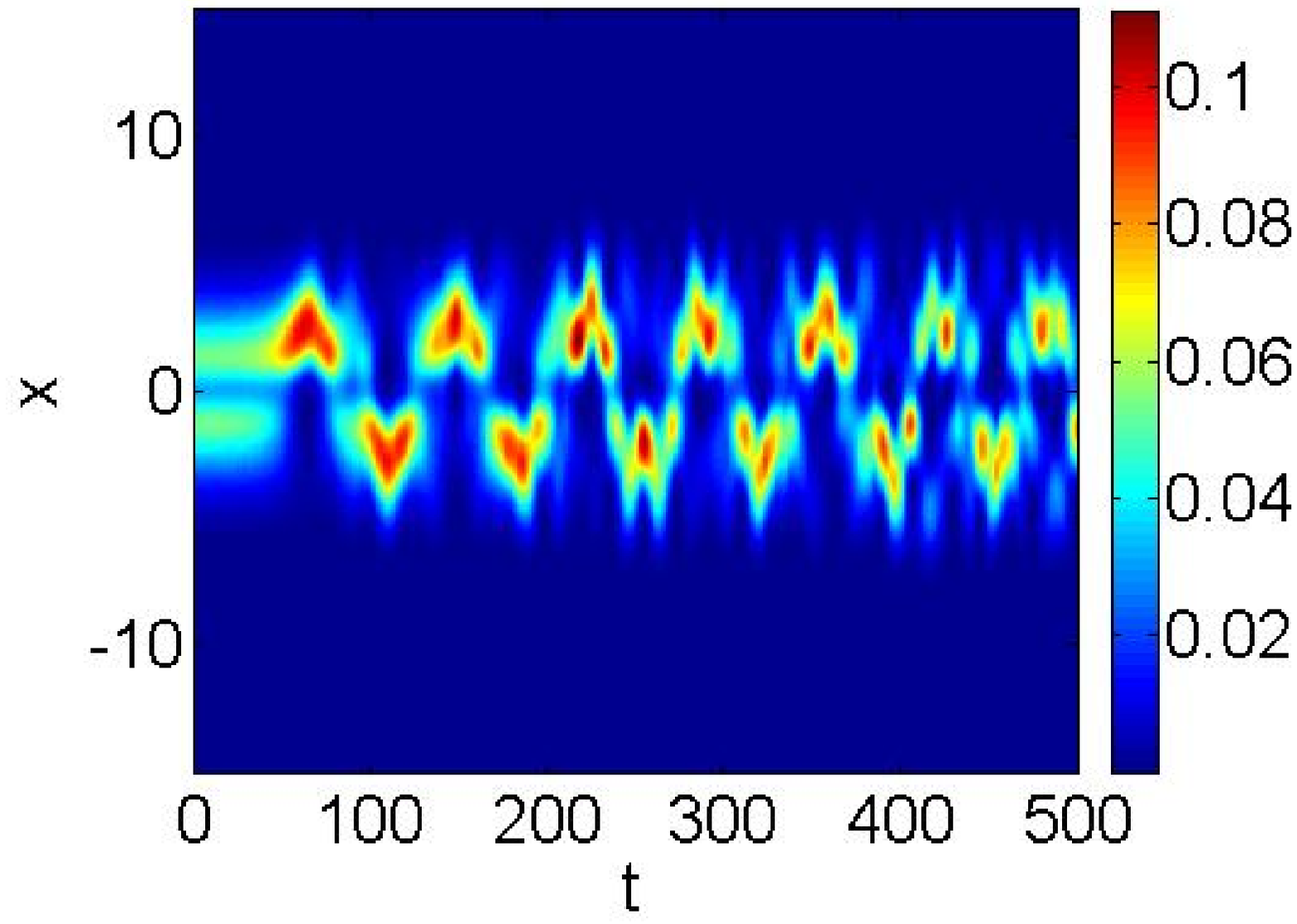} 
\includegraphics[width=.3\textwidth]{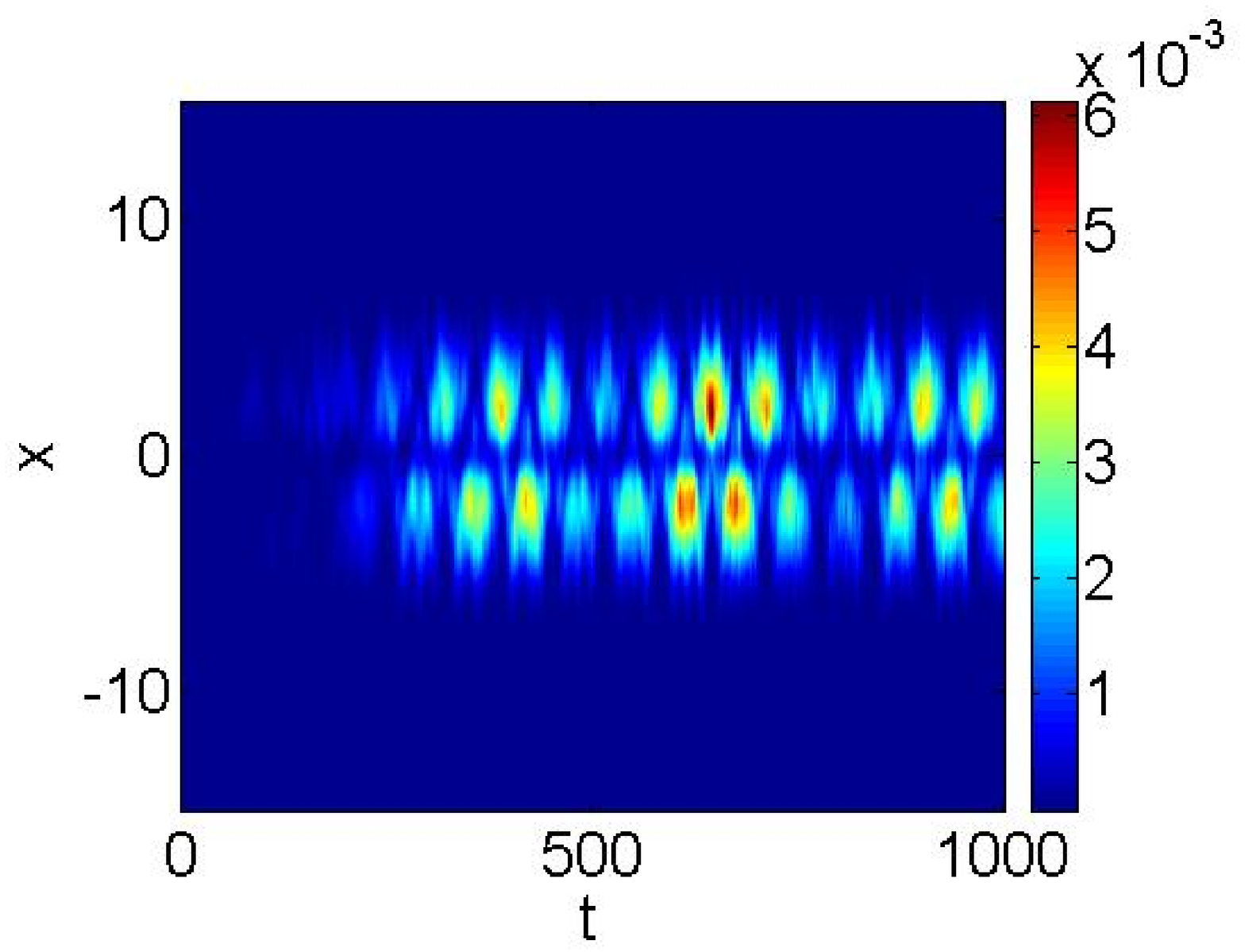}\\
\includegraphics[width=.3\textwidth]{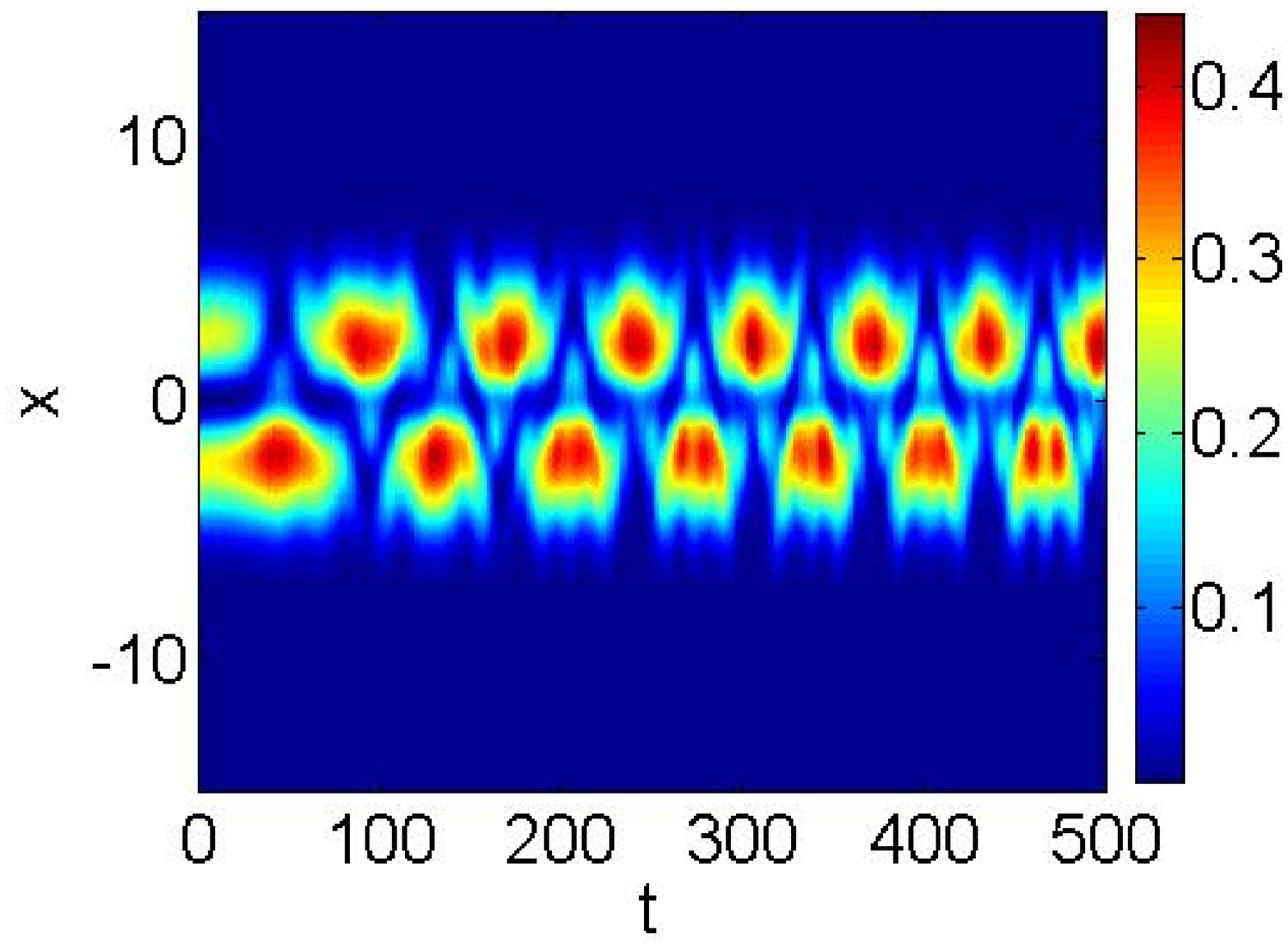}
\includegraphics[width=.3\textwidth]{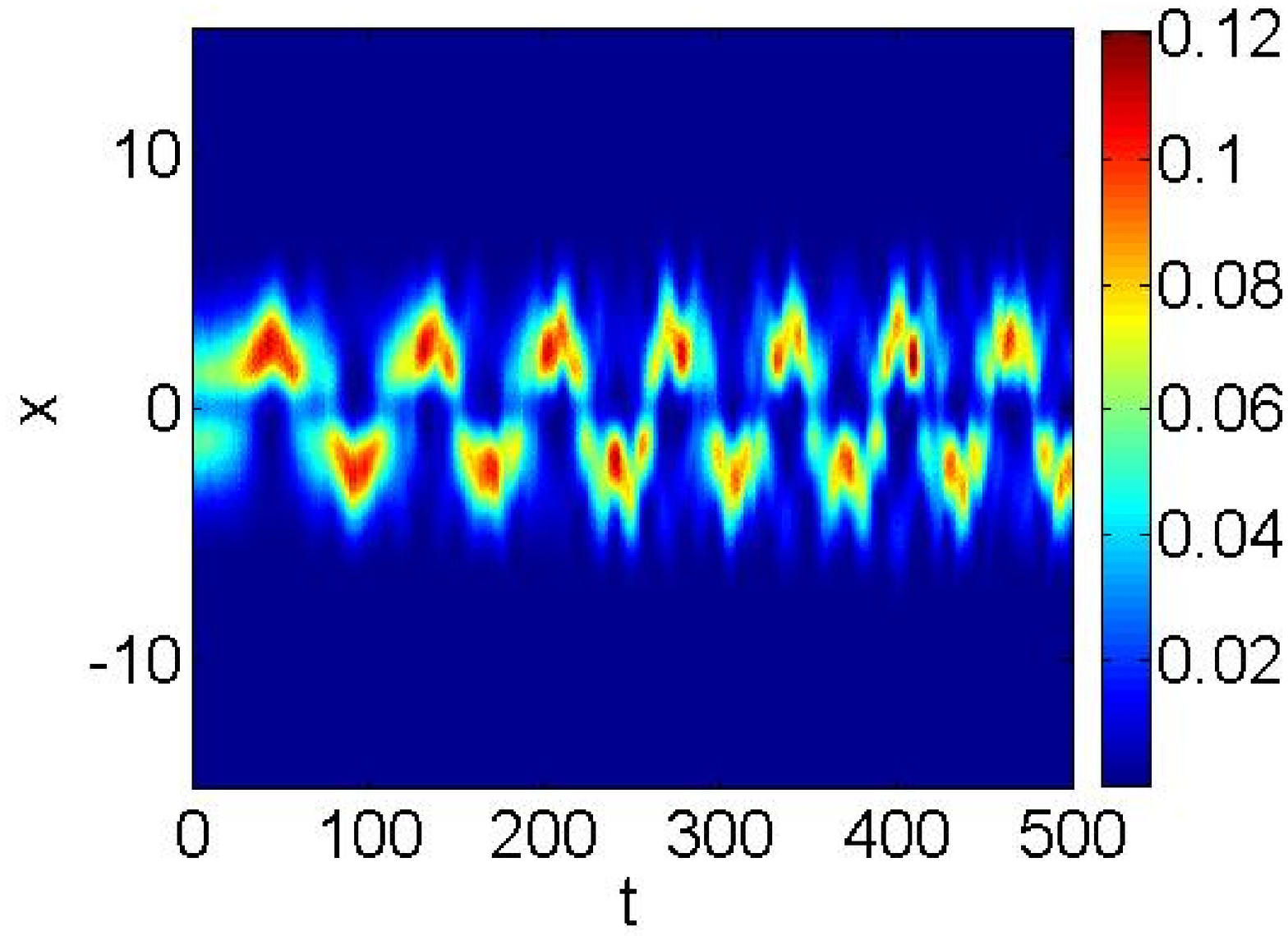}
\includegraphics[width=.3\textwidth]{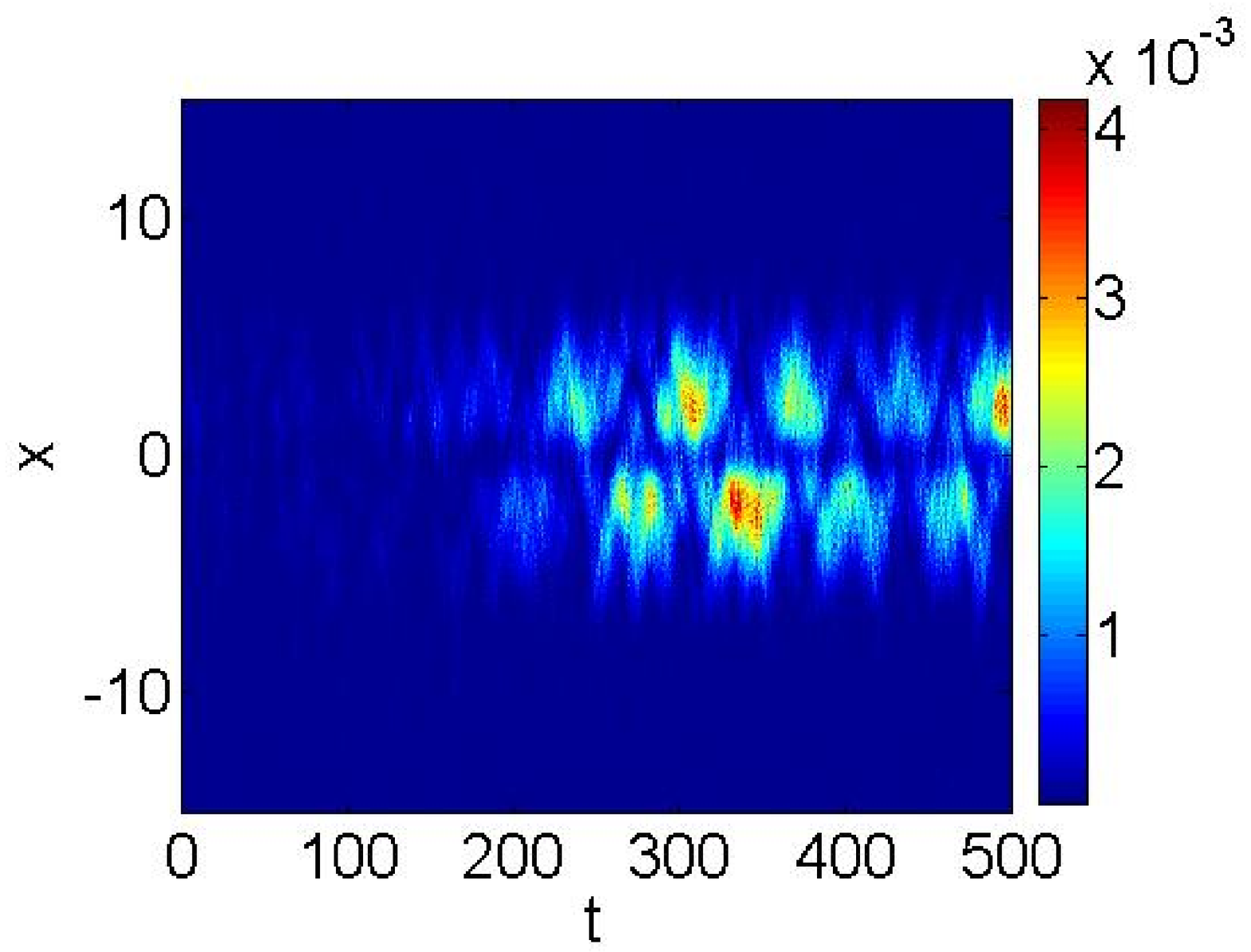}\\
\caption{(Color online) Same as in Fig. \ref{figEvolC28}. 
The top and bottom panels show the simulated evolution of the density of
components 
${1}$ (left), ${0}$ (middle) and ${-1}$ (right) in unstable 
solutions of C2. A random (uniformly distributed) noise of amplitude $10^{-3}$ (top) and $10^{-2}$ (bottom) is involved initially in perturbing the
exact unstable solution, 
compared to the one of amplitude $10^{-4}$ in Fig. \ref{figEvolC28}.}  
\label{figEvolC2_ampl}
\end{figure}

\begin{figure}[tbhp!]
\centering
\includegraphics[width=.3\textwidth]{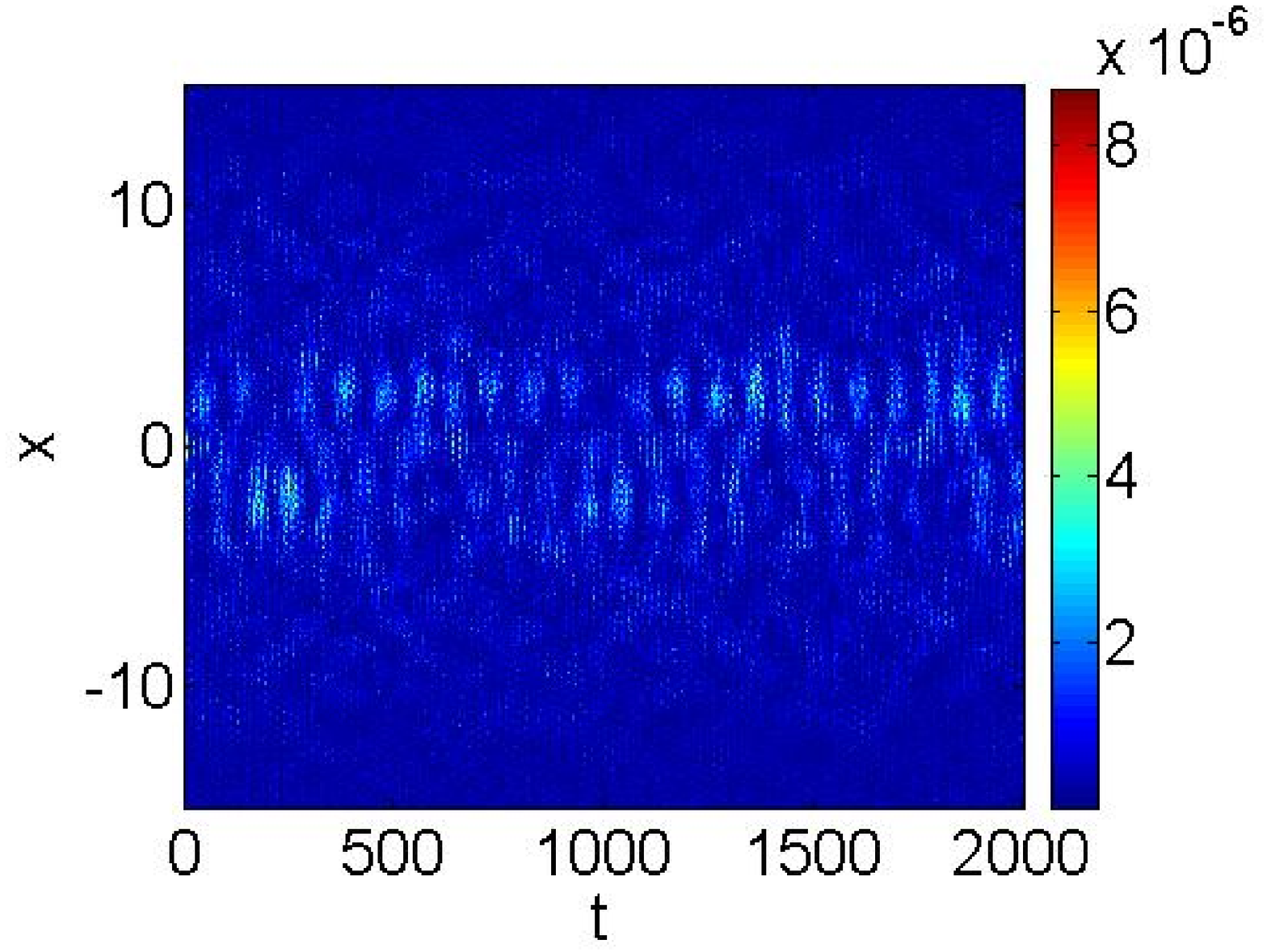}
\includegraphics[width=.3\textwidth]{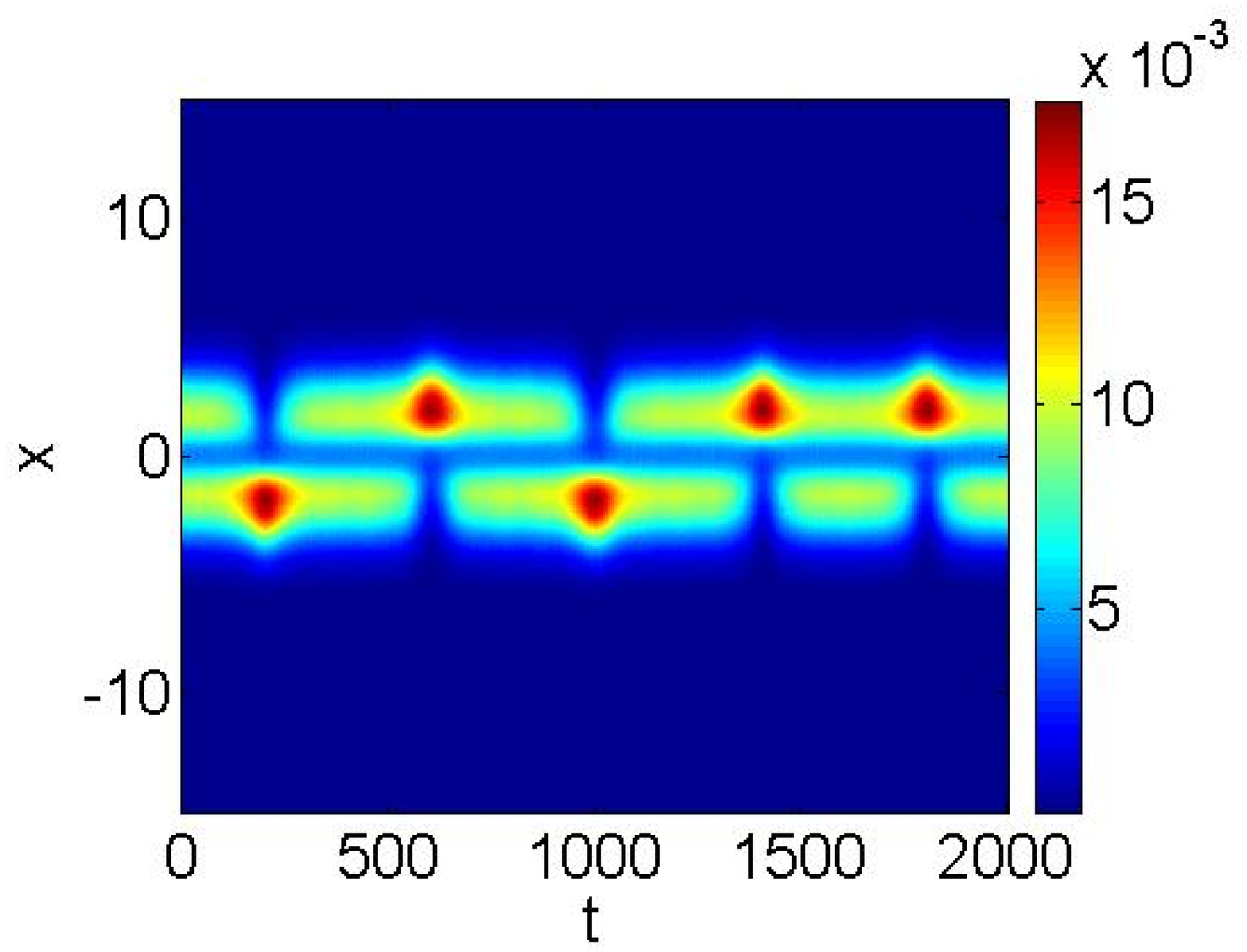} 
\includegraphics[width=.3\textwidth]{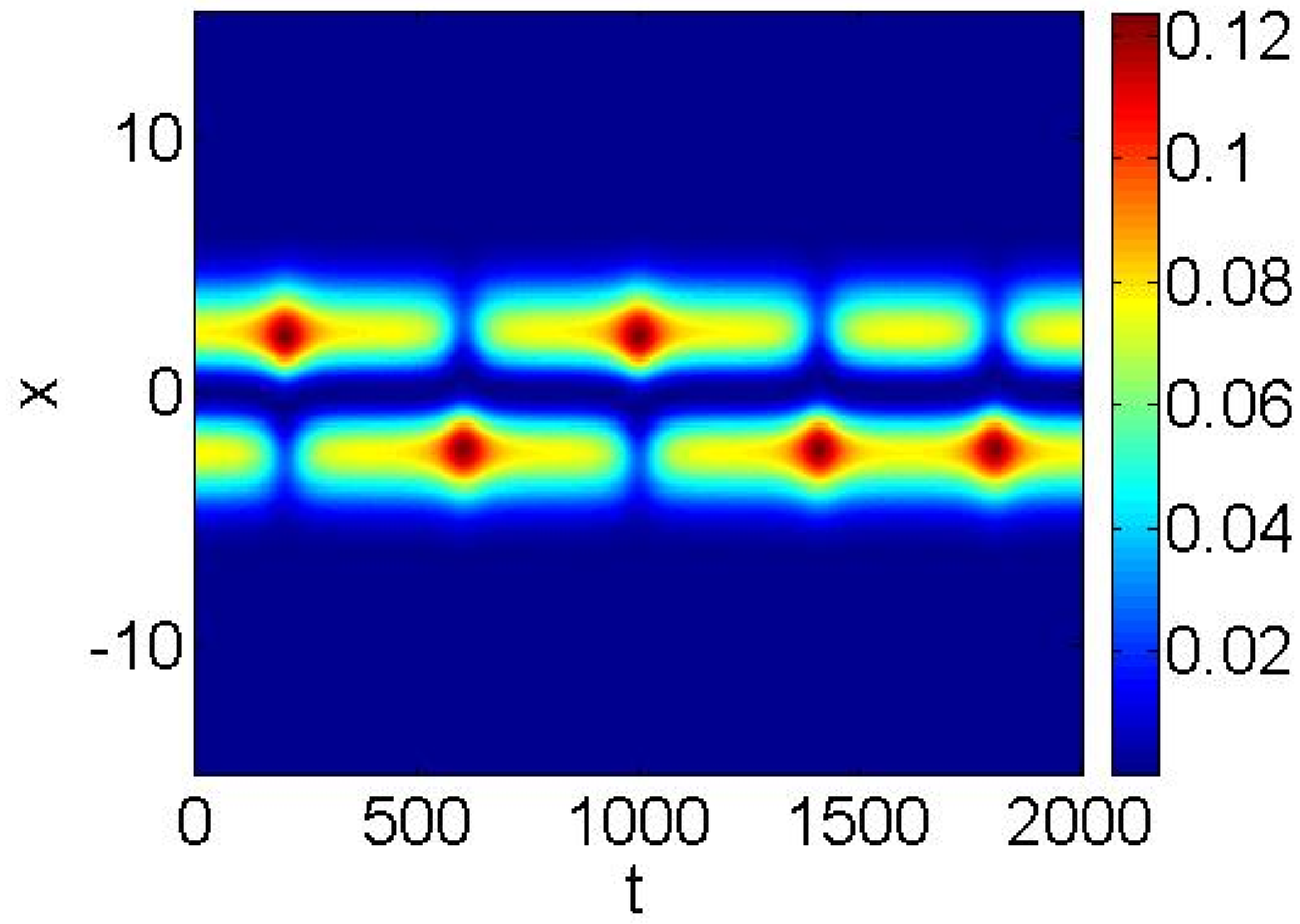}\\
\includegraphics[width=.3\textwidth]{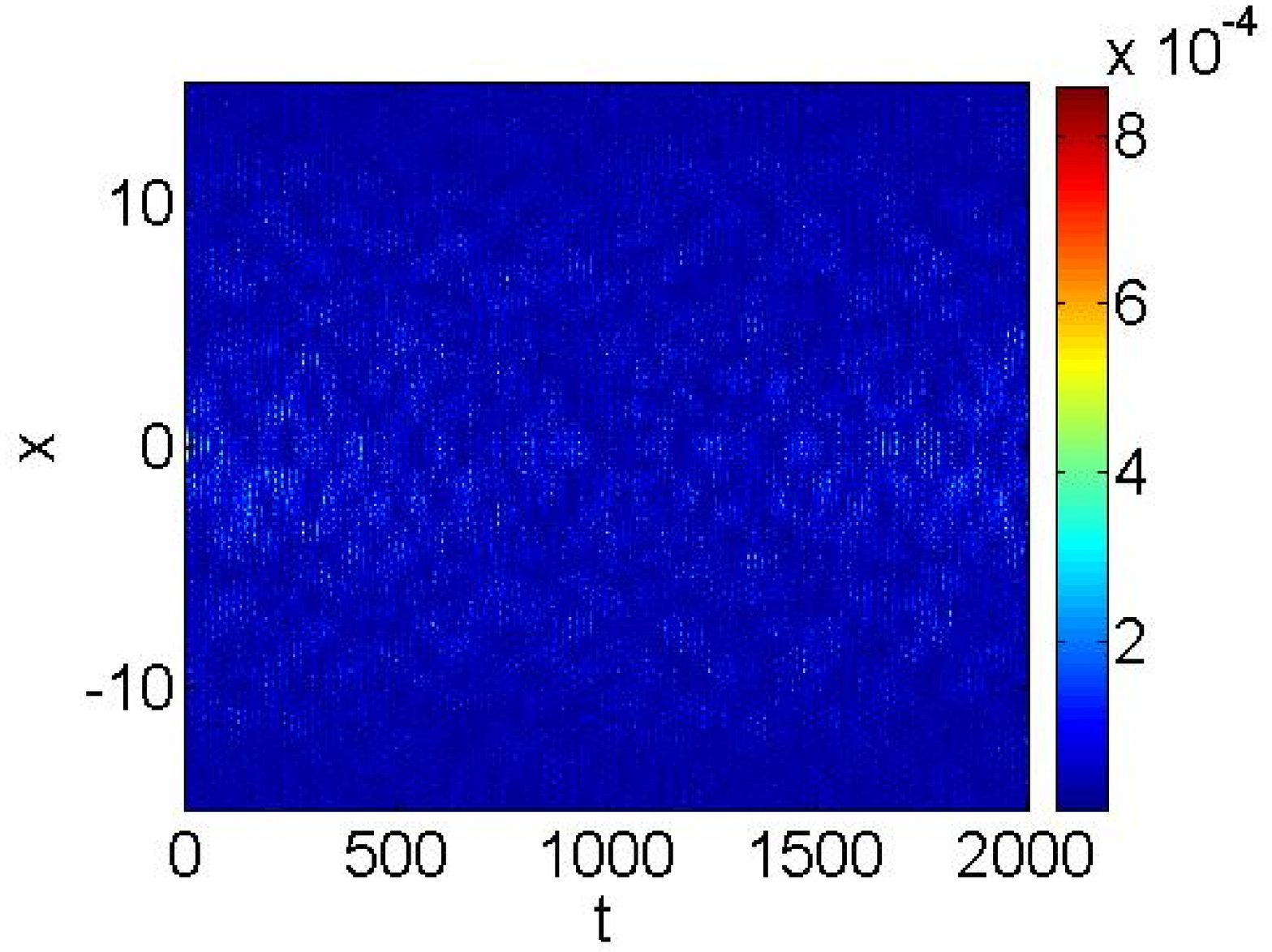}
\includegraphics[width=.3\textwidth]{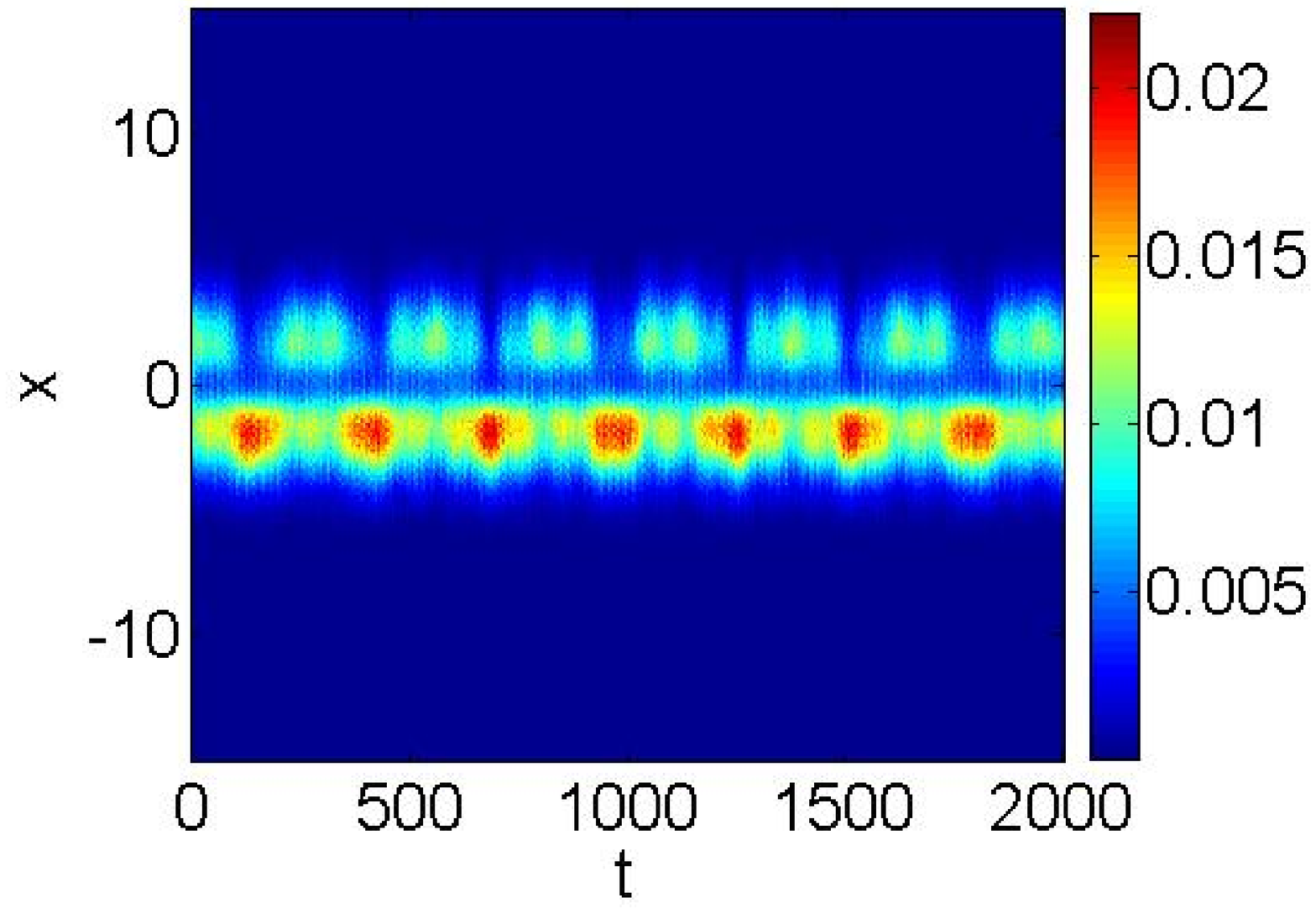}
\includegraphics[width=.3\textwidth]{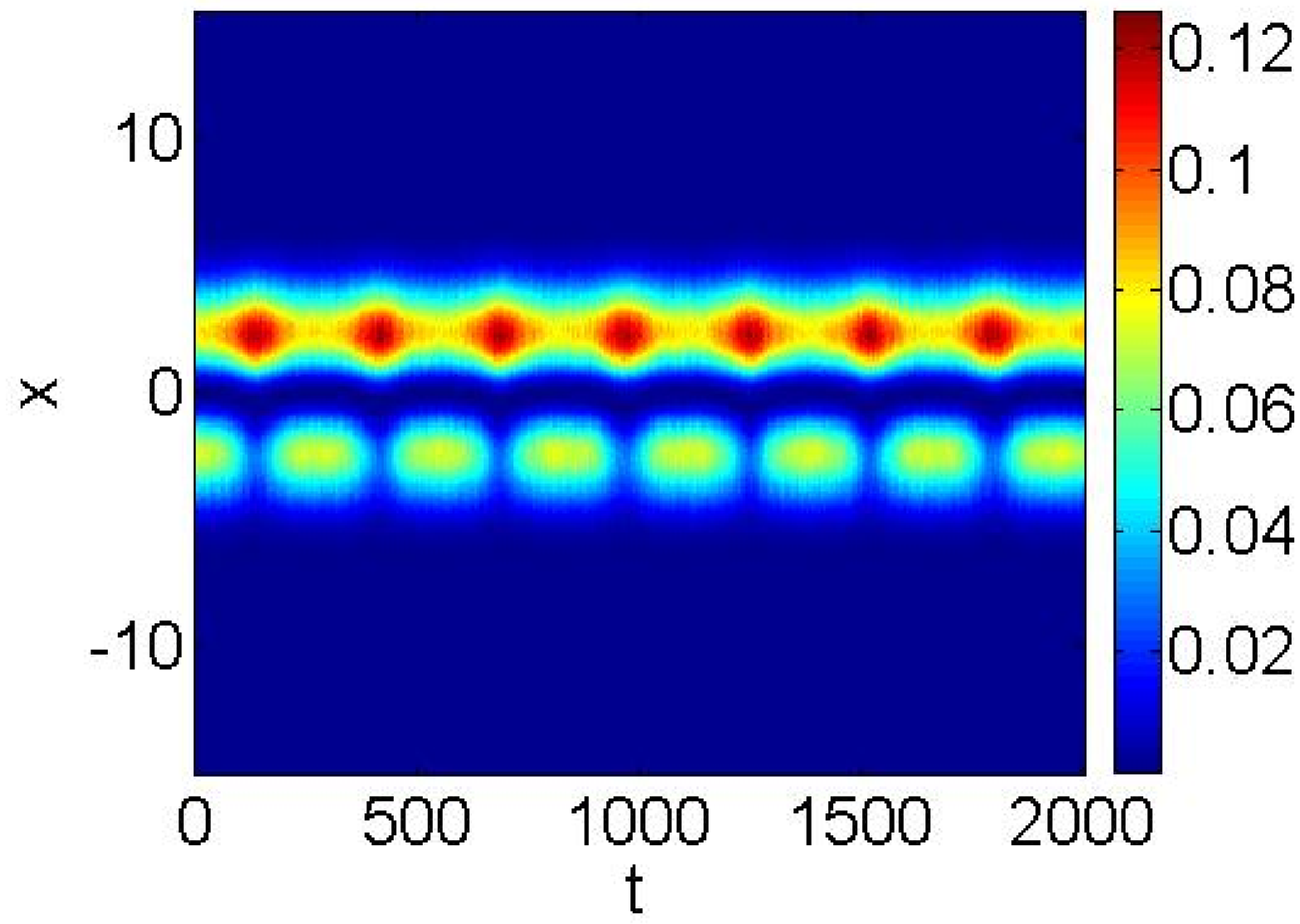}\\
\caption{(Color online) Same as in Fig. \ref{figEvolC28}. 
The top and bottom panels show the simulated evolution of components 
${1}$ (left), ${0}$ (middle) and ${-1}$ (right) in unstable 
solutions of C4, with an initial random noise of amplitude $10^{-3}$ (top) and $10^{-2}$ (bottom), compared to the one of $10^{-4}$ in Fig. \ref{figEvolC46}.}  
\label{figEvolC4_ampl}
\end{figure}

To illustrate the effect of the noise amplitude, we have repeated our
numerical simulations 
with higher noise amplitudes (but with the 
same spatial distribution of the noisy perturbation); see Figs.
\ref{figEvolC2_ampl} and \ref{figEvolC4_ampl}. We can observe that
in such a case the dynamics is principally similar [in all cases,
the symmetry breaking still occurs, although its evolution in time may
differ somewhat for different amplitudes cf. Fig.\ref{figEvolC4_ampl}]. 
The amplitude of the noise
does play a role in the time scale of the manifestation
of the instability, since the larger the initial noise amplitude, the
shorter the time interval until it gets amplified (by the instability)
to an O$(1)$ perturbation.

\begin{figure}[tbhp!]
\centering
\includegraphics[width=.3\textwidth]{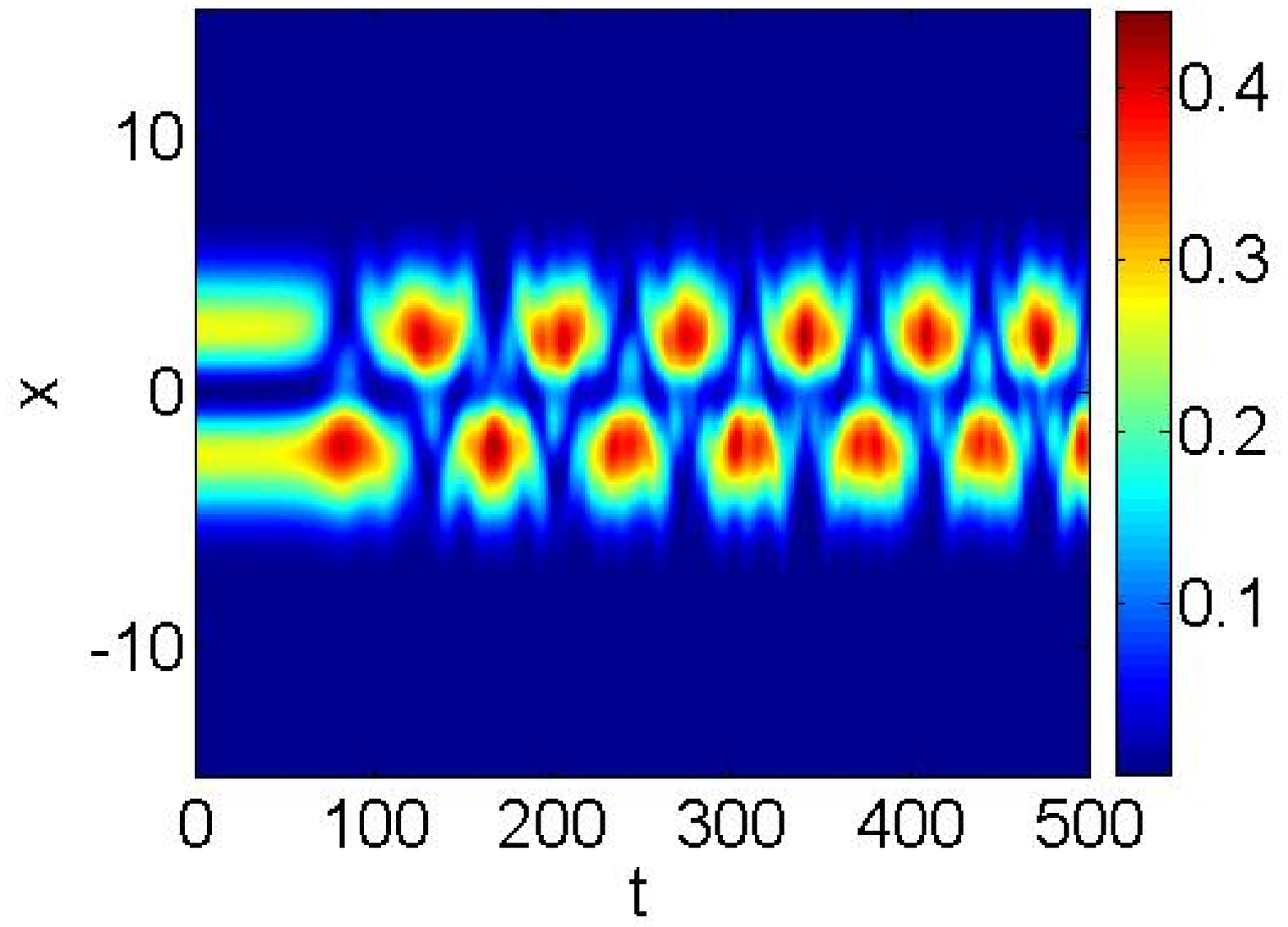}
\includegraphics[width=.3\textwidth]{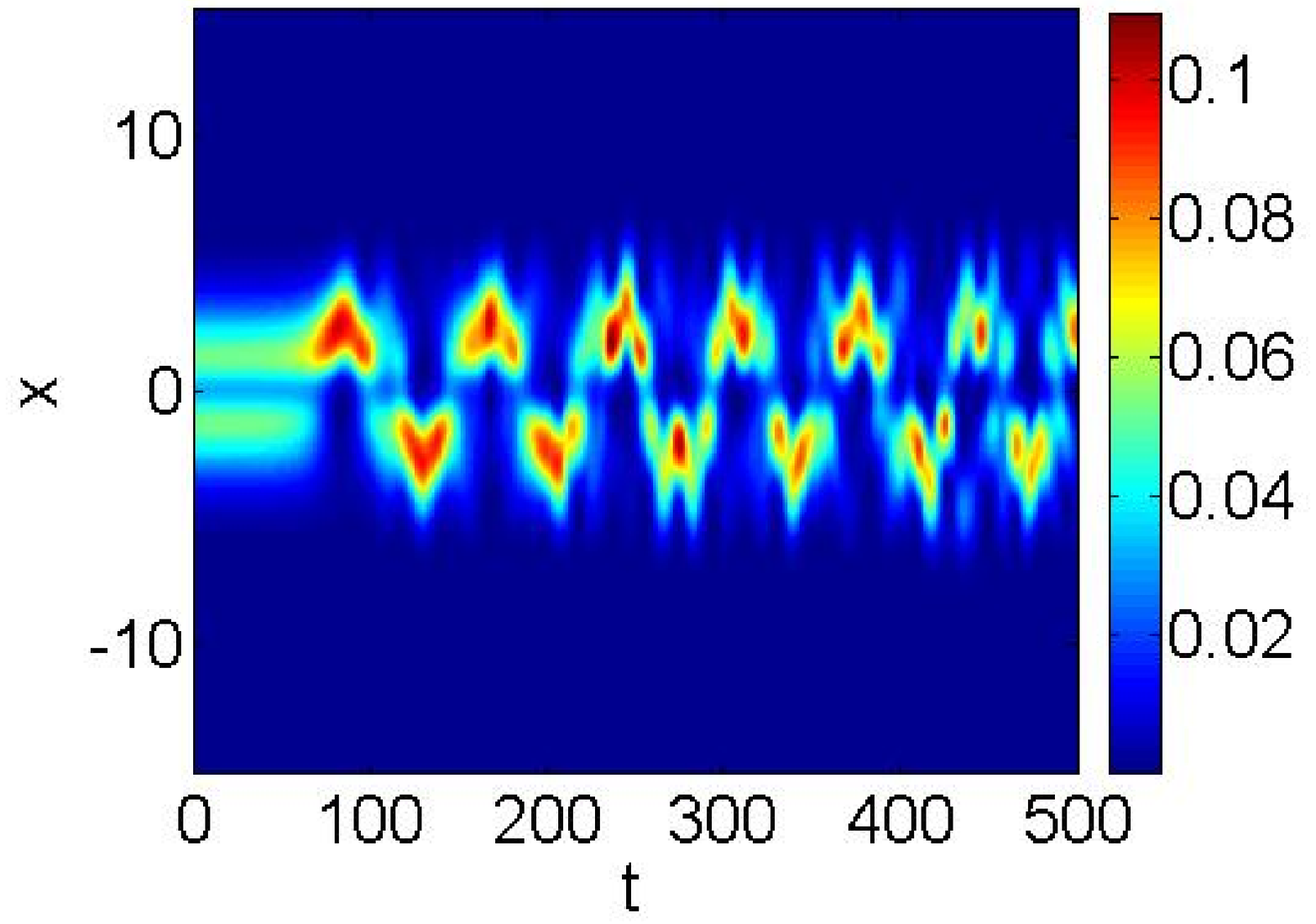} 
\includegraphics[width=.3\textwidth]{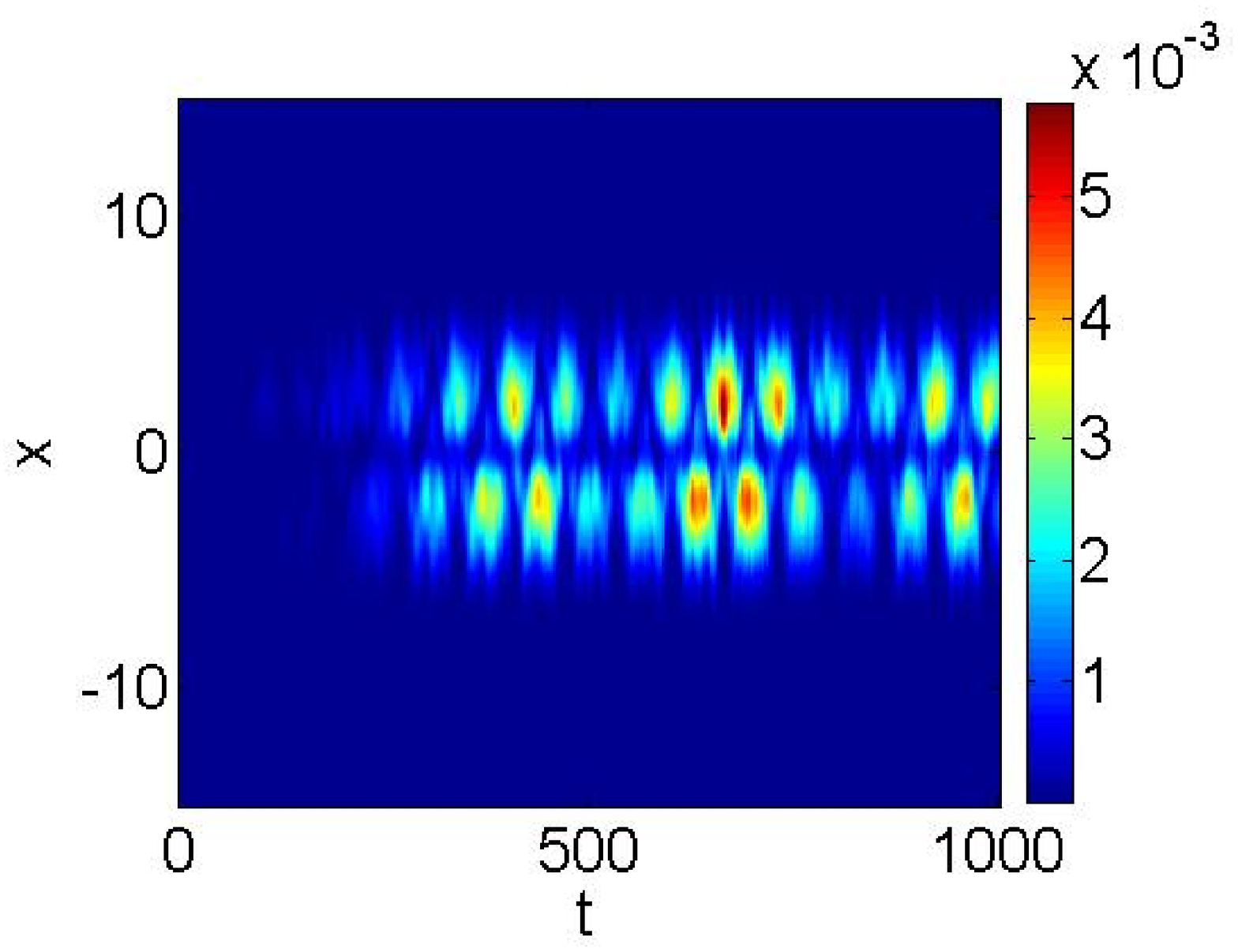}\\
\includegraphics[width=.3\textwidth]{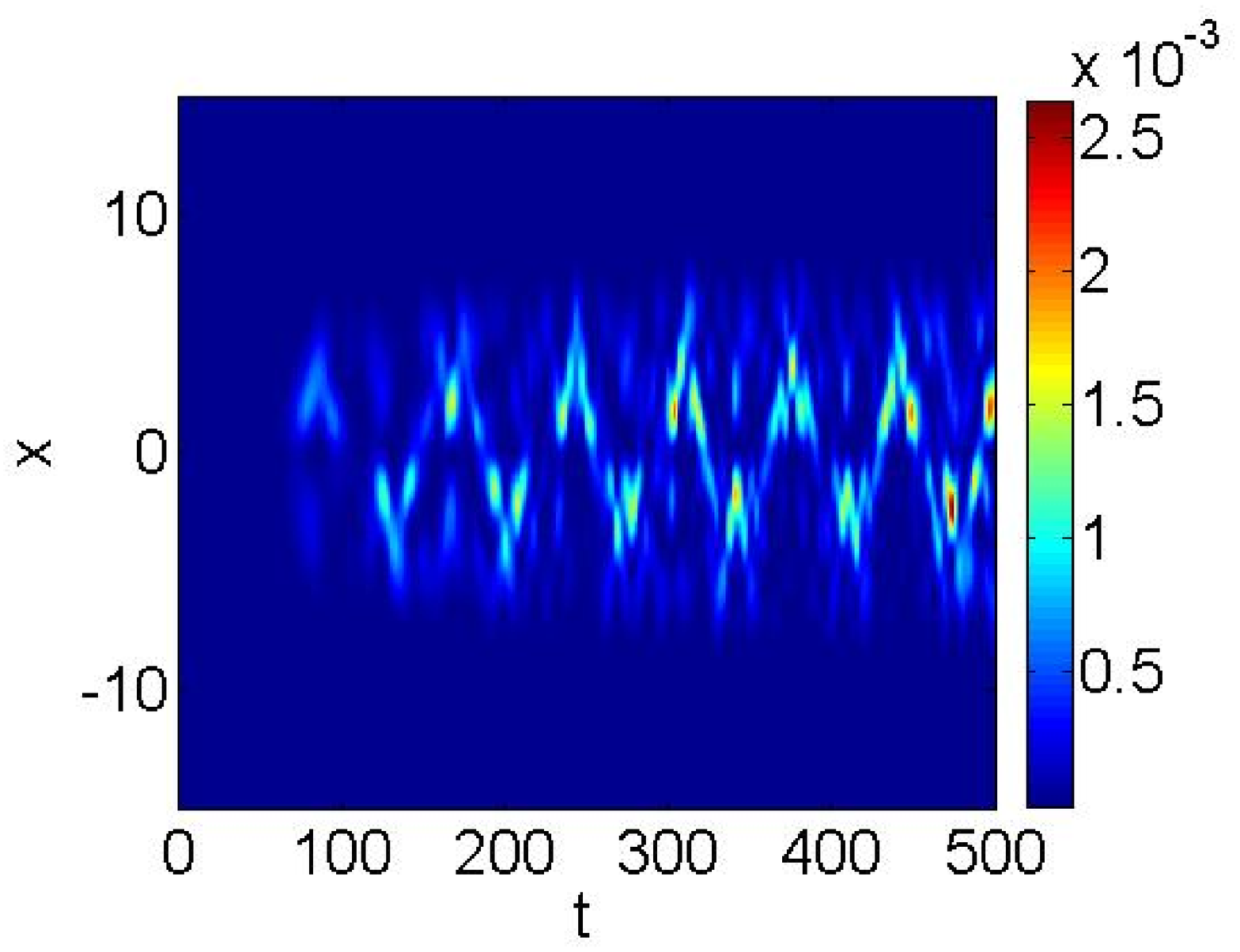}
\includegraphics[width=.3\textwidth]{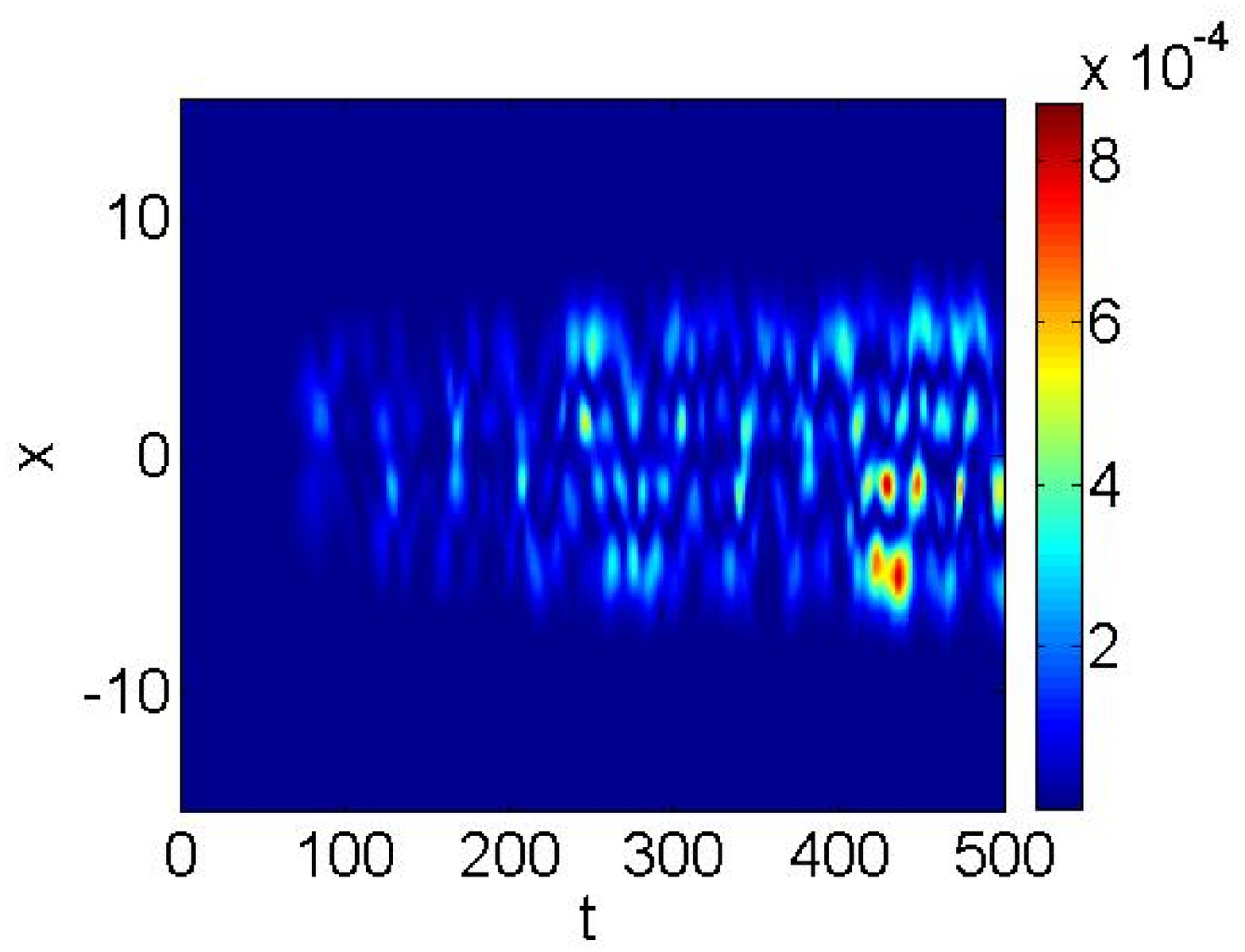}
\includegraphics[width=.3\textwidth]{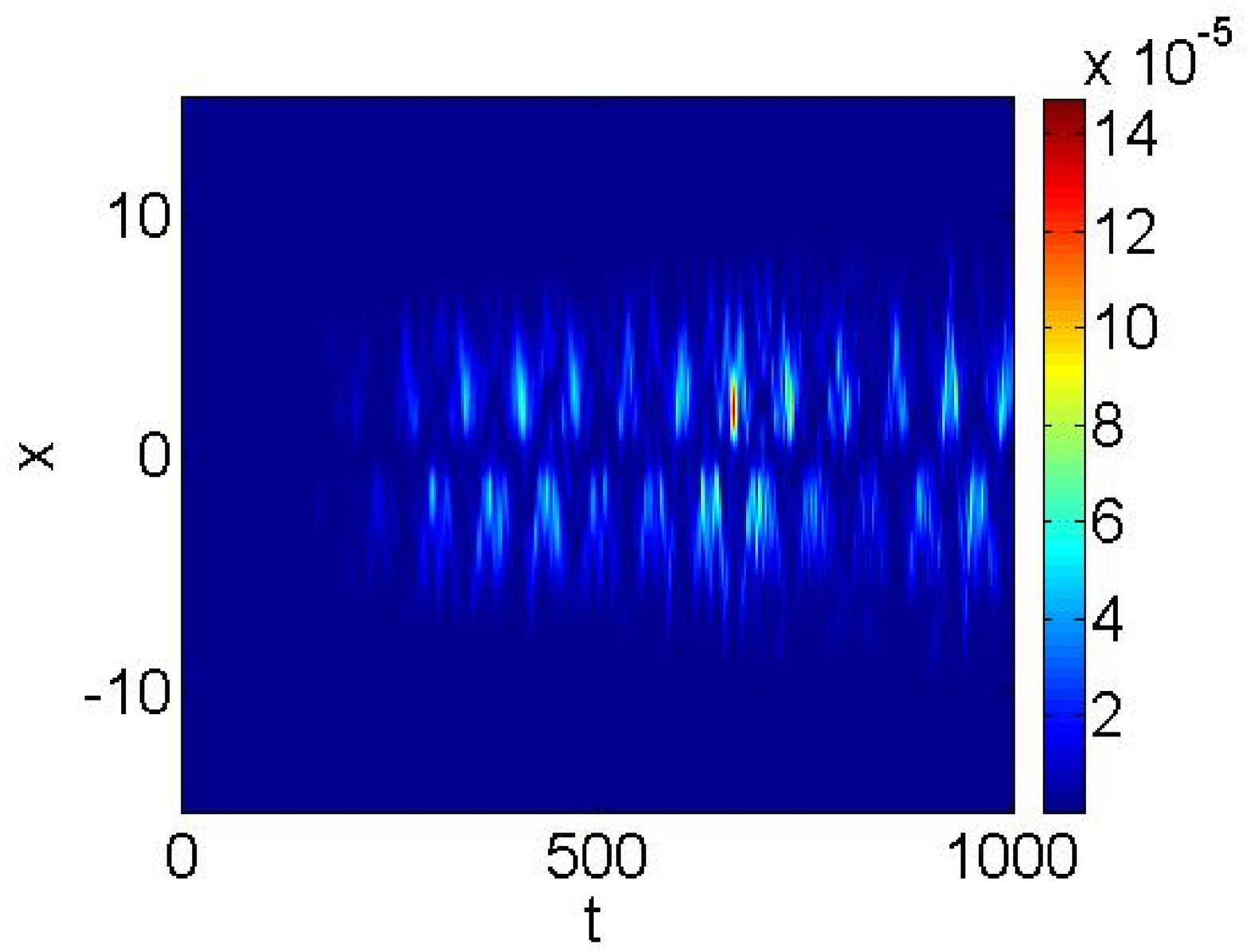}\\
\caption{(Color online) Same as in Fig. \ref{figEvolC28}. 
The top panels show the average of 10 simulated evolutions of the density
of components 
${1}$ (left), ${0}$ (middle) and ${-1}$ (right) in unstable 
solutions of C2 with different initial noises of amplitude $10^{-4}$. 
The bottom panel 
is the difference between the top panels in this figure and the 
top panel in Fig. \ref{figEvolC28}.}  
\label{figEvolC2_aver}
\end{figure}

Finally, to illustrate that the realizations of the noise (for the same 
noise amplitude) are not substantially different, we also did the
following numerical experiment. We took 
ten different (spatial) realizations of
the randomly distributed perturbation, but all of them with the same
amplitude and then averaged the result. The relevant findings are
presented in Fig. \ref{figEvolC2_aver}. The top panel of the figure
shows the result of the average of the 10 random realizations of the
perturbation (to be compared with the corresponding panel of Fig. 
\ref{figEvolC28}).
The difference between the two panels is shown in the bottom of Fig. 
\ref{figEvolC2_aver},
and is clearly minimal. Hence, for all 
ten realizations of essentially
the same noise amplitude, the manifestation of the instability was
essentially similar. This validates our claim above that the
key element in this deterministic dynamical evolution is the projection
of the perturbation to the dominant unstable eigenmode which dictates
the dynamics of the instability development and its symmetry breaking
manifestations.

Lastly, although, in the present work, we have used the two mode reduction
as merely a tool for identifying the static solutions (and their
bifurcations) for the full spinor system, let us briefly comment
in passing about the dynamical properties of the two-mode model.
Similarly to what has been suggested in \cite{Bergeman_2mode}, we
find (results not shown here) that the dynamical results of the 
two-mode approximation are closer to the ones of the full GP
equation for lower nonlinearities (i.e., closer
to the linear limit) and higher potentials (i.e., a more pronounced
double well structure).

\section{Conclusions}

In this work, we considered the statics and dynamics of a $F=1$ spinor condensate confined 
in a double well potential. We illustrated that the two-mode Galerkin-type approximation,
previously developed for one-component \cite{todd,george} and two-component
\cite{chenyu} settings can be extended to this genuinely three-component
setting. The advantage of such a methodology is that it can offer 
considerable insight on the full set of stationary states that the
system can exhibit, not only as pure states involving one-component, 
or two-component combinations (involving the $\psi_1$ and $\psi_{-1}$ components), but even
fully three-component spinorial states. An additional strength of the
method is that it does not rely on the single mode approximation (SMA) that
necessitates the same spatial profile among all the 
hyperfine components,
but rather it permits to fully explore non-SMA 
states that the system clearly and abundantly possesses. In fact, most of these
states are observed to be dynamically stable in at least a fraction
of parameter space of their existence (i.e., effectively for appropriate
atom numbers) and hence should be accessible to relevant experiments
with spinor condensates in double well potentials. We have observed that
the two-mode Galerkin-type approximation is very efficient in unraveling the full
bifurcation diagram of the possible states. Finally, we have illustrated
the dynamics of either two-component or three-component spinorial 
states in direct simulations, observing typically the emergence
of the symmetry breaking instability, leading to a stronger population
of one or the other well, and subsequent recurrence of such asymmetric
patterns.

There are many directions that open up with respect to this analytical
description of the double well system. On the one hand, one
can extend this two-mode Galerkin-type approximation 
to a higher-number of mode description (that, for instance, will involve 
higher excited states) so that one can describe the possible situations for
higher numbers of atoms/chemical potentials. On the other hand,
one can extend the present theory to the mean-field description
of a $F=2$ spinor condensate 
following, e.g., the description of \cite{uedaspin2} (see also references therein). 
Finally, yet another interesting direction (even for the $F=1$ case) 
is to extend the present considerations to the simplest
higher-dimensional case, involving four wells arranged along
the nodes of a square and attempting a corresponding four-mode
reduction for each of the components. In that setting, it would
be especially interesting to identify multi-pole
structures and topological states, such as vortices, among others.
Such studies are currently in progress and will be reported elsewhere. 

\vspace{5mm}
{\bf Acknowledgements}. PGK acknowledges numerous useful discussions with
Elena Ostrovskaya and Yuri Kivshar. He also gratefully acknowledges
support from NSF-DMS-0505663, NSF-DMS-0619492, NSF-CAREER, NSF-DMS-0806762 
and from the
Alexander von Humboldt Foundation through a Research Fellowship.
The work of DJF was partially supported by the Special Research Account 
of the University of Athens.

\appendix 

\section{Special Case Examples of Solutions} 

In this Appendix we consider some special cases, in which 
the solutions of the algebraic equations of our two-mode reduction
are analytically tractable (or, in any case, reduce to previously 
addressed problems with a lower number of components). More specifically, 
there exist {\it one-component} states which can be found 
in the case where, e.g., $c_{a,b}^{(\pm 1)}=0$. 
In this case, we can readily identify a purely {\it symmetric} state with
\begin{eqnarray}
(c_a^{(0)})^2= \frac{\mu_0-\omega_a}{\nu_s \Gamma_a}, 
\quad c_b^{(0)}=0,
\label{eq13}
\end{eqnarray}
as well as a purely {\it anti-symmetric} state with
\begin{eqnarray}
c_a^{(0)}=0, \quad (c_b^{(0)})^2=\frac{\mu_0-\omega_b}{\nu_s \Gamma_b}. 
\label{eq14}
\end{eqnarray}
Finally, there is also a mixed or {\it asymmetric} state with
\begin{eqnarray}
(c_a^{(0)})^2=\frac{(\mu_0-\omega_a) \Gamma_b-3 (\mu_0-\omega_b) \Gamma_{ab}}{\nu_s (\Gamma_a \Gamma_b-9 \Gamma_{ab}^2)},
\quad 
(c_b^{(0)})^2=\frac{(\mu_0-\omega_b) \Gamma_a-3 (\mu_0-\omega_a) \Gamma_{ab}}{\nu_s (\Gamma_a \Gamma_a-9 \Gamma_{ab}^2)}.
\label{eq15}
\end{eqnarray}
The mixed branch bifurcates beyond a critical value of $\mu_0$ 
from the symmetric state if $\nu_s<0$, or from the antisymmetric
state if $\nu_s>0$ \cite{george}. 
It is important to note here that such pure 
and mixed states exist also in components ${\pm 1}$, with the only
difference in their definition (except for the $0 \rightarrow {\pm 1}$
in the indices above) that $\nu_s \rightarrow \nu_s+\nu_a$.

In addition to the above setting, there is another case 
where these pure and mixed mode states appear (now in
all three components), namely in the spinor BEC description 
in the framework of SMA 
\cite{pulaw,ofy2,osgur} (see also discussion in \cite{dabr}). 
In that case, 
$\mu_0=\mu_1=\mu_{-1}=\mu$ and $u_j=s_j u(x)$, where 
$u$ satisfies the single component equation
\begin{eqnarray}
\mu u= {\cal L} u + \nu_s u^3,
\label{eq16}
\end{eqnarray}
%
while the coefficients $s_j$ containing
the spin degree of freedom are given by, 
%
\begin{eqnarray}
s_1 &=& \cos^2(\frac{\beta}{2}), \quad s_0=\sqrt{2} \cos(\frac{\beta}{2})
\sin(\frac{\beta}{2}), \quad s_{-1}=\sin^2(\frac{\beta}{2})
\label{eq17}
\\
s_1 &=& -\frac{1}{\sqrt{2}} \sin(\beta), \quad s_0=\cos(\beta), 
\quad s_{-1}=\frac{1}{\sqrt{2}} \sin(\beta),
\label{eq18}
\end{eqnarray}
where $\beta$ is a free parameter.  
In the context of SMA, which 
is generally valid if the condensate width is far smaller
than the spin healing length (so that the terms proportional to $\nu_a$
do not substantially influence the dynamics), 
again the possible solutions reduce to the above pure and mixed modes. 

Finally, yet another special case is the one corresponding 
to the existence of {\it two-component} states. In paricular, in the case 
$u_0=0$, the three-component system 
degenerates to the two-component setting recently considered in \cite{chenyu}, 
with self-phase modulation proportional to $\nu_s+\nu_a$
and cross-phase modulation proportional to $\nu_s-\nu_a$.
Based on the analysis of \cite{chenyu}, two-component states, both 
symmetric and 
antisymmetric, but also mixed two-component states are expected to exist. 

\end{document}